\documentclass[fleqn,usenatbib]{mnras}

\usepackage{graphicx}
\usepackage{amsmath,amssymb}
\usepackage{xcolor}
\usepackage{subcaption}
\usepackage{booktabs}
\usepackage{float}

\usepackage{soul}

\setstcolor{red}
\sethlcolor{green!20}
\soulregister\cite7
\soulregister\citep7
\soulregister\citet7

\title[RR Lyrae atmospheric parameters with Mephisto]{Atmospheric Parameter Determination of RR Lyrae Stars With Mephisto Multi-band Photometry}

\author[X. Zou et al.]{
Xingzhu Zou,$^{1,2}$ Maosheng Xiang,$^{3,4}$ Tanda Li,$^{5,6}$ Lanya Mou,$^{7,8}$ Haibo Yuan,$^{5,6}$ Xinlei Chen,$^{1,2}$
\newauthor
Yu Pan,$^{1,2}$ Kaushik Chatterjee,$^{1,2}$ Guowang Du,$^{1,2}$ Yuan Fang,$^{1,2}$ Helong Guo,$^{9,10}$ Brajesh Kumar,$^{1,2}$
\newauthor
Edoardo P. Lagioia,$^{1,2}$ Chenxu Liu,$^{1,2}$ Xiangkun Liu,$^{1,2}$ Tao Wang,$^{1,2}$ Yuan-Pei Yang,$^{1,2}$
\newauthor
Jinghua Zhang,$^{1,2}$ Shiyan Zhong,$^{1,2}$ Xufeng Zhu,$^{1,2}$ and Xiaowei Liu$^{1,2}$\thanks{E-mail: x.liu@ynu.edu.cn}\\
\parbox[t]{\textwidth}{%
$^{1}$South-Western Institute for Astronomy Research, Yunnan University, Kunming, Yunnan 650504, People's Republic of China\newline
$^{2}$Yunnan Key Laboratory of Survey Science, Yunnan University, Kunming, Yunnan 650500, People's Republic of China\newline
$^{3}$CAS Key Laboratory of Optical Astronomy, National Astronomical Observatories, Chinese Academy of Sciences, Beijing 100101, People's Republic of China\newline
$^{4}$C3 Institute for Frontiers in Astronomy and Astrophysics, Beijing Normal University, Beijing 102206, People's Republic of China\newline
$^{5}$Institute for Frontiers in Astronomy and Astrophysics, Beijing Normal University, Beijing 102206, People's Republic of China\newline
$^{6}$School of Physics and Astronomy, Beijing Normal University, No. 19 Xinjiekouwai St., Haidian District, Beijing 100875, People's Republic of China\newline
$^{7}$School of Astronomy and Space Science, University of Chinese Academy of Sciences, Beijing 100049, China\newline
$^{8}$National Astronomical Observatories, Chinese Academy of Sciences, Beijing 100101, China\newline
$^{9}$Research Center of Astronomy, QingHai University, Xining, 810016, PR China\newline
$^{10}$Department of Physics and Astronomy, QingHai University, Xining, 810016, PR China
}
}

\date{Accepted XXX. Received YYY; in original form ZZZ}
\pubyear{2026}

\begin{document}
\label{firstpage}
\pagerange{\pageref{firstpage}--\pageref{lastpage}}
\maketitle

\begin{abstract}
Phase-resolved atmospheric parameters of pulsating stars, including RR Lyrae stars and Cepheids, are essential for studying stellar structure and pulsation physics and for establishing accurate stellar benchmarks for Galactic structure and assembly history. We present a photometric framework that simultaneously infers $T_{\rm eff}$, $\log g$, and [Fe/H] from Mephisto multi-epoch, multi-band ($u, v, g, r, i, z$) photometry. A supervised neural-network forward model is built by training on synthetic colors derived from theoretical stellar atmosphere models. It reproduces the synthetic colors to a precision of 0.01 per cent. The model is applied to observed multi-epoch colors to derive phase-dependent atmospheric parameters.
We apply the method to six RRab and four RRc stars in three globular clusters observed with Mephisto. We find typical per-phase uncertainties of about 62 K in effective temperature ($T_{\rm eff}$), 0.10 dex in surface gravity ($\log g$), and 0.16 dex in metallicity ([Fe/H]). Repeated measurements across the pulsation cycle reduce the typical formal uncertainty of the final [Fe/H] estimates to roughly 0.03 dex. A sensitivity analysis indicates that the five Mephisto colors provide distinct and complementary constraints on the three stellar atmospheric parameters: $(u-v)$ provides robust constraints on $\log g$, $(v-g)$ is informative for estimating [Fe/H], and $(u-v)$, $(v-g)$, $(g-r)$, $(r-i)$, and $(i-z)$ all provide information on $T_{\rm eff}$. The results demonstrate that Mephisto multi-epoch, multi-band photometry can deliver precise, cost-effective phase-resolved atmospheric parameters for RR Lyrae stars, offering an efficient way to construct large samples of stellar benchmarks for studies of stellar pulsation physics and Galactic structure and evolution.

\end{abstract}


\section{Introduction}

RR Lyrae (RRL) stars are old, metal-poor, low-mass, radially pulsating variables located on the horizontal branch \citep{Smith1995,Catelan2009}. They are ubiquitous in old stellar populations, being found in most Galactic globular clusters \citep{Clement2001}, throughout the Galactic halo \citep{Sesar2017}, in the Galactic bulge \citep{Udalski1997}, and in nearby galaxies \citep{Soszynski2009,Clementini2010}. Owing to their well-established period--luminosity--metallicity (PLZ) relation \citep{Catelan2004,marconi2015,bhardwaj2023}, RR Lyrae stars have long served as fundamental distance indicators. Beyond their importance for the cosmic distance scale, they are also valuable tracers of old stellar populations and Galactic structure \citep{CatelanSmith2015,Monelli2022,DOrazi2024,sun_identifying_2024}.

Accurate knowledge of their atmospheric parameters is essential to fully exploit their astrophysical utility. Metallicity ([Fe/H]) is particularly critical, as it enters the PLZ calibration \citep{marconi2015,bhardwaj2023} and thus affects distance determinations. Metallicity determinations for large samples of RR Lyrae stars are also important for constraining the chemical enrichment history of the Milky Way, in particular that of the Galactic halo \citep{Monelli2022,DOrazi2024}.
Furthermore, RR Lyrae atmospheres undergo dramatic changes over the pulsation cycle, with effective temperature ($T_{\rm eff}$) and surface gravity ($\log g$) varying significantly in response to the rapidly evolving physical conditions \citep{Bono1994,Kolenberg2010,Fossati2014}. Simultaneous determination of $T_{\rm eff}$, $\log g$, and [Fe/H] across the pulsation cycle 
therefore provides valuable constraints on their pulsation mechanisms and offers a critical benchmark for stellar atmosphere and pulsation models \citep{Feuchtinger1999,Marconi2009,Smolec2008,Paxton2019ApJS}. 

Traditionally, the atmospheric parameters of RR Lyrae stars have been derived primarily from spectroscopic observations. High-resolution spectroscopy can provide precise measurements of effective temperature, surface gravity, metallicity, and detailed elemental abundances \citep{Clementini1995,Nemec2013,Pancino2015,Chadid2017}. However, obtaining high signal-to-noise (SNR) spectra for rapidly varying targets such as RR Lyrae pulsators is very costly of large telescope-time and requires careful handling of their brightness and atmospheric variations, including possible effects of shock waves \citep{Deng2014, Gillet2019}.
As a result, high-resolution spectroscopic studies have been confined to relatively small samples of RR Lyrae stars \citep{Crestani2021}. In recent years, large-scale spectroscopic surveys \citep{Yanny2009, Majewski2017, Deng2012RAA, Liu2014, DeSilva2015, 2023ApJ...947...37C} have extended low- and medium-resolution spectroscopy to much larger RR Lyrae samples \citep{Liu2020,Crestani2021,Fabrizio2021,wang_stellar_2024, 2025arXiv250402924M}. Nevertheless, the lower spectral resolution  combined with the effects of atmospheric shock waves at certain pulsation phase, introduces substantial additional uncertainties to the derived parameters.

Photometric observations provide an efficient way to estimate stellar atmospheric parameters. For RR~Lyrae stars, metallicity can be estimated from empirical relations involving the pulsation period and the Fourier parameters of their light curves  \citep{Jurcsik1996,Morgan2007,Mullen2021,dekany_photometric_2022,Li2023,Muraveva2025}. Alternatively, multi-band photometry provides atmospheric information through stellar colors, with sensitivities to $T_{\rm eff}$, $\log g$, and [Fe/H] varying substantially with the wavelength coverage and the particular filter system adopted \citep[e.g.][]{Yuan2015,HuangY2019,Casagrande2019,YangL2022,Lu2024,lu_stellar_2025}. By combining colors of complementary sensitivities to different parameters, multi-band photometry can constrain $T_{\rm eff}$, $\log g$, and [Fe/H] simultaneously for large samples of stars. Compared with spectroscopy, photometry is considerably more efficient for obtaining homogeneous parameter estimates for large stellar samples with minimum selection effects. This advantage becomes especially critical for RR Lyrae stars, whose atmospheric parameters vary over time, so their reliable parameter determinations require multi-epoch observations.

The Multi-channel Photometric Survey Telescope (Mephisto) is a dedicated wide-field optical survey telescope designed for simultaneous multi-band photometric observations \citep{Liu2019, Yuan2020}. It employs three optical channels with six broad-band filters ($u$, $v$, $g$, $r$, $i$, and $z$), enabling simultaneous imaging in three bands and providing broad wavelength coverage across the whole optical wavelength range (X.~K.~Liu et al. 2026). In particular, the near-ultraviolet $u$ and $v$ bands are highly sensitive to stellar atmospheric properties, with the $u$ band probing the Balmer jump and the $v$ band containing numerous metal absorption features. Furthermore, the simultaneous acquisition of multi-band photometry minimizes color uncertainties introduced by the rapid brightness variations of RR Lyrae stars, making Mephisto particularly well suited for phase-resolved atmospheric-parameter determinations of short-period pulsating variables such as RR Lyrae stars.

In this work, we develop a photometric framework for estimating atmospheric parameters of RR Lyrae stars using Mephisto multi-band observations. Synthetic colors generated from theoretical stellar atmosphere models are used to train a neural-network model that predicts Mephisto colors for a wide range of stellar atmospheric parameters appropriate for RR Lyrae stars. The atmospheric parameters of RR Lyrae stars are subsequently determined by fitting the observed colors to the model predictions. By constructing phase-resolved color curves, atmospheric parameters are independently inferred at individual pulsation phases, yielding the phase-dependent evolution of $T_{\rm eff}$, $\log g$, and [Fe/H]. The final estimates of metallicity can be further improved by combining the phase-dependent estimates through a weighted average. We focus here on establishing and demonstrating the methodology using a sample of ten RR Lyrae stars with well-sampled Mephisto light curves, while results for a large sample of RR Lyrae stars in a number of Galactic globular clusters study will be presented in a subsequent work. This study demonstrates the feasibility of deriving atmospheric parameters from multi-band photometry and provides a methodological foundation for future large-scale studies of RR Lyrae and other short-period pulsating variables with Mephisto and similar multi-band photometric facilities.

This paper is structured as follows. Section~2 describes the methodology, including the construction of the theoretical dataset of stellar atmospheres, the neural network forward model, and the atmospheric parameter inference procedure. Section~3 presents the application of this framework to ten RR Lyrae stars observed with Mephisto, including the observation and data reduction, the construction of multi-band color curves, and the estimation of phase-dependent atmospheric parameters. Section~4 presents the results and discussion, focusing on the phase-dependent variations of the atmospheric parameters, the sensitivity of Mephisto colors to these parameters, and a comparison of the derived metallicities with literature values. Finally, Section~5 summarizes the main conclusions of this work.

\section{Method}

\subsection{Synthetic colors} \label{sec:Syntheticcolors}
To establish the relationship between Mephisto colors and stellar atmospheric parameters, we construct a synthetic photometric dataset based on synthetic spectra generated from the Kurucz ATLAS12 stellar atmosphere models \citep{Kurucz1979,Kurucz2005}. We generat a uniform set of synthetic spectra covering the parameter space typical for RR Lyrae stars, namely, $T_{\rm eff}=5600$--9600~K with a step size of 100~K, $\log g=1.6$--4.0 dex with a step size of 0.2 dex, and $[\mathrm{Fe/H}]=-2.5$ to $-0.5$ dex with a step size of 0.2 dex. We adopt an enhanced $\alpha$-element (O, Mg, Si, Ca, Ti) architecture for the model spectra by fixing  $[\alpha/\mathrm{Fe}]=0.4$. The resulted dataset contains 5863 synthetic spectra covering the wavelength range of 2000--12000~\AA. The spectra are generated at a resolution of $R=500,000$ and subsequently degraded to $R=600$.

The synthetic spectra are convolved with the transmission curves of the six Mephisto bands ($u$, $v$, $g$, $r$, $i$, and $z$) to obtain synthetic magnitudes in each band. Five base color indices, $(u-v)$, $(v-g)$, $(g-r)$, $(r-i)$, and $(i-z)$, are subsequently computed from the synthetic magnitudes. These synthetic colors are used to train the neural network model that predicts Mephisto colors for a given set stellar atmospheric parameters, $T_{\rm eff}$, $\log g$, and [Fe/H]. Any other color of alternative band combination can be expressed as linear combinations of these five base color indices and therefore do not contain independent information. 

\begin{figure*}
    \centering
    \includegraphics[width=\textwidth]{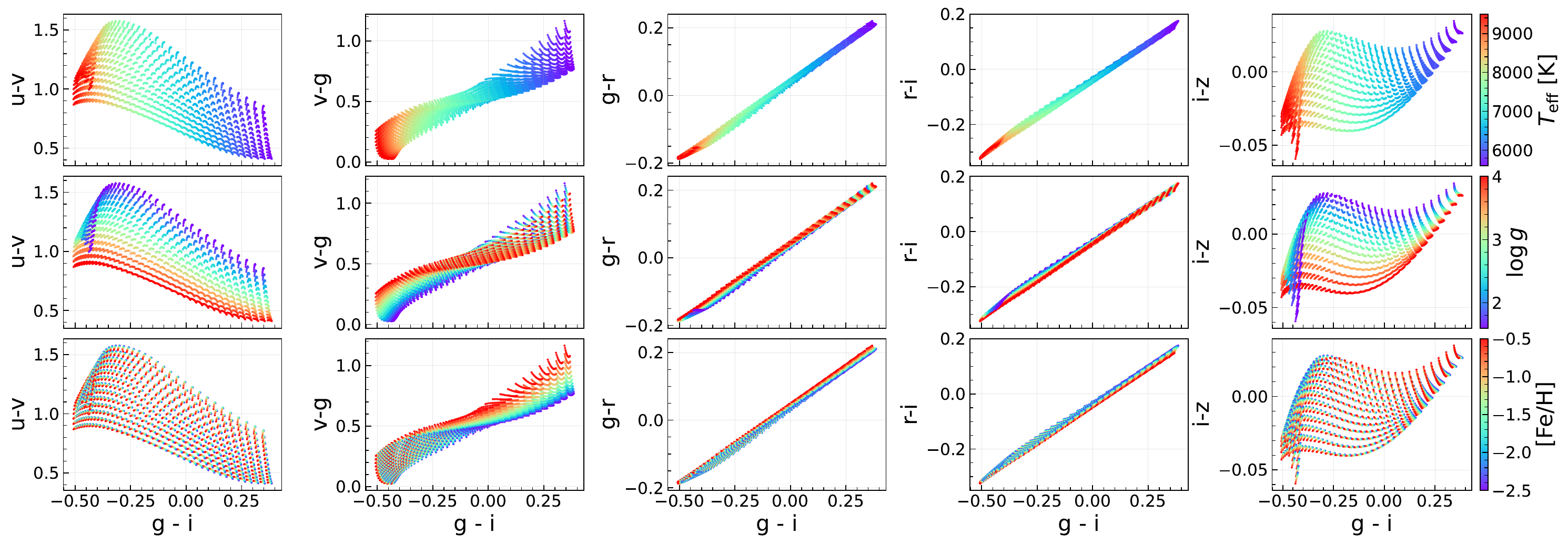}
    \caption{
    Synthetic stellar loci in the $(g-i)$ versus $(u-v)$, $(v-g)$, $(g-r)$, $(r-i)$, and $(i-z)$ color planes. From top to bottom, the loci are coded according to values of $T_{\rm eff}$, $\log g$, and [Fe/H], respectively. The different color gradients illustrate the varying sensitivities of the Mephisto color combinations to the three basic atmospheric parameters.
    }
    \label{fig:color_sensitivity}
\end{figure*}

To investigate the sensitivity of Mephisto colors to stellar atmospheric parameters, we examine the distributions of the synthetic spectra in color--color space. Figure~\ref{fig:color_sensitivity} presents the synthetic stellar loci in the planes of $(g-i)$ versus other selected colors. The three rows are color-coded by varying values of $T_{\rm eff}$, $\log~g$, and [Fe/H], respectively. Different color combinations exhibit different sensitivities to the atmospheric parameters. In particular, $(u-v)$ and $(i-z)$ exhibit the strongest sensitivity to $\log g$, while $(v-g)$ is especially sensitive to metallicity. Effective temperature produces systematic variations for most color combinations. The inclusion of the near ultraviolet filters $u$ and $v$ (the combinations of the two together is roughly equivalent to the SDSS u band) significantly enhances the sensitivity of the Mephisto photometric system to the basic stellar atmospheric parameters. 

\subsection{Neural Network Forward Model}

To enable efficient inference of stellar atmospheric parameters from the observed Mephisto colors, we first construct a neural-network forward model that generates Mephisto colors for any given set of stellar atmospheric parameters. Compared to direct linear interpolation of the theoretical grid, the neural network provides fast and continuous color predictions throughout the wide parameter space of concern.

\begin{figure}
    \centering
    \includegraphics[width=\columnwidth]{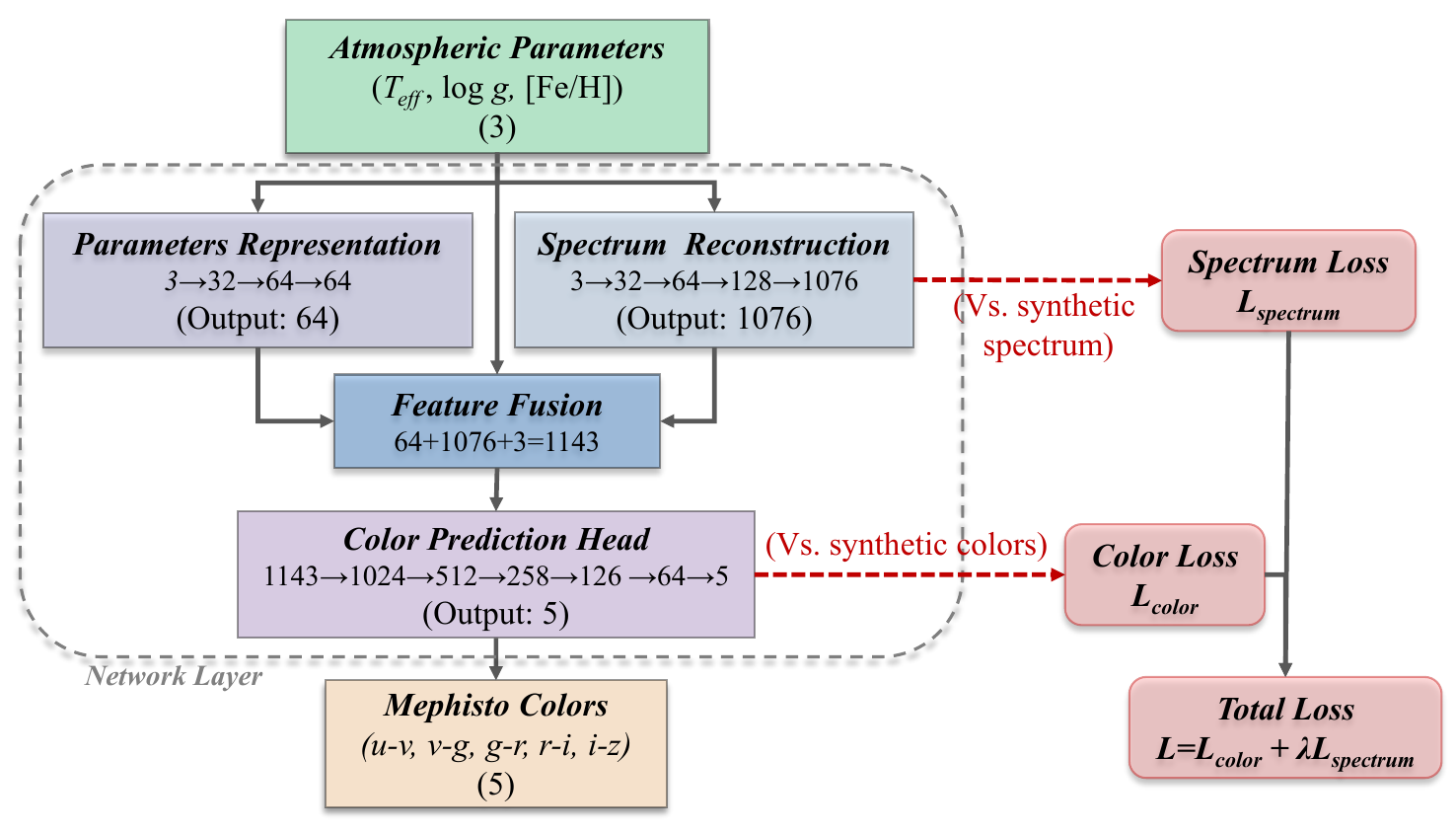}
    \caption{
    Architecture of the adopted neural-network forward model. The atmospheric parameters ($T_{\rm eff}$, $\log g$, and [Fe/H]) are fed into two parallel branches: a parameter representation and a spectrum reconstruction branch. Their learned features are fused to predict the five Mephisto color indices. The network is trained with a joint loss consisting of the color prediction loss and the auxiliary spectrum reconstruction loss.
    }
    \label{fig:nn_arch}
\end{figure}

The overall architecture of the neural-network model is illustrated in Figure~\ref{fig:nn_arch}. The input consists of the three atmospheric parameters, $T_{\rm eff}$, $\log g$, and [Fe/H]. The network contains two parallel branches: a \textbf{parameter representation} branch and a \textbf{spectrum reconstruction} branch. The parameter representation branch extracts a compact nonlinear representation of the atmospheric parameters, while the spectrum reconstruction branch predicts the corresponding synthetic spectrum. Because the stellar spectrum provides the physical basis of the integrated photometric colors and contains richer information than the colors themselves, the spectrum reconstruction task encourages the network to preserve the underlying spectral characteristics rather than learning only the final color indices. The outputs of the two branches are subsequently fused to predict the five Mephisto color indices. The numbers shown in Figure~\ref{fig:nn_arch} indicate the dimensions of the corresponding network layers, and all hidden layers employ ReLU activation functions.

The outputs of the two branches are concatenated with the input atmospheric parameters and passed to a fusion network that predicts the five Mephisto color indices, $(u-v)$, $(v-g)$, $(g-r)$, $(r-i)$, and $(i-z)$. During training, the network is optimized by jointly minimizing the color prediction loss and the spectrum reconstruction loss,

\begin{equation}
L=L_{\rm color}+\lambda L_{\rm spectrum},
\end{equation}

where $\lambda$ is the weight of the auxiliary spectrum reconstruction loss. During inference, only the predicted Mephisto colors are used for the inversion of atmospheric parameters (Section~\ref{sec:AtmosphericParameterInference}).

The neural network is trained using the Adam optimizer with an initial learning rate of $2 \times 10^{-4}$, 
which is reduced by a factor of 0.5 every 1000 epochs via a step learning rate scheduler. The synthetic dataset 
is randomly divided into training and test subsets with a ratio of 8:2. 

We evaluate the prediction performance on the independent test set using the mean residuals and their standard deviations.The residual distributions show that the mean residuals for all colors are smaller than $10^{-4}$~mag. The corresponding standard deviations are $\sigma_{u-v} = 0.0005$, $\sigma_{v-g} = 0.0003$, $\sigma_{g-r} = 0.0002$, $\sigma_{r-i} = 0.0002$, and $\sigma_{i-z} = 0.0001$~mag, respectively.

The auxiliary supervision provided by the spectral branch plays a crucial role in achieving the high accuracy and precision of our color forward model. To quantify this contribution, we perform an ablation study in which this branch is removed while keeping the rest of the architecture. The resulting single-loss model produces color residuals with standard deviations roughly an order of magnitude larger than the full model, confirming that physically informed spectral supervision substantially enhances the accuracy and robustness of the model.

\subsection{Atmospheric Parameter Inference}\label{sec:AtmosphericParameterInference}

Given the five observed colors of an RR Lyrae star, the atmospheric parameters are inferred by minimizing the $\chi^2$ between the observed and model colors,

\begin{equation}
\label{eq}
\chi^2 = \sum_{i=1}^{5}
\frac{\left(c_{\mathrm{model},i} - c_{\mathrm{dered},i}\right)^2}
{\sigma_{\mathrm{model},i}^2 + \sigma_{\mathrm{dered},i}^2},
\end{equation}

where $c_{\mathrm{model},i}$ is the $i$-th color predicted by our neural-network model, and $c_{\mathrm{dered},i}$ is the corresponding extinction-corrected Mephisto color. The model error $\sigma_{\mathrm model}$ is adopted as the standard deviation of the color residuals for the Kurucz test set.

Covariance matrix of the parameter estimates are calculated using the Jacobian provided by PyTorch automatic differentiation,
\[
\mathrm{Cov}=(J^{T}J)^{-1}\chi^2_{\rm red},
\]
where the $\chi^2_{\rm red}$ is the $\chi^2$ divided by the degree of freedom, which is 2 in our case. From this covariance, we derive a $1\sigma$ error for the parameter estimate by taking the square roots of the diagonal elements.


\subsection{Sensitivity of the Mephisto Colors to stellar Atmospheric Parameters}
\label{sec:Sensitivity}

The neural network color model allows us to evaluate how the five Mephisto base colors respond to changes in $T_{\rm eff}$, $\log g$, and [Fe/H] across the adopted parameter space. This sensitivity analysis characterizes the information content for the stellar parameter estimation.

\begin{figure*}
\centering
\includegraphics[width=\textwidth]{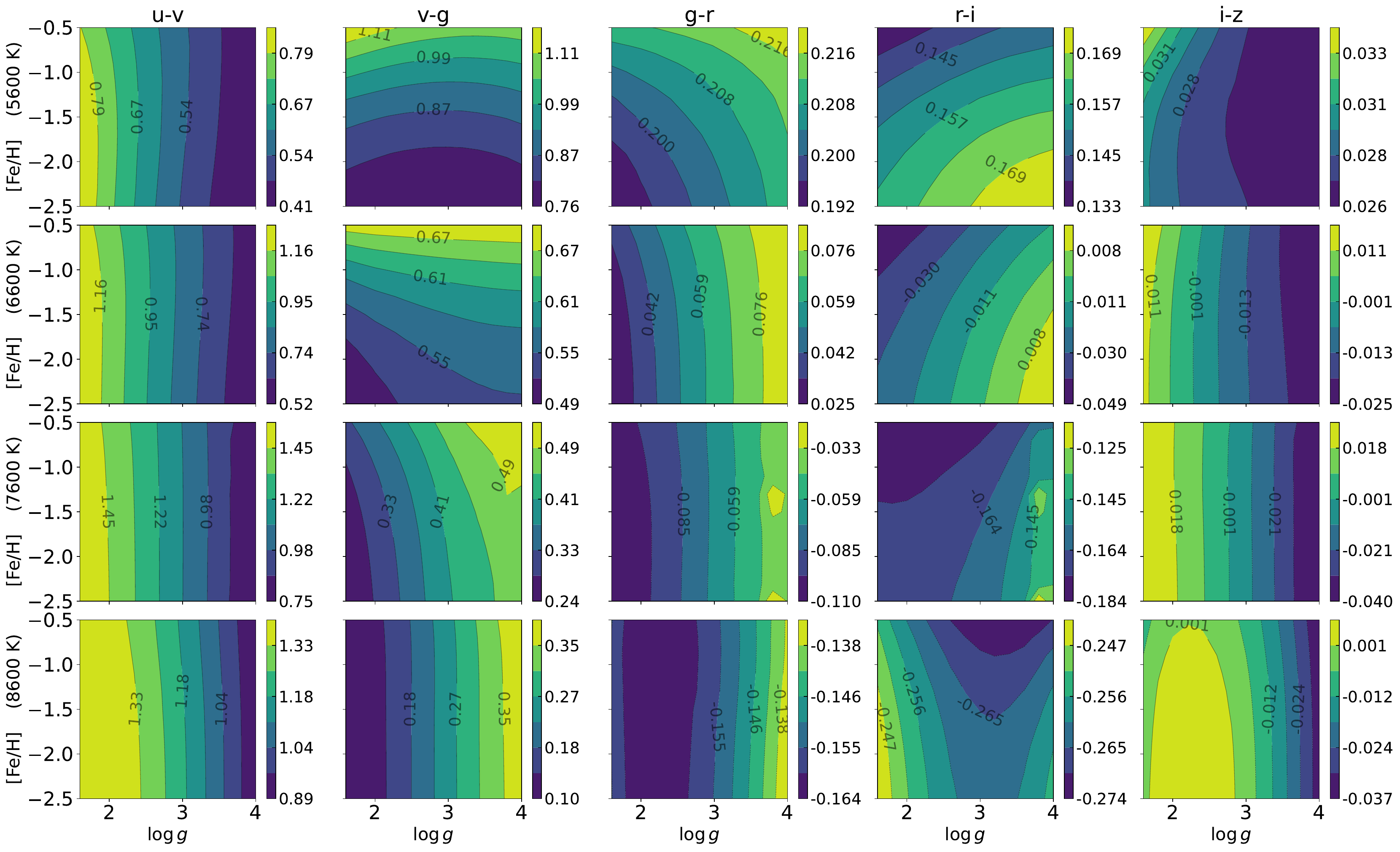}
\caption{
Predicted distributions of the five Mephisto colors in the $\log g$--[Fe/H] plane at four representative effective temperatures. From left to right, the columns show $(u-v)$, $(v-g)$, $(g-r)$, $(r-i)$, and $(i-z)$, while the rows correspond to $T_{\rm eff}=5600$, $6600$, $7600$, and $8600$~K. The contours illustrate the variation of each color as a function of $\log g$ and [Fe/H] for the four representative values of effective temperature.
}
\label{fig:color_parameter_maps}
\end{figure*}

Figure~\ref{fig:color_parameter_maps} shows the color distributions in the $\log g$--[Fe/H] plane at four representative effective temperatures, $T_{\rm eff}=5600$, $6600$, $7600$, and $8600$~K. The five columns correspond to $(u-v)$, $(v-g)$, $(g-r)$, $(r-i)$, and $(i-z)$, respectively. At low temperatures, $(v-g)$ shows a strong dependence on [Fe/H]. This metallicity dependence trend becomes progressively weaker with increasing $T_{\rm eff}$, and essentially vanishes at 8600~K. Color $(u-v)$ retains substantial sensitivity to $\log g$ of all temperatures of concern. These sensitivities are related to the underlying spectral features: the gravity sensitivity of $(u-v)$ is mainly associated with the Balmer discontinuity whose strength is governed by hydrogen opacity and therefore by the electron density, pressure, and ultimately surface gravity $\log~g$, whereas the metallicity sensitivity of $(v-g)$ arises primarily from numerous metal absorption features within the $v$ band. The $(g-r)$, $(r-i)$, and $(i-z)$ colors also exhibit clear sensitivity to $\log g$, although with smaller amplitudes than $(u-v)$. The gravity sensitivity in  $(r-i)$ and $(i-z)$ colors are largely attribute to the Paschen discontinuity, which lies within the $i$ band.

\begin{figure*}
\centering
\includegraphics[width=\textwidth]{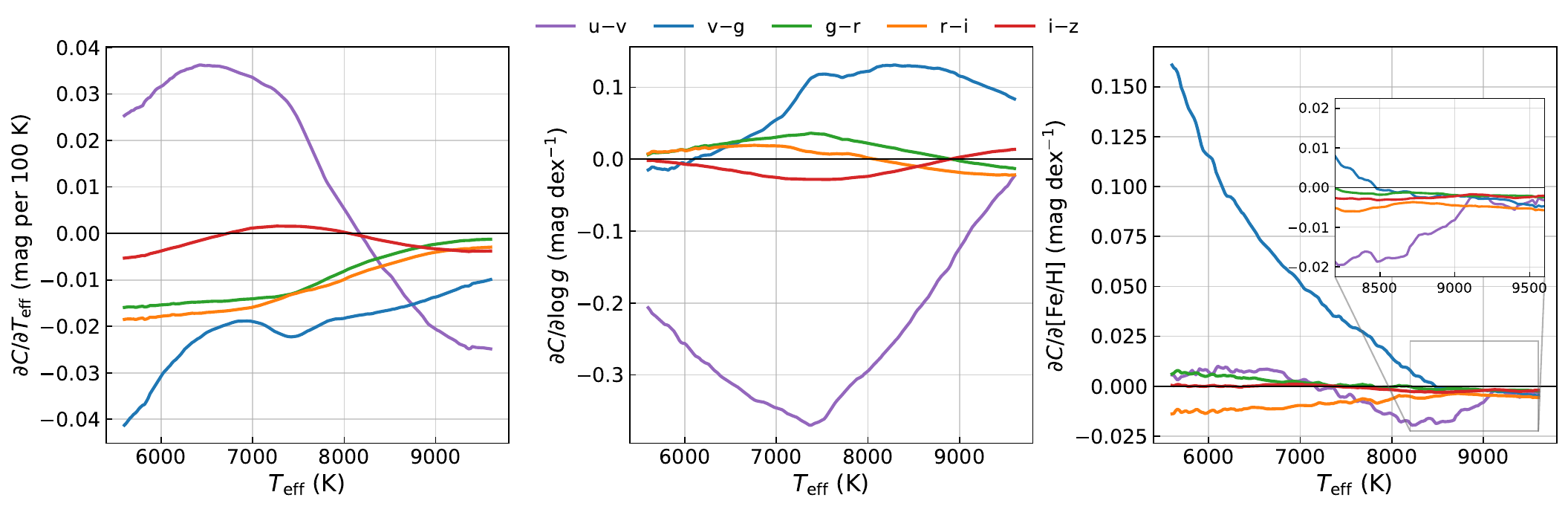}
\caption{Local sensitivities of the five Mephisto colors to the atmospheric parameters as a function of $T_{\rm eff}$. The left, middle, and right panels show $\partial C/\partial T_{\rm eff}$, $\partial C/\partial\log g$, and $\partial C/\partial[\mathrm{Fe/H}]$, respectively. The temperature derivatives are expressed in mag per 100~K, while the gravity and metallicity derivatives are in mag~dex$^{-1}$. The derivatives are estimated using centered finite differences with step sizes of $\Delta T_{\rm eff}=200$~K and $\Delta\log g=\Delta[\mathrm{Fe/H}]=0.2$~dex. The undifferentiated parameters are fixed at $\log g=2.5$~dex and $[\mathrm{Fe/H}]=-1.5$~dex, as applicable.}
\label{fig:color_parameter_derivatives}
\end{figure*}

We further quantify these sensitivities using finite-difference derivatives of the five colors with respect to each atmospheric parameter. Figure~\ref{fig:color_parameter_derivatives} shows the resulting sensitivities as a function of $T_{\rm eff}$, with the other parameters fixed at $\log g=2.5$~dex and $[\mathrm{Fe/H}]=-1.5$~dex. The sign of each derivative indicates whether a color becomes redder or bluer as the corresponding parameter increases, while its absolute value measures the strength of the local response.

The five Mephisto colors all show measurable responses to variation of $T_{\rm eff}$, although the magnitude and sign of the derivatives vary with temperature. For $\log g$, $(u-v)$ maintains a substantial sensitivity across the whole temperature range, although its response gradually decreases as $T_{\rm eff}$ increases, with $\left|\partial(u-v)/\partial\log g\right|$ reaching nearly $0.4~{\rm mag~dex^{-1}}$. At higher temperatures, $(v-g)$ also shows a measurable response to variation of $\log g$, with $\left|\partial(v-g)/\partial\log g\right|$ reaching up to $\sim0.1~{\rm mag~dex^{-1}}$. Together with the temperature sensitivity of the five colors, those responses provide sufficient information to constrain $T_{\rm eff}$ and $\log g$ jointly.

The metallicity sensitivity shows a more pronounced temperature dependence. At the cooler end of the adopted parameter range, $(v-g)$ provides the strongest response to variation of [Fe/H], with $\partial(v-g)/\partial[\mathrm{Fe/H}]$ decreasing from approximately $0.15~{\rm mag~dex^{-1}}$ at $T_{\rm eff}\sim5600$~K to nearly zero around $T_{\rm eff}\sim8400$~K. The other colors have substantially smaller and more structured metallicity derivatives; for example, $(u-v)$ changes sign over the temperature range. At high effective temperatures, the absolute metallicity derivatives of all five colors decrease to only a few $10^{-3}~{\rm mag~dex^{-1}}$. Consequently, changes in [Fe/H] produce only small displacements in the Mephisto color space in this temperature regime, making the metallicity constraint more susceptible to small photometric uncertainties and to degeneracies with $T_{\rm eff}$ and $\log g$.

This temperature dependence is mainly caused by the weakening of metal absorption features in the $v$ band due to increased ionization at high temperature. Consequently, [Fe/H] becomes more difficult to constrain at high temperatures, even when $T_{\rm eff}$ and $\log g$ remain well constrained. This effect is particularly relevant to the phase-dependent [Fe/H] measurements in Section~\ref{sec:AtmosphericParameters}, where the temperature deduced for the hottest phases show much larger scattering with bigger uncertainties.

\section{Application to Mephisto RR Lyrae} 

\subsection{Mephisto Observations and Data Reduction}

The photometric data of RR Lyrae stars used in this work were obtained with Mephisto during the pilot surveys of seven globular clusters (X.~Z.~Zou et al. 2026, hereafter Paper I). Galactic globular clusters are well-studied stellar systems that provide valuable laboratories for investigating old stellar populations and stellar evolution \citep{Renzini1988}. Their relatively well-constrained chemical properties also provide useful references for validating the metallicities derived from the current photometric method.
The classifications and pulsation periods of the RR~Lyrae stars in these clusters were adopted from Gaia DR3 \citep{clementini_gaia_2022}. We identified approximately 500 RR~Lyrae stars in total in the seven clusters. To demonstrate the methodology, we focus here on ten representative RR~Lyrae stars with high SNR ratios and well-sampled multi-band light curves, including both RRab and RRc types, in three clusters: M3, M15, and NGC~5466. A comprehensive analysis of the full RR~Lyrae sample will be presented in a forthcoming paper. Basic information of the selected RR~Lyrae stars, including their host clusters, pulsation periods, types, and the number of observations in the individual Mephisto bands, is summarized in Table~\ref{tab:rrl_sample}. 

\begin{table*}
\centering
\caption{Basic information of selected RR~Lyrae stars}
\label{tab:rrl_sample}
\begin{tabular*}{\textwidth}{@{\extracolsep{\fill}}cclcccccccccc@{}}
\hline\hline
ID & Cluster & Gaia DR3 ID & Type & $P$ (d) & $N_u$ & $N_v$ & $N_g$ & $N_r$ & $N_i$ & $N_z$ \\
\hline
1 & M15 & 1745936435268587136 & RRab & 0.5757 & 213 & 190 & 690 & 622 & 200 & 189 \\
2 & M3 & 1454783149183752576 & RRab & 0.5237 & 251 & 261 & 467 & 471 & 166 & 163 \\
3 & M3 & 1454796927433681152 & RRab & 0.5013 & 251 & 260 & 467 & 472 & 165 & 165 \\
4 & M3 & 1454874717884070400 & RRab & 0.6735 & 250 & 259 & 466 & 470 & 166 & 164 \\
5 & M3 & 1454798782864780288 & RRab & 0.5148 & 251 & 261 & 467 & 471 & 166 & 164 \\
6 & NGC5466 & 1452625632488460288 & RRab & 0.7034 & 228 & 224 & 220 & 203 & 125 & 117 \\
7 & M3 & 1454878493160635136 & RRc & 0.3141 & 250 & 259 & 428 & 393 & 165 & 164 \\
8 & M3 & 1454781053240023680 & RRc & 0.3498 & 251 & 261 & 467 & 470 & 162 & 164 \\
9 & NGC5466 & 1452625765630871296 & RRc & 0.4218 & 228 & 224 & 220 & 203 & 124 & 117 \\
10 & NGC5466 & 1452626216604020992 & RRc & 0.3416 & 225 & 221 & 217 & 202 & 123 & 117 \\
\hline
\end{tabular*}
\end{table*}

Multi-epoch observations of the selected clusters were obtained in the six Mephisto photometric bands at irregular intervals during the Mephisto pilot surveys (December 2022 to September 2025). Owing to the three-channel camera design of Mephisto, the observations were carried out simultaneously in either the $u$, $g$, and $i$ bands or the $v$, $r$, and $z$ bands. The exposure times were optimized for each cluster and filter to achieve sufficient SNR for RR~Lyrae stars while avoiding saturation of the brightest sources. Since the red camera covering the $i$ and $z$ bands was installed of a later stage (after December 2023) of instrument commissioning, the number of observations in these two bands are smaller than those obtained in the other four bands. Further details of the observations and data reduction are provided in  Paper I.

The raw images were processed using the standard Mephisto reduction pipeline (Y.~Fang et al. 2026), including image preprocessing, artifact removal, astrometric calibration, and photometric calibration. The typical calibration precisions for bright stars of approximately $0.010$, $0.011$, $0.010$, $0.009$, $0.011$, and $0.012$~mag in the $u$, $v$, $g$, $r$, $i$, and $z$ bands, respectively (J.~H.~Zhang et al. 2026). Since the ten RR~Lyrae stars presented in this work are sufficiently isolated, not affected by the field crowding, aperture photometry with aperture correction was adopted. For isolated sources, the aperture-corrected photometry provides better consistency among repeated observations while maintaining accurate photometric calibration.
The calibrated magnitudes were subsequently corrected for interstellar extinction using the reddening values $E(B-V)$, adopted from the \citet{Harris1996} catalog (2010 edition) together with the extinction law of \citet{Fitzpatrick1999} with $R_V=3.1$. The resulting data products are calibrated, extinction-corrected multi-band light curves for the selected RR Lyrae stars.

\subsection{Construction of Multi-band Color Curves} \label{sec:ColorCurves}

\begin{figure*}
    \centering
    \begin{subfigure}[t]{0.49\textwidth}
        \centering
        \includegraphics[width=\linewidth]{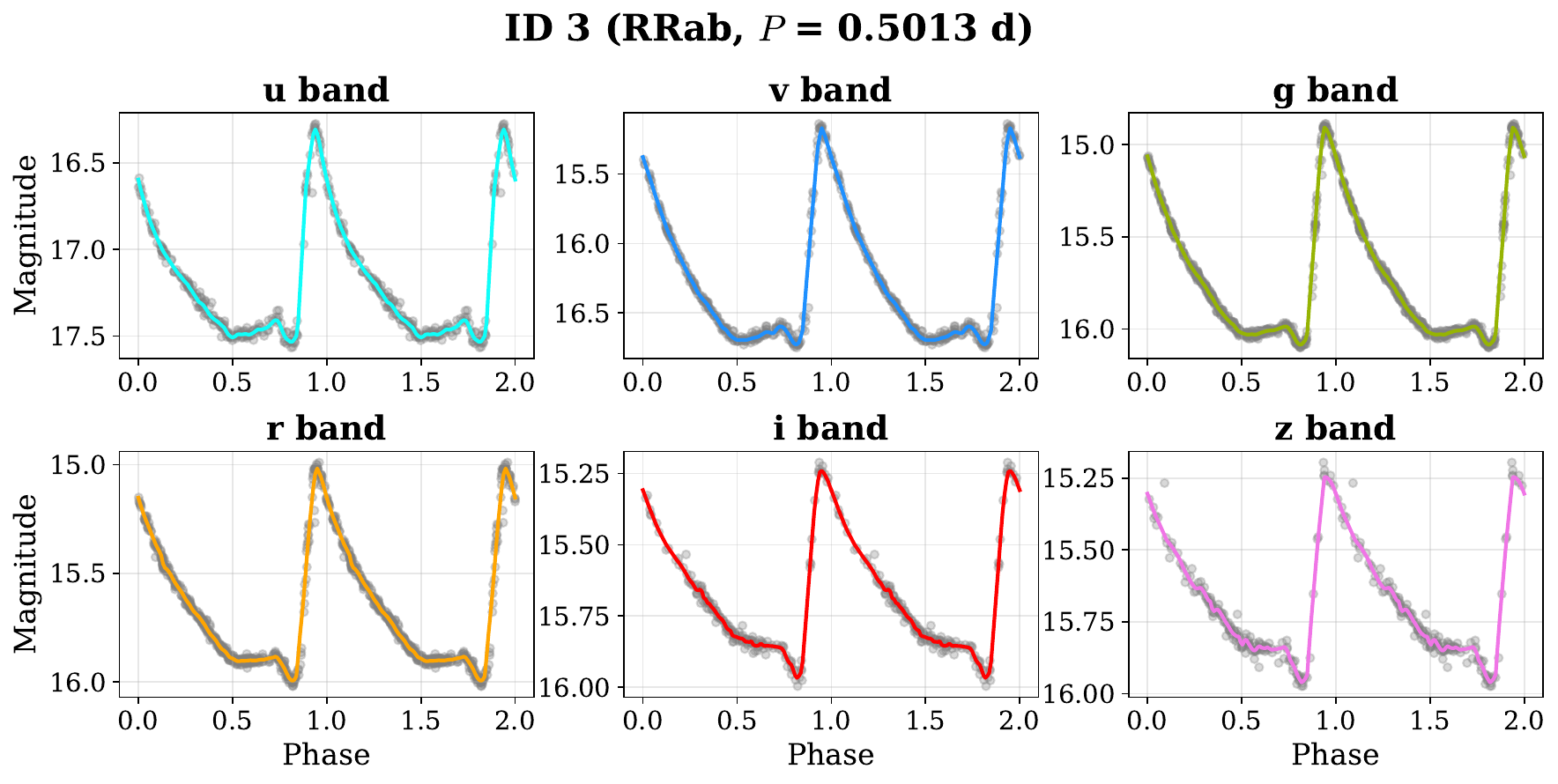}
    \end{subfigure}
    \hfill
    \begin{subfigure}[t]{0.49\textwidth}
        \centering
        \includegraphics[width=\linewidth]{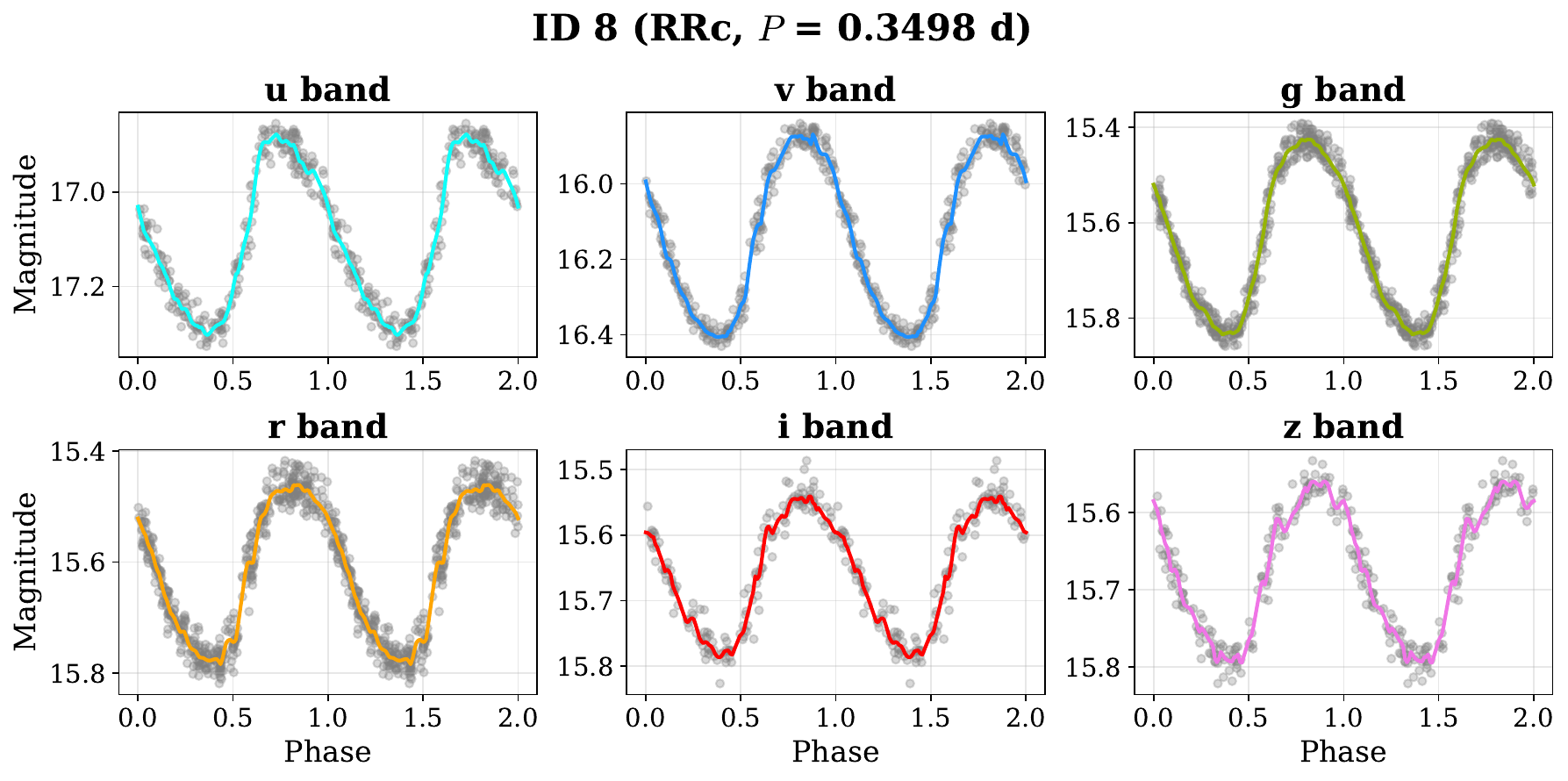}
    \end{subfigure}

    \caption{Representative multi-band light-curve templates constructed using the LOWESS smoothing method. Each figure contains six panels corresponding to the $u$, $v$, $g$, $r$, $i$, and $z$ bands. Gray points denote the extinction-corrected observations, while the colored solid curves show the corresponding LOWESS templates used to interpolate the light curves onto a common phase grid.}
    \label{fig:lightcurve_templates}
\end{figure*}

The atmospheric parameter inference described in Section~2 requires phase-dependent multi-band color indices as input. Although Mephisto provides simultaneous observations in three photometric bands ($u$, $g$, and $i$ or $v$, $r$, and $z$), the five color indices adopted in this work involve magnitudes from both observing configurations. Consequently, the corresponding magnitudes are generally measured at different pulsation phases and cannot be directly combined to derive accurate colors. To obtain color indices at common pulsation phases, continuous light-curves were first constructed independently for each photometric band and subsequently used to derive phase-dependent color curves.
For each photometric band, the extinction-corrected light curve was folded using the pulsation period adopted from the Gaia catalog. The folded light curves were then smoothed using the locally weighted scatterplot smoothing (LOWESS) algorithm \citep{Cleveland01121979}, which is robust against irregular sampling and outliers. The resulting smooth light-curves (LOWESS templates) were interpolated onto a uniform phase grid consisting of 100 equally spaced phase points. Figure~\ref{fig:lightcurve_templates} shows the multi-band light-curve templates for one RRab star (IDs 3) and one RRc star (IDs 8) as representatives. The resulting templates are used to derive the phase-dependent color curves. For visualization purposes, one pulsation cycle is repeated over two consecutive cycles.

The uncertainty of the smoothed light-curves at the individual phase point are estimated from the residuals between the observed magnitudes and the corresponding LOWESS templates. A robust scatter estimate based on the median absolute deviation (MAD) was adopted,

\[
\sigma
=
1.4826
\times
{\rm median}
(|r_i-\mathrm{median}(r)|),
\]

where r denotes the residuals of the observation with respect to the LOWESS template.

\begin{figure*}
    \centering
    \includegraphics[width=\textwidth]{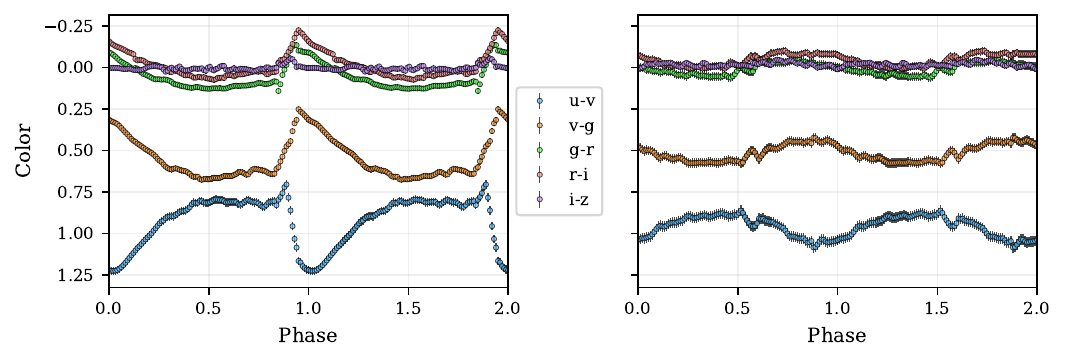}
    \caption{
    Phase-dependent color curves for two representative RR~Lyrae stars (IDs 3 and 8).
    The five curves represent the color indices
    $(u-v)$, $(v-g)$, $(g-r)$, $(r-i)$, and $(i-z)$ derived from the
    multi-band light-curve templates.
    }
    \label{fig:color_curves}
\end{figure*}

The phase-dependent color curves were then constructed by subtracting the template magnitudes of adjacent bands at each phase grid point, yielding the five color indices, $(u-v)$, $(v-g)$, $(g-r)$, $(r-i)$, and $(i-z)$. Their uncertainties were estimated by propagating the template uncertainties. The resulting phase-dependent color curves for the two representative RR Lyrae stars are shown in Figure~\ref{fig:color_curves}. 

\subsection{Atmospheric Parameter Estimation}

The phase-dependent color curves derived in the previous Section \ref{sec:ColorCurves} were used as input to the atmospheric parameter inference modle described in Section~2. At each phase grid point, the five color indices were fitted simultaneously to determine the three basic atmospheric parameters, $T_{\rm eff}$, $\log g$, and [Fe/H]. Repeating this procedure over the full phase grid yielded the phase-dependent atmospheric parameter curves for each RR~Lyrae star.

Across the sample, the median uncertainties at individual phase points are approximately 62~K and 0.10~dex for $T_{\rm eff}$ and $\log g$, respectively. These values represent the typical precision of the inferred $T_{\rm eff}$ and $\log g$ at individual pulsation phases.

\begin{figure*}
    \centering
    \begin{minipage}{0.49\textwidth}
        \centering
        \includegraphics[width=\linewidth]{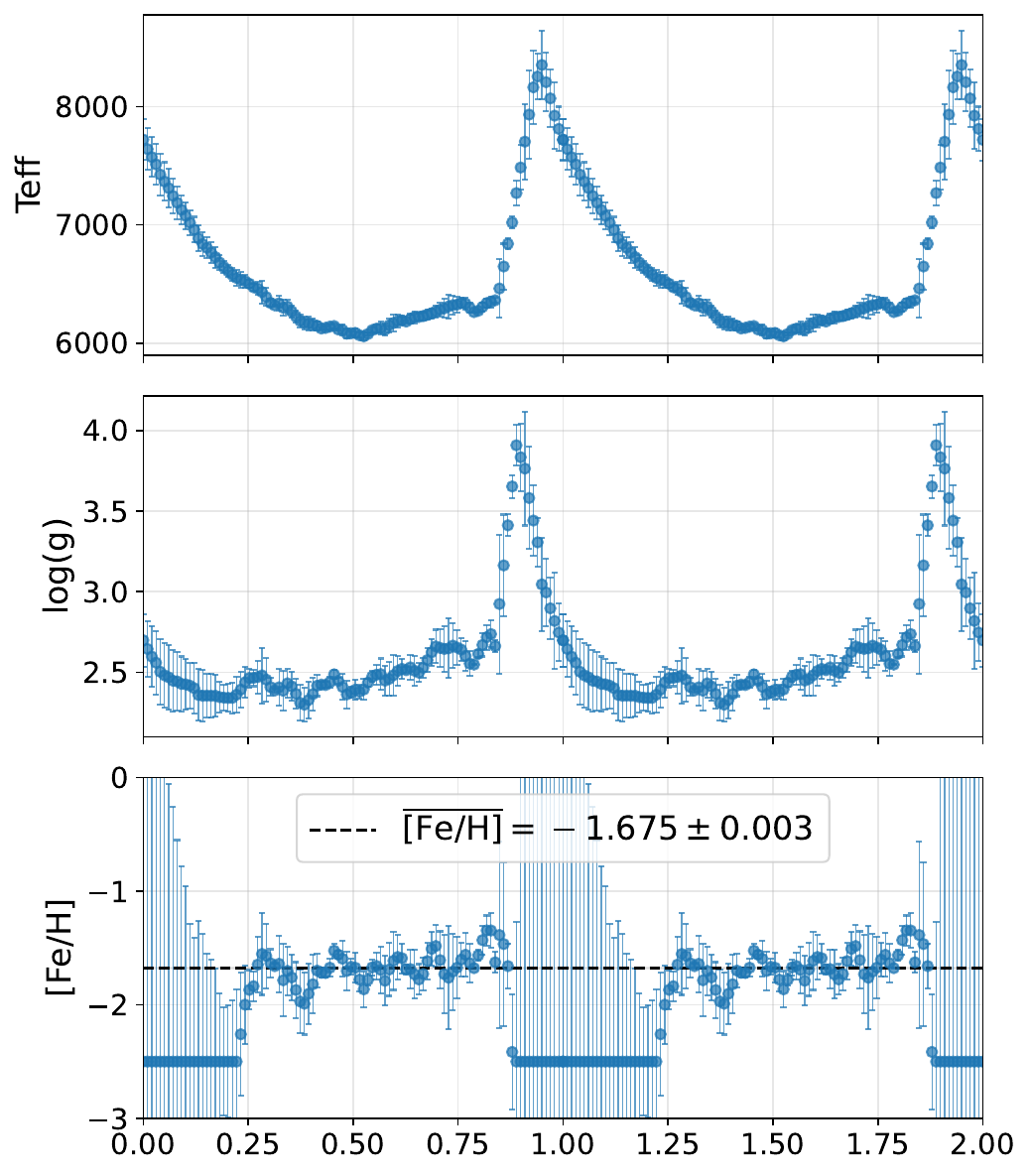}
    \end{minipage}
    \hfill
    \begin{minipage}{0.49\textwidth}
        \centering
        \includegraphics[width=\linewidth]{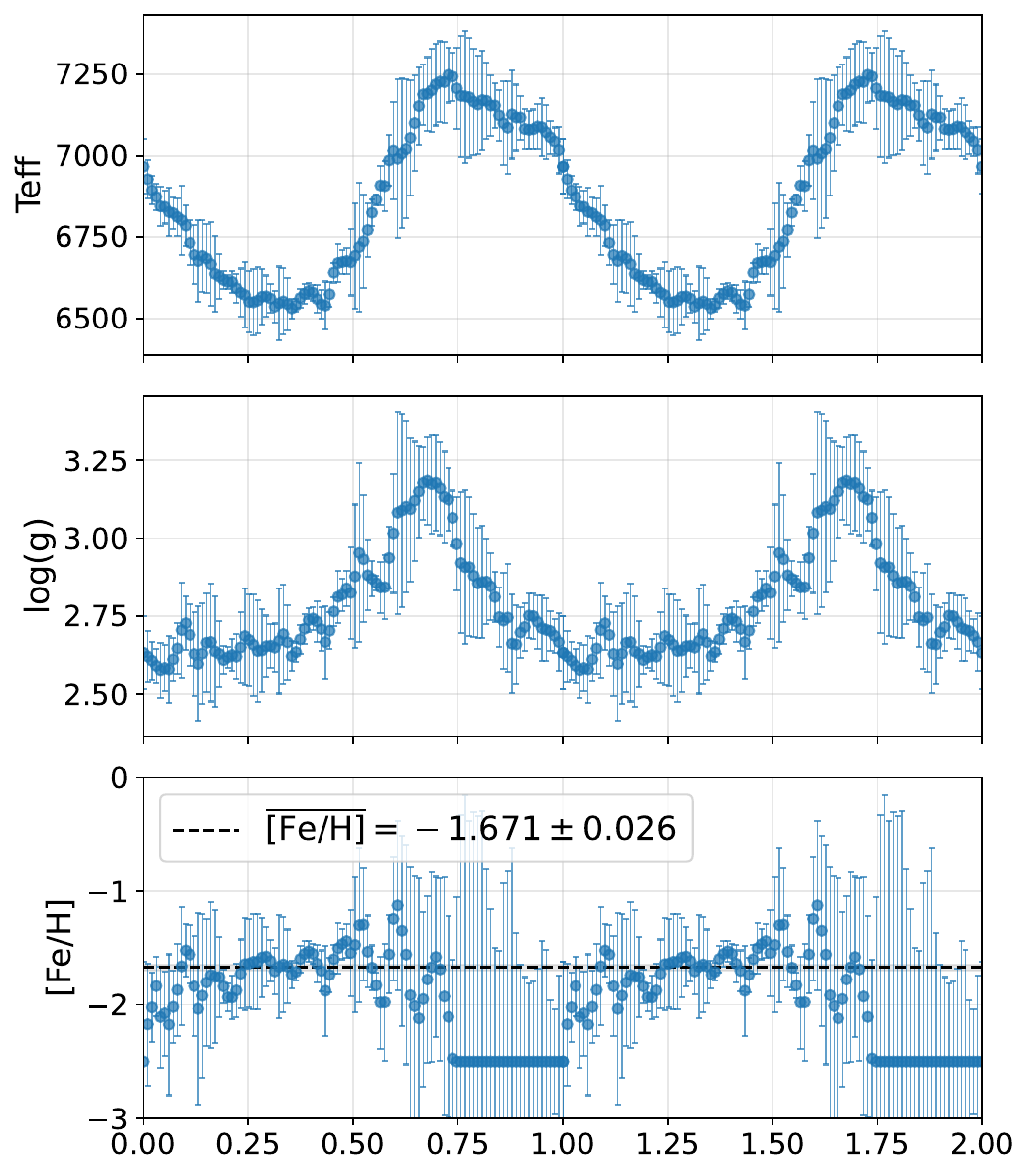}
    \end{minipage}
    
    \caption{
    Phase-dependent atmospheric parameters inferred for two out of the ten representative RR~Lyrae stars (IDs 3 and 8).
    From top to bottom, the panels show the variations of $T_{\rm eff}$, $\log g$, and [Fe/H] as a function of pulsation phase, with the corresponding uncertainties.
    The horizontal line and shaded region in the [Fe/H] panel indicate the inverse-variance weighted mean metallicity and its uncertainty (the shaded region is too narrow to be visible), respectively.
    }
    \label{fig:param_curve}
\end{figure*}

Figure~\ref{fig:param_curve} presents the inferred atmospheric parameters for two representative RR~Lyrae stars (IDs 3 and 8) as a function of pulsation phase. The three panels show the phase-dependent variations of $T_{\rm eff}$, $\log g$, and [Fe/H], respectively, with their corresponding uncertainties. Unlike $T_{\rm eff}$ and $\log g$, the metallicity of an RR~Lyrae star is not expected to vary over the observational timespan. The phase-resolved [Fe/H] estimates are therefore combined to obtain an inverse-variance weighted mean for each star. In the bottom panel, the horizontal line and shaded region indicate the weighted mean and its formal uncertainty, respectively. The uncertainties are 0.003 and 0.026 dex for IDs 3 and 8, respectively, making the shaded regions too narrow to be clearly visible. The light curves and LOWESS templates, color curves, and atmospheric-parameter curves for the other eight stars (IDs~1, 2, 4--7, 9, and 10) are presented in Appendix~\ref{app:other_stars}.

For each RR~Lyrae star, only phase points satisfying the adopted quality criteria were retained for the metallicity estimation. We required that the data yield successful fits, with the reduced $\chi^2$ values falling within $5$ robust standard deviations of the median, and the resulting [Fe/H] values falling within the adopted parameters space (\ref{sec:Syntheticcolors}). The retained phase-dependent [Fe/H] measurements were then combined to give an inverse-variance weighted mean,
\begin{equation}
\label{eq:weighted_mean}
\overline{[\mathrm{Fe/H}]}
=
\frac{\displaystyle\sum_{i=1}^{N}
w_i[\mathrm{Fe/H}]_i}
{\displaystyle\sum_{i=1}^{N}w_i},
\end{equation}
where $w_i=1/\sigma_i^2$ is the inverse-variance weight and $\sigma_i$ denotes the uncertainty of the metallicity inferred at the $i$-th phase point. The formal uncertainty of the weighted mean was calculated as
\begin{equation}
\label{eq:weighted_mean_error}
\sigma_{\overline{[\mathrm{Fe/H}]}}
=
\left(\sum_{i=1}^{N}\frac{1}{\sigma_i^2}\right)^{-1/2}.
\end{equation}
The weighted-mean metallicity was adopted as the final [Fe/H] estimate of the RR~Lyrae star concerned.

The resulting metallicity estimates are summarized in Table~\ref{tab:feh_results}, which lists the number of retained phase points, the inverse-variance weighted mean metallicity $\overline{[\mathrm{Fe/H}]}$, its formal uncertainty $\sigma_{\overline{[\mathrm{Fe/H}]} }$, and the weighted standard deviation of the retained phase-dependent [Fe/H] measurements, $\mathrm{Std}([\mathrm{Fe/H}])$. The latter is used to characterize the phase-to-phase scatter in the inferred metallicity. We adopt the median $\mathrm{Std}([\mathrm{Fe/H}])$ across the sample, approximately $0.16$~dex, as an empirical estimate of the typical uncertainty in the individual phase-dependent [Fe/H] measurements. There is no clear correlation between $\sigma_{\overline{[\mathrm{Fe/H}]}}$ and $\mathrm{Std}([\mathrm{Fe/H}])$, as they depend in a complex manner on several factors including the observational uncertainties, the number of observational epochs, metallicity, and effective temperature.

\begin{table}
\centering
\caption{
Final metallicity estimates for the analyzed RR~Lyrae stars.
$N_{\rm valid}$ denotes the number of phase points retained after the quality selection.
$\overline{[\mathrm{Fe/H}]}$ and $\sigma_{\overline{[\mathrm{Fe/H}]}}$ are the inverse-variance weighted mean metallicity and its formal uncertainty $\sigma_{\overline{[\mathrm{Fe/H}]}}$, respectively.
The last column gives the weighted standard deviation of the retained phase-dependent [Fe/H] measurements $\mathrm{Std}([\mathrm{Fe/H}])$, characterizing the phase-to-phase scatter.
}
\label{tab:feh_results}
\begin{tabular}{cccccc}
\hline
ID & Type & $N_{\rm valid}$ & $\overline{[\mathrm{Fe/H}]}$ & $\sigma_{\overline{[\mathrm{Fe/H}]}}$ & $\mathrm{Std}([\mathrm{Fe/H}])$ \\
\hline
1 & RRab & 66 & $-1.995$ & $0.029$ & 0.216 \\
2 & RRab & 71 & $-1.604$ & $0.006$ & 0.063 \\
3 & RRab & 64 & $-1.675$ & $0.003$ & 0.026 \\
4 & RRab & 99 & $-1.481$ & $0.008$ & 0.100 \\
5 & RRab & 62 & $-1.643$ & $0.009$ & 0.198 \\
6 & RRab & 64 & $-2.005$ & $0.031$ & 0.171 \\
7 & RRc & 50 & $-1.628$ & $0.063$ & 0.223 \\
8 & RRc & 70 & $-1.671$ & $0.026$ & 0.157 \\
9 & RRc & 26 & $-2.168$ & $0.063$ & 0.143 \\
10 & RRc & 33 & $-1.953$ & $0.051$ & 0.301 \\
\hline
\end{tabular}
\end{table}

\section{Results and Discussion} 

\subsection{Phase-dependent Atmospheric Parameters}\label{sec:AtmosphericParameters}

The inferred atmospheric parameters $T_{\rm eff}$ and $\log g$ show clear and systematic variations over the pulsation cycle (Figure~\ref{fig:param_curve}), with median peak-to-peak amplitudes across the ten-star sample of approximately $1020$~K and $0.80$~dex, respectively. The peak-to-peak amplitudes range from $422$ to $2198$~K for $T_{\rm eff}$ and from $0.39$ to $1.61$~dex for $\log g$. The variations demonstrate that the Mephisto multi-band photometry can trace the changing atmospheric state of an RR~Lyrae star throughout the pulsation cycle. The median values and full phase-dependent ranges for individual stars are summarized in Table~\ref{tab:teff_logg}.

\begin{table}
\centering
\caption{Median $T_{\rm eff}$ and $\log g$, together with their phase-dependent ranges, for the analyzed RR~Lyrae stars.}
\label{tab:teff_logg}
\begin{tabular}{cclcccc}
\toprule
\multicolumn{3}{c}{} & \multicolumn{2}{c}{$T_{\rm eff}$ (K)} & \multicolumn{2}{c}{$\log g$ (dex)} \\
\cmidrule(lr){4-5} \cmidrule(lr){6-7}
ID & Type & & Median & Range & Median & Range \\
\midrule
1  & RRab &  & 6290 & 5983--7495 & 2.48 & 2.25--3.42 \\
2  & RRab &  & 6252 & 6017--7877 & 2.42 & 2.28--3.14 \\
3  & RRab &  & 6321 & 6057--8255 & 2.47 & 2.30--3.91 \\
4  & RRab &  & 6182 & 6048--6470 & 2.45 & 2.34--2.73 \\
5  & RRab &  & 6256 & 6029--7686 & 2.45 & 2.29--3.41 \\
6  & RRab &  & 6017 & 5858--6863 & 2.20 & 2.00--3.46 \\
7  & RRc  &  & 7125 & 6685--7725 & 2.89 & 2.70--3.35 \\
8  & RRc  &  & 6834 & 6531--7248 & 2.71 & 2.58--3.19 \\
9  & RRc  &  & 6635 & 6249--7103 & 2.54 & 2.32--2.93 \\
10 & RRc  &  & 6844 & 6448--7430 & 2.65 & 2.40--3.14 \\
\bottomrule
\end{tabular}
\end{table}

\begin{figure}
\centering
\includegraphics[width=\columnwidth]{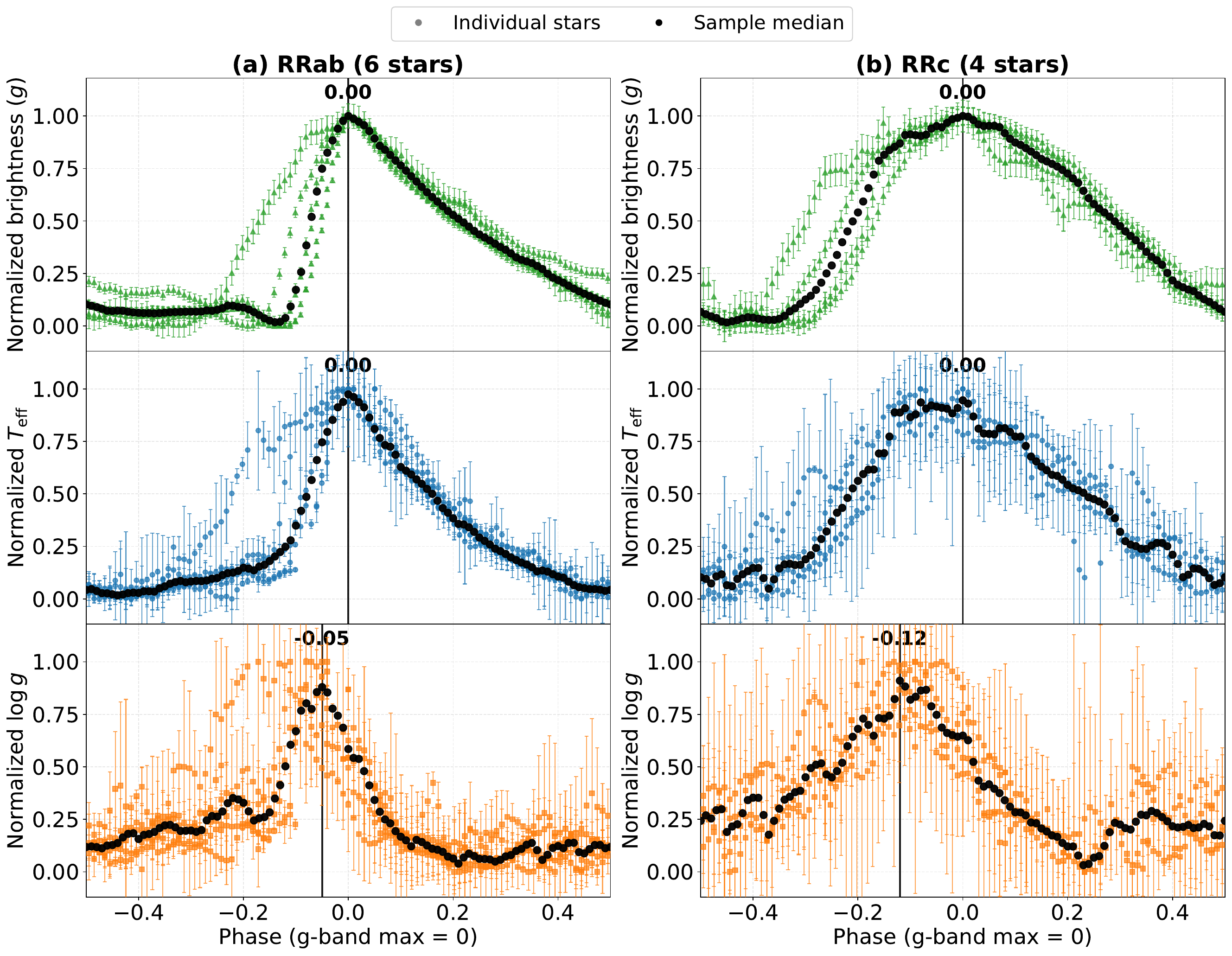}
\caption{Phase-aligned and normalized pulsation curves for the six RRab (left) and four RRc (right) stars. From top to bottom, the panels show the $g$-band light curve, $T_{\rm eff}$, and $\log g$, respectively. The black curves show the median trends, with the maximum marked by a vertical line along with the corresponding phase value.}
\label{fig:phase_aligned_atmosphere}
\end{figure}

A notable feature is the relative phase shifts between $T_{\rm eff}$, $\log g$, and the light curve. As shown in Figure~\ref{fig:phase_aligned_atmosphere}, the maximum $T_{\rm eff}$ occurs close to maximum light, consistent with the results of previous hydrodynamical modeling and high-resolution spectroscopy \citep{Kolenberg2010,Fossati2014}. The phase-dependent $\log g$ follows a similar overall trend, with its maximum generally occurring slightly earlier than that of $T_{\rm eff}$ in both RRab and RRc stars.

The phase shifts are clearly related to the dynamical behavior of the pulsating atmosphere. Previous studies indicate that the minimum stellar radius occurs before the maximum light \citep{Kolenberg2010,Fossati2014}. However, surface gravity inferred for a pulsating atmosphere represents the effective value, contributed by the mass of the star as well as by acceleration of the pulsating atmosphere. The observed phase shifts of $\log g$ therefore likely result from the combined effects of the variations in stellar radius and the dynamical response of the atmosphere. Clearly, the phase-dependent atmospheric parameters as presented here should provide useful observational constraints on the theoretical models of pulsating RR~Lyrae stars. Such a quantitative analysis will be presented in future work. 

The uncertainties of the inferred [Fe/H] increase markedly for phases of high $T_{\rm eff}$. This primarily arises because the sensitivity of the Mephisto colors to [Fe/H] declines considerably at higher temperatures, as discussed in Section~\ref{sec:Sensitivity}. Another possible contributing factor is the dynamical nature of RR~Lyrae atmospheres during these phases of high temperature. In particular, shock waves, atmospheric compression, and strong infalling motions can modify the atmospheric structure and emergent flux distribution, potentially affecting the colors and hence the inferred [Fe/H] \citep{Gillet2019,wang_stellar_2024}.

Among the retained phase points, the inferred phase-dependent [Fe/H] measurements remain largely constant, as expected for a chemically stable stellar atmosphere, although some phase-to-phase scattering can be seen. The scattering is unlikely to be caused by the intrinsic changes in the stellar atmospheric metallicity, but rather arises from a combination of uncertainties in the photometric measurements and calibration, and in the construction of the phase-dependent color curves. In addition, the limitations of the forward-model inversion and potentially phase-dependent atmospheric effects may also contribute to the scattering.

We assume that the observed scattering in the phase-dependent measurements of metallicity mainly arise from the measurements and calibration uncertainties, such that their weighted mean provide a reasonable estimate of the true [Fe/H]. We therefore adopt $\overline{[\mathrm{Fe/H}]}$ as the final estimate of [Fe/H] and $\sigma_{\overline{[\mathrm{Fe/H}]}}$ as its uncertainty. Under this assumption, combining measurements over multiple phases reduces the random measurement uncertainty, from approximately $0.16$~dex for a single sampled phase point to approximately $0.03$~dex for the phase-averaged [Fe/H].

\subsection{Comparison with the Literature Metallicities}

\begin{figure*}
    \centering
    \includegraphics[width=\textwidth]{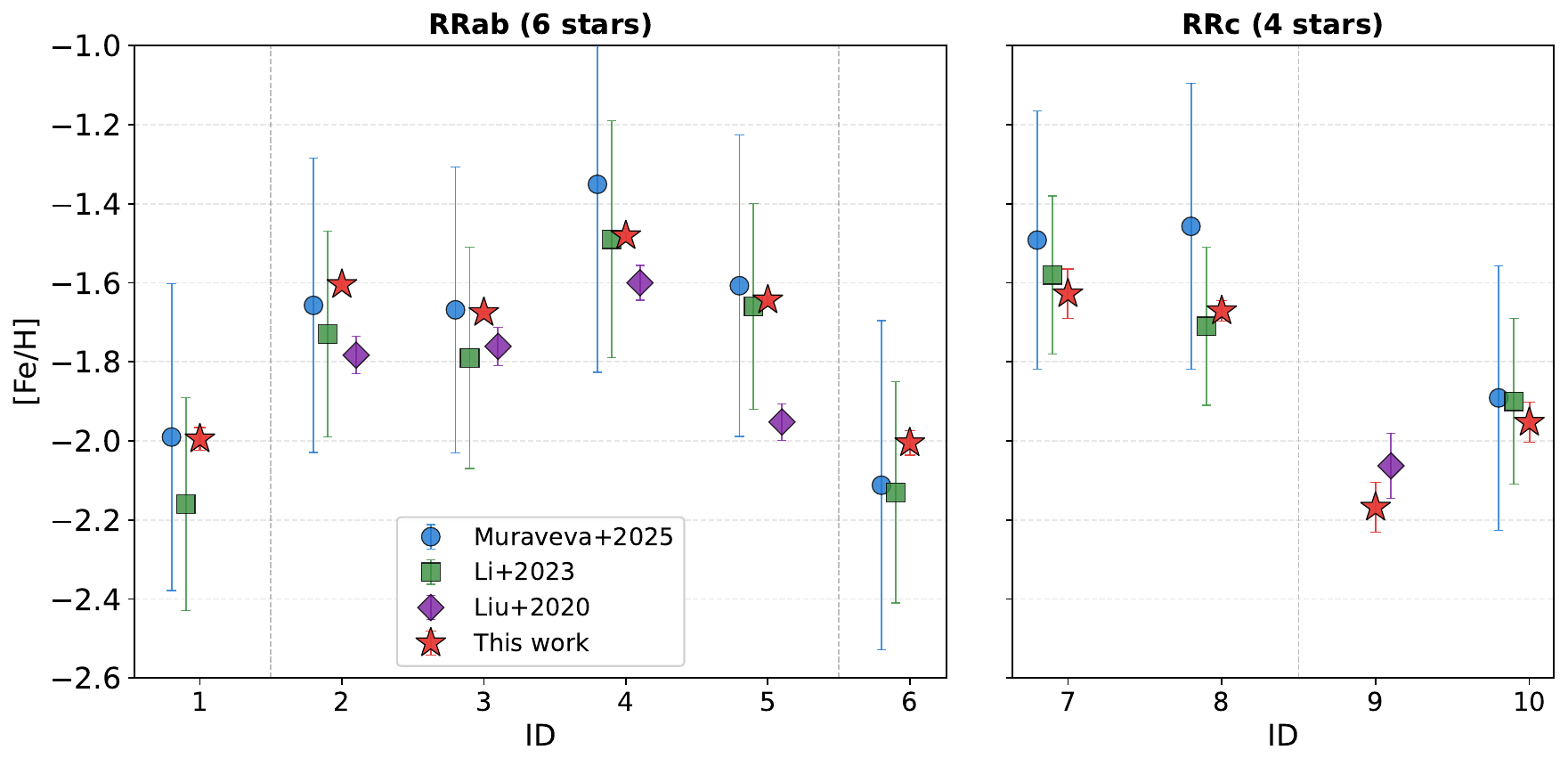}
    \caption{Comparison between the final metallicities derived in this work and literature values for the selected RR~Lyrae stars. The left and right panels show RRab and RRc stars, respectively. Blue circles, green squares, and purple diamonds denote the values from \citet{Muraveva2025}, \citet{Li2023}, and \citet{Liu2020}, respectively; red stars represent the results of this work.The error bars for the results of this work represent $\sigma_{\overline{[\mathrm{Fe/H}]}}$.}
    \label{fig:feh_comparison}
\end{figure*}

To assess the reliability of the final metallicities derived in this work, we compare the inverse-variance weighted [Fe/H] estimates with the literature values from \citet{Muraveva2025}, \citet{Li2023}, and \citet{Liu2020}. Figure~\ref{fig:feh_comparison} shows the comparison for six RRab and four RRc stars. The metallicities reported by \citet{Muraveva2025} and \citet{Li2023} were derived from Gaia light-curves, whereas \citet{Liu2020} estimated metallicities from LAMOST and SEGUE spectra. These independent photometric and spectroscopic measurements provide an external assessment of the Mephisto multi-band photometric results.

For the RRab stars, the metallicities derived in this work are in good
overall agreement with the literature measurements. In particular, the
relative metallicity ranking is consistently recovered: IDs~1 and 6
are identified as the most metal-poor stars in the sample, whereas
ID~4 is relatively metal-rich. Our results are generally consistent
with the values of \citet{Muraveva2025} and \citet{Li2023} within their quoted uncertainties. For the four RRab stars in common with
\citet{Liu2020}, our measurements are modestly (0.1-0.2~dex) more metal-rich, while preserving the same overall pattern.  

The RRc stars show the same broad agreement with the literature measurements, although some differences are present. The formal uncertainties in the weighted-mean [Fe/H] estimates are generally larger for the RRc stars than for the RRab stars. This is partly attributable to the higher temperatures of RRc stars, as the colors become less sensitive to metallicity at higher temperatures (Figure~\ref{fig:color_parameter_maps}). The RRc stars have a median $T_{\rm eff}$ of approximately 6840~K, compared with 6250~K for the RRab stars. In addition, the relatively metal-poor stars in the sample tend to have larger uncertainties because their colors provide weaker constraints on [Fe/H].

We further examine the metallicity distribution of RR Lyrae stars within individual globular clusters. In M3, five of the six stars show closely consistent metallicities, with a mean [Fe/H] of $-1.64$ and a star-to-star scatter of only 0.03 dex, while M3 ID4 exhibits a noticeably higher metallicity of $-1.48$. The three RR Lyrae stars in NGC~5466 also show good internal consistency, with a mean [Fe/H] of $-2.04$ and a scatter of 0.11 dex. The metal-rich outlier in M3 is potentially interesting in the context of the multiple stellar populations known in globular clusters, although its origin cannot be established from the present data alone. A detailed investigation of the metallicity distribution of RR Lyrae stars and its possible connection to multiple stellar populations will be explored in future work.

\subsection{Future Work}

The present study establishes a photometric framework for phase-resolved atmospheric-parameter estimation of RR~Lyrae stars using Mephisto multi-band photometry and demonstrates its potential for tracing the changing atmospheric properties throughout the pulsation cycle. Applying this framework to a larger sample of RR~Lyrae stars in Galactic globular clusters will enable more robust statistical characterization of their phase-dependent atmospheric parameters and provide a homogeneous dataset for investigating the metallicity distributions and stellar populations of their host clusters. A larger sample will also allow the phase relation between $T_{\rm eff}$ and $\log g$ identified in this work to be quantified statistically and examined in greater detail.

Improved observing strategies could further improve the construction of phase-dependent six-band color curves. In particular, alternating observations between the $ugi$ and $vrz$ filter sets would provide more closely spaced measurements in all six bands, reducing the temporal separation between different filters and the resulting uncertainties from light-curve template interpolation. This would provide more reliable quasi-simultaneous multi-band colors throughout the pulsation cycle and improve the input data for atmospheric-parameter inference.

Further improvements in photometric calibration and overall system performance are also expected to improve the precision of the inferred atmospheric parameters. The present results are based on the preliminary calibration of the Mephisto pilot surveys (X.~K.~Liu et al. 2026; Y. Fang et al. 2026), with typical calibration precisions for bright stars of approximately $0.010$, $0.011$, $0.010$, $0.009$, $0.011$, and $0.012$~mag in the $u$, $v$, $g$, $r$, $i$, and $z$ bands, respectively.
Recently, significant progress has been made in photometric calibration (J.~H.~Zhang et al. 2026). Future high-S/N observations will provide multi-band light curves with photometric and calibration errors smaller than 0.01, 0.008, 0.005, 0.005, 0.005~mag in $u$, $v$, $g$, $r$, $i$, and $z$ bands. These data are promising to yield atmospheric parameters for RR Lyrae stars with a precision of $\sim$35~K in $T_{\rm eff}$, 0.063~dex in $\log~g$, and 0.14~dex in [Fe/H] for a ssingle sampled phase point.  


With these improvements, the method can be extended to larger samples of RR~Lyrae stars, including both cluster member stars and field stars in the Galactic halo, providing a valuable photometric approach to studying the atmospheric properties and pulsation physics of RR~Lyrae stars, as well as the structure and assembly history of the Milky Way.

\section{Conclusions}
Precise stellar atmospheric parameters for RR~Lyrae stars provide important constraints of their pulsation physics and the Galactic structure. In this work, we have developed a photometric framework for inferring the phase-dependent atmospheric parameters of RR~Lyrae stars from the Mephisto multi-band photometry. Synthetic colors generated from ATLAS12 model spectra were used to train a neural-network forward model, and the observed phase-dependent colors were fitted to infer $T_{\rm eff}$, $\log g$, and [Fe/H] throughout the pulsation cycle. Applying the method to six RRab and four RRc stars with well-sampled light curves, we obtain the following main results:

\begin{enumerate}
\item The inferred $T_{\rm eff}$ and $\log g$ curves show clear phase-dependent variations. The maximum $T_{\rm eff}$ occurs approximately at maximum light, while the maximum $\log g$ occurs slightly earlier in the pulsation cycle. The results are consistent with previous hydrodynamical and spectroscopic studies, which find that the minimum radius occurs shortly before maximum light \citep{Kolenberg2010,Fossati2014}. Applying the current method to a large sample of RR Lyrae stars should provide valuable constraints on the structure and pulsation physics of RR~Lyrae stars.

\item The deduced phase-dependent [Fe/H] value constant, as expected for a fixed chemical composition. The typical uncertainty of a single-phase [Fe/H] measurement is approximately $0.16$~dex. Assuming that the phase-dependent measurements mainly represent random fluctuations around the true metallicity, combining measurements over multiple phases reduces the uncertainty to approximately $0.03$~dex for the phase-averaged [Fe/H]. The resulting metallicities show overall agreement with independent literature determinations based on Gaia light curves and low-resolution spectroscopy. Further improvements can be achieved through optimized multi-band observing strategies and improved photometric calibration.

\item The sensitivity analysis demonstrates that the five independent Mephisto colors have distinct and complementary responses to $T_{\rm eff}$, $\log g$, and [Fe/H]. In particular, $(u-v)$ is relatively sensitive to $\log g$, while $(v-g)$ provides strong metallicity sensitivity except at high effective temperatures. The different parameter dependences of the five colors provide complementary information for the joint inference of the three atmospheric parameters, supporting the effectiveness of the five-color combination for phase-dependent atmospheric parameter estimation. The metallicity sensitivity decreases toward higher effective temperatures, consistent with the large uncertainties of [Fe/H] estimates obtained for the hottest pulsation phases.

\item Multi-band photometry offers a new approach to determining the atmospheric parameters of RR~Lyrae stars. Compared with spectroscopic observations, it enables efficient and repeated measurements throughout the pulsation cycle, making phase-resolved parameter estimation feasible for large samples of stara. The present method simultaneously constrains $T_{\rm eff}$, $\log g$, and [Fe/H], providing a comprehensive framework for phase-resolved atmospheric-parameter studies of large RR~Lyrae samples in future time-domain surveys.

\end{enumerate}

The results demonstrate the potential of Mephisto multi-band photometry for phase-resolved atmospheric studies of RR~Lyrae stars. By combining simultaneous multi-band observations with repeated measurements over the pulsation cycle, the method provides an efficient way to characterize the evolving atmospheric properties of RR~Lyrae stars without requiring spectroscopic observations at every phase. Future applications to larger samples of cluster and field RR~Lyrae stars can provide homogeneous measurements of $T_{\rm eff}$, $\log g$, and [Fe/H] across different stellar populations. Combined with Gaia astrometry and kinematics, such measurements can be used to investigate the properties and evolution of RR~Lyrae populations and to explore the metallicity distribution and chemo-dynamical structure of the Galactic halo, including its substructure and accretion history.

\section*{Acknowledgements}
This work is also supported by the National Key Research and Development
Program of China (2024YFA1611601). Mephisto was developed at and is operated by the South-Western Institute for Astronomy Research of Yunnan University (SWIFAR-YNU), with funding from the “Yunnan University Development Plan for World-Class University” and the “Yunnan University
Development Plan for World-Class Astronomy Discipline”. The authors acknowledge support from the “Science \& Technology Champion Project” (202005AB160002) and from two “Team Projects”—the “Innovation Team” (202105AE160021) and the “Top Team” (202305AT350002), all funded by the “Yunnan Revitalization Talent Support Program”. This work also acknowledges support from the NSFC of China under grant No. 12603161 and National Natural Science Foundation of China (NSFC) (grants 12373031, 12541303, 12588202),  China Manned Spaced Program with grant no.CMS-CSST-2025-A12.

\section*{Data Availability}
The phase-resolved Mephisto light curves and atmospheric parameter estimates for the ten RR Lyrae stars analyzed in this work are publicly available at Zenodo:
\url{https://doi.org/10.5281/zenodo.23661544}.

\bibliographystyle{mnras}
\bibliography{Mephisto_RRL_AtmosParams}

@INPROCEEDINGS{Liu2019,
       author = {{Liu}, X.},
        title = "{Multi-channel Photometric Survey Telescope - Mephisto}",
    booktitle = {Galactic Archaeology in the Gaia Era},
         year = 2019,
        month = jan,
          eid = {14},
        pages = {14},
       adsurl = {https://ui.adsabs.harvard.edu/abs/2019gage.confE..14L}
}

@ARTICLE{HuangY2019,
       author = {{Huang}, Y. and {Chen}, B.-Q. and {Yuan}, H.-B. and {Zhang}, H.-W. and {Xiang}, M.-S. and {Wang}, C. and {Wang}, H.-F. and {Wolf}, C. and {Liu}, G.-C. and {Liu}, X.-W.},
        title = "{Milky Way Tomography with the SkyMapper Southern Survey. I. Atmospheric Parameters and Distances of One Million Red Giants}",
      journal = {\apjs},
         year = 2019,
          doi = {10.3847/1538-4365/ab1f72},
archivePrefix = {arXiv},
       eprint = {1901.08033},
 primaryClass = {astro-ph.GA},
       adsurl = {https://ui.adsabs.harvard.edu/abs/2019ApJS..243....7H}
}

@ARTICLE{YangL2022,
       author = {{Yang}, Lin and {Yuan}, Haibo and {Xiang}, Maosheng and {Duan}, Fuqing and {Huang}, Yang and {Liu}, Jifeng and {Beers}, Timothy C. and {Galarza}, Carlos Andr{\'e}s and {Daflon}, Simone and {Fern{\'a}ndez-Ontiveros}, Juan Antonio and {Cenarro}, Javier and {Crist{\'o}bal-Hornillos}, David and {Hern{\'a}ndez-Monteagudo}, Carlos and {L{\'o}pez-Sanjuan}, Carlos and {Mar{\'\i}n-Franch}, Antonio and {Moles}, Mariano and {Varela}, Jes{\'u}s and {V{\'a}zquez Rami{\'o}}, H{\'e}ctor and {Alcaniz}, Jailson and {Dupke}, Renato and {Ederoclite}, Alessandro and {Sodr{\'e}}, Laerte and {Angulo}, Raul E.},
        title = "{J-PLUS: Stellar parameters, C, N, Mg, Ca, and [{\ensuremath{\alpha}}/Fe] abundances for two million stars from DR1}",
      journal = {\aap},
         year = 2022,
          doi = {10.1051/0004-6361/202142724},
archivePrefix = {arXiv},
       eprint = {2112.07304},
 primaryClass = {astro-ph.SR},
       adsurl = {https://ui.adsabs.harvard.edu/abs/2022A&A...659A.181Y}
}

@ARTICLE{Bono1994,
       author = {{Bono}, G. and {Stellingwerf}, R.~F.},
        title = "{Pulsation and Stability of RR Lyrae Stars. I. Instability Strip}",
      journal = {\apjs},
         year = 1994,
        month = jul,
       volume = {93},
        pages = {233},
          doi = {10.1086/192054},
       adsurl = {https://ui.adsabs.harvard.edu/abs/1994ApJS...93..233B}
}

@book{Smith1995,
  author    = {Smith, Horace A.},
  title     = {RR Lyrae Stars},
  publisher = {Cambridge University Press},
  year      = {1995},
  address   = {Cambridge}
}

@BOOK{CatelanSmith2015,
       author = {{Catelan}, M. and {Smith}, H.~A.},
        title = "{Pulsating Stars}",
         year = 2015,
       adsurl = {https://ui.adsabs.harvard.edu/abs/2015pust.book.....C}
}

@article{Catelan2009,
  author  = {Catelan, M{\'a}rcio},
  title   = {Horizontal Branch Stars: The Interplay between Observations and Theory, and Insights into the Formation of the Galaxy},
  journal = {Astrophysics and Space Science},
  year    = {2009},
  volume  = {320},
  pages   = {261--309},
  doi     = {10.1007/s10509-009-9987-8}
}

@article{DOrazi2024,
  author  = {D'Orazi, Valentina and Storm, Nicholas and Casey, Andrew R. and Braga, Vittorio F. and Zocchi, Alice and Bono, Giuseppe and Fabrizio, Michele and Sneden, Christopher and Massari, Davide and Giribaldi, Riano E. and Bergemann, Maria and Campbell, Simon W. and Casagrande, Luca and de Grijs, Richard and De Silva, Gayandhi and Lugaro, Maria and Zucker, Daniel B. and Bragaglia, Angela and Feuillet, Diane and Fiorentino, Giuliana and Chaboyer, Brian and Dall'Ora, Massimo and Marengo, Massimo and Martinez-Vazquez, Clara E. and Matsunaga, Noriyuki and Monelli, Matteo and Mullen, Joseph P. and Nataf, David and Tantalo, Maria and Thevenin, Frederic and Vitello, Fabio R. and Kudritzki, Rolf-Peter and Bland-Hawthorn, Joss and Buder, Sven and Freeman, Kenneth and Kos, Janez and Lewis, Geraint F. and Lind, Karin and Martell, Sarah and Sharma, Sanjib and Stello, Dennis and Zwitter, Tomaz},
  title   = {The GALAH survey: tracing the Milky Way's formation and evolution through RR Lyrae stars},
  journal = {Monthly Notices of the Royal Astronomical Society},
  year    = {2024},
  volume  = {531},
  number  = {1},
  pages   = {137--162},
  doi     = {10.1093/mnras/stae1149}
}

@article{Sesar2017,
  author  = {Sesar, Branimir and Hernitschek, Nina and Mitrovi{\'c}, Sandra and Ivezi{\'c}, {\v{Z}}eljko and Rix, Hans-Walter and Cohen, Judith G. and Bernard, Edouard J. and Grebel, Eva K. and Martin, Nicolas F. and Schlafly, Edward F. and Burgett, William S. and Draper, Peter W. and Flewelling, Heather and Kaiser, Nick and Kudritzki, Rolf P. and Magnier, Eugene A. and Metcalfe, Nigel and Tonry, John L. and Waters, Christopher},
  title   = {Machine-learned Identification of RR Lyrae Stars from Sparse, Multi-band Data: The PS1 Sample},
  journal = {The Astronomical Journal},
  year    = {2017},
  volume  = {153},
  pages   = {204},
  doi     = {10.3847/1538-3881/aa661b}
}

@article{Udalski1997,
  author  = {Udalski, Andrzej and Olech, Arkadiusz and Szyma{\'n}ski, Micha{\l} and Kaluzny, Janusz and Kubiak, Mariusz and Mateo, Mario and Krzeminski, Wojtek and Stanek, Krzysztof Z.},
  title   = {The Optical Gravitational Lensing Experiment. The Catalog of Periodic Variable Stars in the Galactic Bulge. V. Periodic Variables in Fields: MM5-A, MM5-B, MM7-A and MM7-B},
  journal = {Acta Astronomica},
  year    = {1997},
  volume  = {47},
  pages   = {319--344}
}

@article{Soszynski2009,
  author  = {Soszy{\'n}ski, I. and Udalski, A. and Szyma{\'n}ski, M. K. and Kubiak, M. and Pietrzy{\'n}ski, G. and Wyrzykowski, {\L}. and Szewczyk, O. and Ulaczyk, K. and Poleski, R.},
  title   = {The Optical Gravitational Lensing Experiment. The OGLE-III Catalog of Variable Stars. III. RR Lyrae Stars in the Large Magellanic Cloud},
  journal = {Acta Astronomica},
  year    = {2009},
  volume  = {59},
  pages   = {1--18}
}

@article{clement2001,
	title = {Variable Stars in Galactic Globular Clusters},
	volume = {122},
	issn = {0004-6256},
	url = {https://ui.adsabs.harvard.edu/abs/2001AJ....122.2587C},
	doi = {10.1086/323719},
	pages = {2587--2599},
	journal = {The Astronomical Journal},
	author = {Clement, Christine M. and Muzzin, Adam and Dufton, Quentin and Ponnampalam, Thivya and Wang, John and Burford, Jay and Richardson, Alan and Rosebery, Tara and Rowe, Jason and Hogg, Helen Sawyer},
	urldate = {2023-12-04},
	date = {2001-11-01},
  year = {2001},
	note = {{ADS} Bibcode: 2001AJ....122.2587C},
}

@INPROCEEDINGS{Clementini2010,
       author = {{Clementini}, Gisella},
        title = "{RR Lyrae Stars in Dwarf Spheroidal Galaxies}",
    booktitle = {Variable Stars, the Galactic halo and Galaxy Formation},
         year = 2010,
       editor = {{Sterken}, Christiaan and {Samus}, Nikolay and {Szabados}, Laszlo},
        month = feb,
        pages = {107},
          doi = {10.48550/arXiv.1002.1575},
archivePrefix = {arXiv},
       eprint = {1002.1575},
 primaryClass = {astro-ph.CO},
       adsurl = {https://ui.adsabs.harvard.edu/abs/2010vsgh.conf..107C}
}

@ARTICLE{Catelan2004,
       author = {{Catelan}, M. and {Pritzl}, Barton J. and {Smith}, Horace A.},
        title = "{The RR Lyrae Period-Luminosity Relation. I. Theoretical Calibration}",
      journal = {\apjs},
         year = 2004,
        month = oct,
       volume = {154},
       number = {2},
        pages = {633-649},
          doi = {10.1086/422916},
archivePrefix = {arXiv},
       eprint = {astro-ph/0406067},
 primaryClass = {astro-ph},
       adsurl = {https://ui.adsabs.harvard.edu/abs/2004ApJS..154..633C}
}

@article{marconi2015,
	title = {On a New Theoretical Framework for {RR} Lyrae Stars. I. The Metallicity Dependence},
	volume = {808},
	issn = {0004-637X},
	url = {https://ui.adsabs.harvard.edu/abs/2015ApJ...808...50M/abstract},
	doi = {10.1088/0004-637X/808/1/50},
	pages = {50},
	number = {1},
	journal = {The Astrophysical Journal},
	author = {Marconi, M. and Coppola, G. and Bono, G. and Braga, V. and Pietrinferni, A. and Buonanno, R. and Castellani, M. and Musella, I. and Ripepi, V. and Stellingwerf, R. F.},
	urldate = {2023-11-08},
	date = {2015-07},
  year = {2015},
	langid = {english},
}

@article{bhardwaj2023,
	title = {Precise Empirical Determination of Metallicity Dependence of Near-infrared Period-Luminosity Relations for {RR} Lyrae Variables},
	volume = {944},
	issn = {0004-637X},
	url = {https://ui.adsabs.harvard.edu/abs/2023ApJ...944L..51B},
	doi = {10.3847/2041-8213/acba7f},
	pages = {L51},
	journal = {The Astrophysical Journal},
	author = {Bhardwaj, Anupam and Marconi, Marcella and Rejkuba, Marina and de Grijs, Richard and Singh, Harinder P. and Braga, Vittorio F. and Kanbur, Shashi and Ngeow, Chow-Choong and Ripepi, Vincenzo and Bono, Giuseppe and De Somma, Giulia and Dall'Ora, Massimo},
	urldate = {2023-09-22},
	date = {2023-02-01},
  year = {2023},
	note = {{ADS} Bibcode: 2023ApJ...944L..51B},
}

@ARTICLE{Monelli2022,
       author = {{Monelli}, Matteo and {Fiorentino}, Giuliana},
        title = "{RR Lyrae Stars and Anomalous Cepheids as Population Tracers in Local Group Galaxies}",
      journal = {Universe},
         year = 2022,
        month = mar,
       volume = {8},
       number = {3},
          eid = {191},
        pages = {191},
          doi = {10.3390/universe8030191},
       adsurl = {https://ui.adsabs.harvard.edu/abs/2022Univ....8..191M}
}

@ARTICLE{Feuchtinger1999,
       author = {{Feuchtinger}, M.~U.},
        title = "{A nonlinear convective model of pulsating stars. II. RR Lyrae stars}",
      journal = {\aap},
         year = 1999,
        month = nov,
       volume = {351},
        pages = {103-118},
       adsurl = {https://ui.adsabs.harvard.edu/abs/1999A&A...351..103F}
}

@INPROCEEDINGS{Marconi2009,
       author = {{Marconi}, Marcella},
        title = "{RR Lyrae Pulsation Theory}",
    booktitle = {Stellar Pulsation: Challenges for Theory and Observation},
         year = 2009,
       editor = {{Guzik}, Joyce Ann and {Bradley}, Paul A.},
       series = {American Institute of Physics Conference Series},
       volume = {1170},
        month = sep,
    publisher = {AIP},
        pages = {223-234},
          doi = {10.1063/1.3246450},
archivePrefix = {arXiv},
       eprint = {0909.0900},
 primaryClass = {astro-ph.SR},
       adsurl = {https://ui.adsabs.harvard.edu/abs/2009AIPC.1170..223M}
}

@ARTICLE{Smolec2008,
       author = {{Smolec}, R. and {Moskalik}, P.},
        title = "{Convective Hydrocodes for Radial Stellar Pulsation. Physical and Numerical Formulation}",
      journal = {\actaa},
         year = 2008,
        month = sep,
       volume = {58},
        pages = {193-232},
          doi = {10.48550/arXiv.0809.1979},
archivePrefix = {arXiv},
       eprint = {0809.1979},
 primaryClass = {astro-ph},
       adsurl = {https://ui.adsabs.harvard.edu/abs/2008AcA....58..193S}
}

@ARTICLE{Paxton2019ApJS,
       author = {{Paxton}, Bill and {Smolec}, R. and {Schwab}, Josiah and {Gautschy}, A. and {Bildsten}, Lars and {Cantiello}, Matteo and {Dotter}, Aaron and {Farmer}, R. and {Goldberg}, Jared A. and {Jermyn}, Adam S. and {Kanbur}, S.~M. and {Marchant}, Pablo and {Thoul}, Anne and {Townsend}, Richard H.~D. and {Wolf}, William M. and {Zhang}, Michael and {Timmes}, F.~X.},
        title = "{Modules for Experiments in Stellar Astrophysics (MESA): Pulsating Variable Stars, Rotation, Convective Boundaries, and Energy Conservation}",
      journal = {\apjs},
         year = 2019,
        month = jul,
       volume = {243},
       number = {1},
          eid = {10},
        pages = {10},
          doi = {10.3847/1538-4365/ab2241},
archivePrefix = {arXiv},
       eprint = {1903.01426},
 primaryClass = {astro-ph.SR},
       adsurl = {https://ui.adsabs.harvard.edu/abs/2019ApJS..243...10P}
}

@ARTICLE{Clementini1995,
       author = {{Clementini}, G. and {Carretta}, E. and {Gratton}, R. and {Merighi}, R. and {Mould}, J.~R. and {McCarthy}, J.~K.},
        title = "{The Composition of HB Stars: RR Lyrae Variables}",
      journal = {\aj},
         year = 1995,
        month = nov,
       volume = {110},
        pages = {2319},
          doi = {10.1086/117692},
archivePrefix = {arXiv},
       eprint = {astro-ph/9507056},
 primaryClass = {astro-ph},
       adsurl = {https://ui.adsabs.harvard.edu/abs/1995AJ....110.2319C}
}

@ARTICLE{Nemec2013,
       author = {{Nemec}, James M. and {Cohen}, Judith G. and {Ripepi}, Vincenzo and {Derekas}, Aliz and {Moskalik}, Pawel and {Sesar}, Branimir and {Chadid}, Merieme and {Bruntt}, Hans},
        title = "{Metal Abundances, Radial Velocities, and Other Physical Characteristics for the RR Lyrae Stars in The Kepler Field}",
      journal = {\apj},
         year = 2013,
        month = aug,
       volume = {773},
       number = {2},
          eid = {181},
        pages = {181},
          doi = {10.1088/0004-637X/773/2/181},
archivePrefix = {arXiv},
       eprint = {1307.5820},
 primaryClass = {astro-ph.SR},
       adsurl = {https://ui.adsabs.harvard.edu/abs/2013ApJ...773..181N}
}

@ARTICLE{Pancino2015,
       author = {{Pancino}, E. and {Britavskiy}, N. and {Romano}, D. and {Cacciari}, C. and {Mucciarelli}, A. and {Clementini}, G.},
        title = "{Chemical abundances of solar neighbourhood RR Lyrae stars}",
      journal = {\mnras},
         year = 2015,
        month = mar,
       volume = {447},
       number = {3},
        pages = {2404-2419},
          doi = {10.1093/mnras/stu2616},
archivePrefix = {arXiv},
       eprint = {1412.4580},
 primaryClass = {astro-ph.SR},
       adsurl = {https://ui.adsabs.harvard.edu/abs/2015MNRAS.447.2404P}
}

@ARTICLE{Chadid2017,
       author = {{Chadid}, Merieme and {Sneden}, Christopher and {Preston}, George W.},
        title = "{Spectroscopic Comparison of Metal-rich RRab Stars of the Galactic Field with their Metal-poor Counterparts}",
      journal = {\apj},
         year = 2017,
        month = feb,
       volume = {835},
       number = {2},
          eid = {187},
        pages = {187},
          doi = {10.3847/1538-4357/835/2/187},
archivePrefix = {arXiv},
       eprint = {1611.02368},
 primaryClass = {astro-ph.SR},
       adsurl = {https://ui.adsabs.harvard.edu/abs/2017ApJ...835..187C}
}

@ARTICLE{Crestani2021,
       author = {{Crestani}, J. and {Fabrizio}, M. and {Braga}, V.~F. and {Sneden}, C. and {Preston}, G. and {Ferraro}, I. and {Iannicola}, G. and {Bono}, G. and {Alves-Brito}, A. and {Nonino}, M. and {D'Orazi}, V. and {Inno}, L. and {Monelli}, M. and {Storm}, J. and {Altavilla}, G. and {Chaboyer}, B. and {Dall'Ora}, M. and {Fiorentino}, G. and {Gilligan}, C. and {Grebel}, E.~K. and {Lala}, H. and {Lemasle}, B. and {Marengo}, M. and {Marinoni}, S. and {Marrese}, P.~M. and {Mart{\'\i}nez-V{\'a}zquez}, C.~E. and {Matsunaga}, N. and {Mullen}, J.~P. and {Neeley}, J. and {Prudil}, Z. and {da Silva}, R. and {Stetson}, P.~B. and {Th{\'e}venin}, F. and {Valenti}, E. and {Walker}, A. and {Zoccali}, M.},
        title = "{On the Use of Field RR Lyrae as Galactic Probes. II. A New {\ensuremath{\Delta}}S Calibration to Estimate Their Metallicity}",
      journal = {\apj},
         year = 2021,
        month = feb,
       volume = {908},
       number = {1},
          eid = {20},
        pages = {20},
          doi = {10.3847/1538-4357/abd183},
archivePrefix = {arXiv},
       eprint = {2012.02284},
 primaryClass = {astro-ph.SR},
       adsurl = {https://ui.adsabs.harvard.edu/abs/2021ApJ...908...20C}
}

@ARTICLE{Yanny2009,
       author = {{Yanny}, Brian and {Rockosi}, Constance and {Newberg}, Heidi Jo and {Knapp}, Gillian R. and {Adelman-McCarthy}, Jennifer K. and {Alcorn}, Bonnie and {Allam}, Sahar and {Allende Prieto}, Carlos and {An}, Deokkeun and {Anderson}, Kurt S.~J. and {Anderson}, Scott and {Bailer-Jones}, Coryn A.~L. and {Bastian}, Steve and {Beers}, Timothy C. and {Bell}, Eric and {Belokurov}, Vasily and {Bizyaev}, Dmitry and {Blythe}, Norm and {Bochanski}, John J. and {Boroski}, William N. and {Brinchmann}, Jarle and {Brinkmann}, J. and {Brewington}, Howard and {Carey}, Larry and {Cudworth}, Kyle M. and {Evans}, Michael and {Evans}, N.~W. and {Gates}, Evalyn and {G{\"a}nsicke}, B.~T. and {Gillespie}, Bruce and {Gilmore}, Gerald and {Nebot Gomez-Moran}, Ada and {Grebel}, Eva K. and {Greenwell}, Jim and {Gunn}, James E. and {Jordan}, Cathy and {Jordan}, Wendell and {Harding}, Paul and {Harris}, Hugh and {Hendry}, John S. and {Holder}, Diana and {Ivans}, Inese I. and {Ivezi{\v{c}}}, {\v{Z}}eljko and {Jester}, Sebastian and {Johnson}, Jennifer A. and {Kent}, Stephen M. and {Kleinman}, Scot and {Kniazev}, Alexei and {Krzesinski}, Jurek and {Kron}, Richard and {Kuropatkin}, Nikolay and {Lebedeva}, Svetlana and {Lee}, Young Sun and {French Leger}, R. and {L{\'e}pine}, S{\'e}bastien and {Levine}, Steve and {Lin}, Huan and {Long}, Daniel C. and {Loomis}, Craig and {Lupton}, Robert and {Malanushenko}, Olena and {Malanushenko}, Viktor and {Margon}, Bruce and {Martinez-Delgado}, David and {McGehee}, Peregrine and {Monet}, Dave and {Morrison}, Heather L. and {Munn}, Jeffrey A. and {Neilsen}, Jr., Eric H. and {Nitta}, Atsuko and {Norris}, John E. and {Oravetz}, Dan and {Owen}, Russell and {Padmanabhan}, Nikhil and {Pan}, Kaike and {Peterson}, R.~S. and {Pier}, Jeffrey R. and {Platson}, Jared and {Re Fiorentin}, Paola and {Richards}, Gordon T. and {Rix}, Hans-Walter and {Schlegel}, David J. and {Schneider}, Donald P. and {Schreiber}, Matthias R. and {Schwope}, Axel and {Sibley}, Valena and {Simmons}, Audrey and {Snedden}, Stephanie A. and {Allyn Smith}, J. and {Stark}, Larry and {Stauffer}, Fritz and {Steinmetz}, M. and {Stoughton}, C. and {SubbaRao}, Mark and {Szalay}, Alex and {Szkody}, Paula and {Thakar}, Aniruddha R. and {Sivarani}, Thirupathi and {Tucker}, Douglas and {Uomoto}, Alan and {Vanden Berk}, Dan and {Vidrih}, Simon and {Wadadekar}, Yogesh and {Watters}, Shannon and {Wilhelm}, Ron and {Wyse}, Rosemary F.~G. and {Yarger}, Jean and {Zucker}, Dan},
        title = "{SEGUE: A Spectroscopic Survey of 240,000 Stars with g = 14-20}",
      journal = {\aj},
         year = 2009,
        month = may,
       volume = {137},
       number = {5},
        pages = {4377-4399},
          doi = {10.1088/0004-6256/137/5/4377},
archivePrefix = {arXiv},
       eprint = {0902.1781},
 primaryClass = {astro-ph.GA},
       adsurl = {https://ui.adsabs.harvard.edu/abs/2009AJ....137.4377Y}
}

@ARTICLE{Majewski2017,
       author = {{Majewski}, Steven R. and {Schiavon}, Ricardo P. and {Frinchaboy}, Peter M. and {Allende Prieto}, Carlos and {Barkhouser}, Robert and {Bizyaev}, Dmitry and {Blank}, Basil and {Brunner}, Sophia and {Burton}, Adam and {Carrera}, Ricardo and {Chojnowski}, S. Drew and {Cunha}, K{\'a}tia and {Epstein}, Courtney and {Fitzgerald}, Greg and {Garc{\'\i}a P{\'e}rez}, Ana E. and {Hearty}, Fred R. and {Henderson}, Chuck and {Holtzman}, Jon A. and {Johnson}, Jennifer A. and {Lam}, Charles R. and {Lawler}, James E. and {Maseman}, Paul and {M{\'e}sz{\'a}ros}, Szabolcs and {Nelson}, Matthew and {Nguyen}, Duy Coung and {Nidever}, David L. and {Pinsonneault}, Marc and {Shetrone}, Matthew and {Smee}, Stephen and {Smith}, Verne V. and {Stolberg}, Todd and {Skrutskie}, Michael F. and {Walker}, Eric and {Wilson}, John C. and {Zasowski}, Gail and {Anders}, Friedrich and {Basu}, Sarbani and {Beland}, Stephane and {Blanton}, Michael R. and {Bovy}, Jo and {Brownstein}, Joel R. and {Carlberg}, Joleen and {Chaplin}, William and {Chiappini}, Cristina and {Eisenstein}, Daniel J. and {Elsworth}, Yvonne and {Feuillet}, Diane and {Fleming}, Scott W. and {Galbraith-Frew}, Jessica and {Garc{\'\i}a}, Rafael A. and {Garc{\'\i}a-Hern{\'a}ndez}, D. An{\'\i}bal and {Gillespie}, Bruce A. and {Girardi}, L{\'e}o and {Gunn}, James E. and {Hasselquist}, Sten and {Hayden}, Michael R. and {Hekker}, Saskia and {Ivans}, Inese and {Kinemuchi}, Karen and {Klaene}, Mark and {Mahadevan}, Suvrath and {Mathur}, Savita and {Mosser}, Beno{\^\i}t and {Muna}, Demitri and {Munn}, Jeffrey A. and {Nichol}, Robert C. and {O'Connell}, Robert W. and {Parejko}, John K. and {Robin}, A.~C. and {Rocha-Pinto}, Helio and {Schultheis}, Matthias and {Serenelli}, Aldo M. and {Shane}, Neville and {Silva Aguirre}, Victor and {Sobeck}, Jennifer S. and {Thompson}, Benjamin and {Troup}, Nicholas W. and {Weinberg}, David H. and {Zamora}, Olga},
        title = "{The Apache Point Observatory Galactic Evolution Experiment (APOGEE)}",
      journal = {\aj},
         year = 2017,
        month = sep,
       volume = {154},
       number = {3},
          eid = {94},
        pages = {94},
          doi = {10.3847/1538-3881/aa784d},
archivePrefix = {arXiv},
       eprint = {1509.05420},
 primaryClass = {astro-ph.IM},
       adsurl = {https://ui.adsabs.harvard.edu/abs/2017AJ....154...94M}
}

@ARTICLE{Deng2012RAA,
       author = {{Deng}, Li-Cai and {Newberg}, Heidi Jo and {Liu}, Chao and {Carlin}, Jeffrey L. and {Beers}, Timothy C. and {Chen}, Li and {Chen}, Yu-Qin and {Christlieb}, Norbert and {Grillmair}, Carl J. and {Guhathakurta}, Puragra and {Han}, Zhan-Wen and {Hou}, Jin-Liang and {Lee}, Hsu-Tai and {L{\'e}pine}, S{\'e}bastien and {Li}, Jing and {Liu}, Xiao-Wei and {Pan}, Kai-Ke and {Sellwood}, J.~A. and {Wang}, Bo and {Wang}, Hong-Chi and {Yang}, Fan and {Yanny}, Brian and {Zhang}, Hao-Tong and {Zhang}, Yue-Yang and {Zheng}, Zheng and {Zhu}, Zi},
        title = "{LAMOST Experiment for Galactic Understanding and Exploration (LEGUE) {\textemdash} The survey's science plan}",
      journal = {Research in Astronomy and Astrophysics},
         year = 2012,
        month = jul,
       volume = {12},
       number = {7},
        pages = {735-754},
          doi = {10.1088/1674-4527/12/7/003},
archivePrefix = {arXiv},
       eprint = {1206.3578},
 primaryClass = {astro-ph.GA},
       adsurl = {https://ui.adsabs.harvard.edu/abs/2012RAA....12..735D}
}

@INPROCEEDINGS{Liu2014,
       author = {{Liu}, X.-W. and {Yuan}, H.-B. and {Huo}, Z.-Y. and {Deng}, L.-C. and {Hou}, J.-L. and {Zhao}, Y.-H. and {Zhao}, G. and {Shi}, J.-R. and {Luo}, A.-L. and {Xiang}, M.-S. and {Zhang}, H.-H. and {Huang}, Y. and {Zhang}, H.-W.},
        title = "{LSS-GAC - A LAMOST Spectroscopic Survey of the Galactic Anti-center}",
    booktitle = {Setting the scene for Gaia and LAMOST},
         year = 2014,
       editor = {{Feltzing}, Sofia and {Zhao}, Gang and {Walton}, Nicholas A. and {Whitelock}, Patricia},
       series = {IAU Symposium},
       volume = {298},
        month = jan,
        pages = {310-321},
          doi = {10.1017/S1743921313006510},
archivePrefix = {arXiv},
       eprint = {1306.5376},
 primaryClass = {astro-ph.GA},
       adsurl = {https://ui.adsabs.harvard.edu/abs/2014IAUS..298..310L}
}

@ARTICLE{DeSilva2015,
       author = {{De Silva}, G.~M. and {Freeman}, K.~C. and {Bland-Hawthorn}, J. and {Martell}, S. and {de Boer}, E. Wylie and {Asplund}, M. and {Keller}, S. and {Sharma}, S. and {Zucker}, D.~B. and {Zwitter}, T. and {Anguiano}, B. and {Bacigalupo}, C. and {Bayliss}, D. and {Beavis}, M.~A. and {Bergemann}, M. and {Campbell}, S. and {Cannon}, R. and {Carollo}, D. and {Casagrande}, L. and {Casey}, A.~R. and {Da Costa}, G. and {D'Orazi}, V. and {Dotter}, A. and {Duong}, L. and {Heger}, A. and {Ireland}, M.~J. and {Kafle}, P.~R. and {Kos}, J. and {Lattanzio}, J. and {Lewis}, G.~F. and {Lin}, J. and {Lind}, K. and {Munari}, U. and {Nataf}, D.~M. and {O'Toole}, S. and {Parker}, Q. and {Reid}, W. and {Schlesinger}, K.~J. and {Sheinis}, A. and {Simpson}, J.~D. and {Stello}, D. and {Ting}, Y.-S. and {Traven}, G. and {Watson}, F. and {Wittenmyer}, R. and {Yong}, D. and {{\v{Z}}erjal}, M.},
        title = "{The GALAH survey: scientific motivation}",
      journal = {\mnras},
         year = 2015,
        month = may,
       volume = {449},
       number = {3},
        pages = {2604-2617},
          doi = {10.1093/mnras/stv327},
archivePrefix = {arXiv},
       eprint = {1502.04767},
 primaryClass = {astro-ph.GA},
       adsurl = {https://ui.adsabs.harvard.edu/abs/2015MNRAS.449.2604D}
}

@ARTICLE{Liu2020,
       author = {{Liu}, G.-C. and {Huang}, Y. and {Zhang}, H.-W. and {Xiang}, M.-S. and {Ren}, J.-J. and {Chen}, B.-Q. and {Yuan}, H.-B. and {Wang}, C. and {Yang}, Y. and {Tian}, Z.-J. and {Wang}, F. and {Liu}, X.-W.},
        title = "{Probing the Galactic Halo with RR Lyrae Stars. I. The Catalog}",
      journal = {\apjs},
         year = 2020,
        month = apr,
       volume = {247},
       number = {2},
          eid = {68},
        pages = {68},
          doi = {10.3847/1538-4365/ab72f8},
archivePrefix = {arXiv},
       eprint = {2002.01188},
 primaryClass = {astro-ph.SR},
       adsurl = {https://ui.adsabs.harvard.edu/abs/2020ApJS..247...68L}
}

@ARTICLE{Harris1996,
       author = {{Harris}, William E.},
        title = "{A Catalog of Parameters for Globular Clusters in the Milky Way}",
      journal = {\aj},
         year = 1996,
        month = oct,
       volume = {112},
        pages = {1487},
          doi = {10.1086/118116},
       adsurl = {https://ui.adsabs.harvard.edu/abs/1996AJ....112.1487H}
}

@ARTICLE{Fitzpatrick1999,
       author = {{Fitzpatrick}, Edward L.},
        title = "{Correcting for the Effects of Interstellar Extinction}",
      journal = {\pasp},
         year = 1999,
        month = jan,
       volume = {111},
       number = {755},
        pages = {63-75},
          doi = {10.1086/316293},
archivePrefix = {arXiv},
       eprint = {astro-ph/9809387},
 primaryClass = {astro-ph},
       adsurl = {https://ui.adsabs.harvard.edu/abs/1999PASP..111...63F}
}

@ARTICLE{Renzini1988,
       author = {{Renzini}, Alvio and {Fusi Pecci}, Flavio},
        title = "{Tests of evolutionary sequences using color-magnitude diagrams of globular clusters.}",
      journal = {\araa},
         year = 1988,
        month = jan,
       volume = {26},
        pages = {199-244},
          doi = {10.1146/annurev.aa.26.090188.001215},
       adsurl = {https://ui.adsabs.harvard.edu/abs/1988ARA&A..26..199R}
}

@misc{clementini_gaia_2022,
	title = {Gaia {DR}3: Specific processing and validation of all-sky {RR} Lyrae and Cepheid stars -- The {RR} Lyrae sample},
	url = {http://arxiv.org/abs/2206.06278},
	shorttitle = {Gaia {DR}3},
	number = {{arXiv}:2206.06278},
	publisher = {{arXiv}},
	author = {Clementini, G. and Ripepi, V. and Garofalo, A. and Molinaro, R. and Muraveva, T. and Leccia, S. and Rimoldini, L. and Holl, B. and de Fombelle, G. Jevardat and Sartoretti, P. and Marchal, O. and Audard, M. and Nienartowicz, K. and Andrae, R. and Marconi, M. and Szabados, L. and Evans, D. W. and Lecoeur-Taibi, I. and Mowlavi, N. and Musella, I. and Eyer, L.},
	urldate = {2023-03-20},
	date = {2022-06-13},
    year = {2022},
	langid = {english},
	eprinttype = {arxiv},
	eprint = {2206.06278 [astro-ph]},
}

@article{Fossati2014,
	title = {An in-depth spectroscopic analysis of {RR} Lyr Variations over the pulsation cycle★},
	volume = {445},
	issn = {1365-2966, 0035-8711},
	url = {http://academic.oup.com/mnras/article/445/4/4094/1753542/An-indepth-spectroscopic-analysis-of-RR-Lyr},
	doi = {10.1093/mnras/stu2044},
	pages = {4094--4104},
	number = {4},
	journal = {Monthly Notices of the Royal Astronomical Society},
	author = {Fossati, L. and Kolenberg, K. and Shulyak, D. V. and Elmasli, A. and Tsymbal, V. and Barnes, T. G. and Guggenberger, E. and Kochukhov, O.},
	urldate = {2023-03-09},
  year = {2014},
	date = {2014-12-21},
	langid = {english},
}

@article{Kolenberg2010,
	title = {An in-depth spectroscopic analysis of the Blazhko star {RR} Lyrae: I. Characterisation of the star: abundance analysis and fundamental parameters},
	volume = {519},
	issn = {0004-6361, 1432-0746},
	url = {http://www.aanda.org/10.1051/0004-6361/201014471},
	doi = {10.1051/0004-6361/201014471},
	pages = {A64},
	journal = {A\&A},
	author = {Kolenberg, K. and Fossati, L. and Shulyak, D. and Pikall, H. and Barnes, T. G. and Kochukhov, O. and Tsymbal, V.},
	urldate = {2023-03-09},
	date = {2010-09},
  year = {2010},
	langid = {english},
}

@article{Muraveva2025,
  author = {Muraveva, Tatiana and Giannetti, Andrea and Clementini, Gisella and Garofalo, Alessia and Monti, Lorenzo},
  title = {Metallicity of {RR} Lyrae Stars from the {Gaia} Data Release 3 Catalogue Computed with Machine Learning Algorithms},
  journal = {Monthly Notices of the Royal Astronomical Society},
  year = {2025},
  volume = {536},
  number = {3},
  pages = {2749--2769},
  doi = {10.1093/mnras/stae2679}
}

@article{Li2023,
  author = {Li, Xin-Yi and Huang, Yang and Liu, Gao-Chao and Beers, Timothy C. and Zhang, Hua-Wei},
  title = {Photometric Metallicity and Distance Estimates for $\sim$136,000 {RR} Lyrae Stars from {Gaia} Data Release 3},
  journal = {The Astrophysical Journal},
  year = {2023},
  volume = {944},
  number = {1},
  pages = {88},
  doi = {10.3847/1538-4357/acadd5}
}

@ARTICLE{Fabrizio2021,
       author = {{Fabrizio}, M. and {Braga}, V.~F. and {Crestani}, J. and {Bono}, G. and {Ferraro}, I. and {Fiorentino}, G. and {Iannicola}, G. and {Preston}, G.~W. and {Sneden}, C. and {Th{\'e}venin}, F. and {Altavilla}, G. and {Chaboyer}, B. and {Dall'Ora}, M. and {da Silva}, R. and {Grebel}, E.~K. and {Gilligan}, C.~K. and {Lala}, H. and {Lemasle}, B. and {Magurno}, D. and {Marengo}, M. and {Marinoni}, S. and {Marrese}, P.~M. and {Mart{\'\i}nez-V{\'a}zquez}, C.~E. and {Matsunaga}, N. and {Monelli}, M. and {Mullen}, J.~P. and {Neeley}, J. and {Nonino}, M. and {Prudil}, Z. and {Salaris}, M. and {Stetson}, P.~B. and {Valenti}, E. and {Zoccali}, M.},
        title = "{On the Use of Field RR Lyrae As Galactic Probes: IV. New Insights Into and Around the Oosterhoff Dichotomy}",
      journal = {\apj},
         year = 2021,
        month = oct,
       volume = {919},
       number = {2},
          eid = {118},
        pages = {118},
          doi = {10.3847/1538-4357/ac1115},
archivePrefix = {arXiv},
       eprint = {2107.00919},
 primaryClass = {astro-ph.SR},
       adsurl = {https://ui.adsabs.harvard.edu/abs/2021ApJ...919..118F}
}

@ARTICLE{Gillet2019,
       author = {{Gillet}, D. and {Mauclaire}, B. and {Lemoult}, T. and {Mathias}, Ph. and {Devaux}, J.-S. and {de France}, Th. and {Garrel}, T.},
        title = "{Dynamical structure of the pulsating atmosphere of RR Lyrae. I. A typical pulsation cycle}",
      journal = {\aap},
         year = 2019,
        month = mar,
       volume = {623},
          eid = {A109},
        pages = {A109},
          doi = {10.1051/0004-6361/201833869},
archivePrefix = {arXiv},
       eprint = {1903.03848},
 primaryClass = {astro-ph.SR},
       adsurl = {https://ui.adsabs.harvard.edu/abs/2019A&A...623A.109G}
}

@ARTICLE{Deng2014,
       author = {{Yang}, F. and {Deng}, L. and {Liu}, C. and {Carlin}, J.~L. and {Newberg}, H. Jo and {Carrell}, K. and {Justham}, S. and {Zhang}, X. and {Bai}, Z. and {Wang}, F. and {Zhang}, H. and {Wang}, K. and {Xin}, Y. and {Xu}, Y. and {Gao}, S. and {Zhang}, Y. and {Li}, J. and {Zhao}, Y.},
        title = "{Hydrogen lines in LAMOST low-resolution spectra of RR Lyrae stars}",
      journal = {\na},
         year = 2014,
        month = jan,
       volume = {26},
        pages = {72-76},
          doi = {10.1016/j.newast.2013.06.002},
archivePrefix = {arXiv},
       eprint = {1306.3754},
 primaryClass = {astro-ph.SR},
       adsurl = {https://ui.adsabs.harvard.edu/abs/2014NewA...26...72Y}
}

@ARTICLE{Jurcsik1996,
       author = {{Jurcsik}, J. and {Kovacs}, G.},
        title = "{Determination of [Fe/H] from the light curves of RR Lyrae stars.}",
      journal = {\aap},
         year = 1996,
        month = aug,
       volume = {312},
        pages = {111-120},
       adsurl = {https://ui.adsabs.harvard.edu/abs/1996A&A...312..111J}
}

@ARTICLE{Morgan2007,
       author = {{Morgan}, Siobahn M. and {Wahl}, Jennifer N. and {Wieckhorst}, Rachel M.},
        title = "{[Fe/H] relations for c-type RR Lyrae variables based upon Fourier coefficients}",
      journal = {\mnras},
         year = 2007,
        month = feb,
       volume = {374},
       number = {4},
        pages = {1421-1426},
          doi = {10.1111/j.1365-2966.2006.11247.x},
archivePrefix = {arXiv},
       eprint = {astro-ph/0506348},
 primaryClass = {astro-ph},
       adsurl = {https://ui.adsabs.harvard.edu/abs/2007MNRAS.374.1421M}
}

@misc{wang_stellar_2024,
	title = {Stellar atmospheric parameters of \${\textbackslash}sim\$ 11,000 {RR} Lyrae stars from {LAMOST} Spectra},
	url = {http://arxiv.org/abs/2404.15411},
	number = {{arXiv}:2404.15411},
	publisher = {{arXiv}},
	author = {Wang, Jiangtao and Shi, Jianrong and Fu, Jianning and Zong, Weikai and Li, Chunqian},
	urldate = {2024-07-12},
	date = {2024-04-23},
  year = {2024},
	langid = {english},
	eprinttype = {arxiv},
	eprint = {2404.15411 [astro-ph]},
}

@misc{sun_identifying_2024,
	title = {Identifying the Galactic Substructures in 5D Space Using All-sky {RR} Lyrae Stars in Gaia {DR}3},
	url = {http://arxiv.org/abs/2411.13122},
	number = {{arXiv}:2411.13122},
	publisher = {{arXiv}},
	author = {Sun, Shenglan and Wang, Fei and Zhang, Huawei and Xue, Xiang-Xiang and Huang, Yang and Zhang, Ruizhi and Rix, Hans-Walter and Li, Xinyi and Liu, Gaochao and Zhang, Lan and Yang, Chengqun and Zhang, Shuo},
	urldate = {2024-11-21},
	date = {2024-11-20},
  year = {2024},
	langid = {english},
	eprinttype = {arxiv},
	eprint = {2411.13122 [astro-ph]},
}

@article{dekany_photometric_2022,
	title = {Photometric Metallicity Prediction of Fundamental-mode {RR} Lyrae Stars in the Gaia Optical and K s Infrared Wave Bands by Deep Learning},
	volume = {261},
	issn = {0067-0049},
	url = {https://ui.adsabs.harvard.edu/abs/2022ApJS..261...33D},
	doi = {10.3847/1538-4365/ac74ba},
	pages = {33},
	journal = {The Astrophysical Journal Supplement Series},
	publisher = {{IOP}},
	author = {Dékány, István and Grebel, Eva K.},
	urldate = {2026-05-28},
	date = {2022-08-01},
	year = {2022},
	note = {{ADS} Bibcode: 2022ApJS..261...33D},
}

@article{Mullen2021,
  author  = {Mullen, Joseph P. and Marengo, Massimo and Mart{\'i}nez-V{\'a}zquez, Clara E. and Neeley, Jillian R. and Bono, Giuseppe and Dall'Ora, Massimo and Chaboyer, Brian and Th{\'e}venin, Fr{\'e}d{\'e}ric and Braga, Vittorio F. and Crestani, Juliana and Fabrizio, Michele and Fiorentino, Giuliana and Gilligan, Christina K. and Monelli, Matteo and Stetson, Peter B.},
  title   = {Metallicity of Galactic RR Lyrae from Optical and Infrared Light Curves. I. Period-Fourier-Metallicity Relations for Fundamental-mode RR Lyrae},
  journal = {The Astrophysical Journal},
  year    = {2021},
  volume  = {912},
  number  = {2},
  pages   = {144},
  doi     = {10.3847/1538-4357/abefd4}
}

@article{Cleveland01121979,
author = {William S. Cleveland},
title = {Robust Locally Weighted Regression and Smoothing Scatterplots},
journal = {Journal of the American Statistical Association},
volume = {74},
number = {368},
pages = {829--836},
year = {1979},
publisher = {Taylor \& Francis},
doi = {10.1080/01621459.1979.10481038},
}

@ARTICLE{Yuan2015,
       author = {{Yuan}, Haibo and {Liu}, Xiaowei and {Xiang}, Maosheng and {Huang}, Yang and {Chen}, Bingqiu},
        title = "{Stellar Loci. I. Metallicity Dependence and Intrinsic Widths}",
      journal = {\apj},
         year = 2015,
        month = feb,
       volume = {799},
       number = {2},
          eid = {134},
        pages = {134},
          doi = {10.1088/0004-637X/799/2/134},
archivePrefix = {arXiv},
       eprint = {1412.1232},
 primaryClass = {astro-ph.SR},
       adsurl = {https://ui.adsabs.harvard.edu/abs/2015ApJ...799..134Y}
}

@ARTICLE{Casagrande2019,
       author = {{Casagrande}, L. and {Wolf}, C. and {Mackey}, A.~D. and {Nordlander}, T. and {Yong}, D. and {Bessell}, M.},
        title = "{SkyMapper stellar parameters for Galactic Archaeology on a grand-scale}",
      journal = {\mnras},
         year = 2019,
        month = jan,
       volume = {482},
       number = {2},
        pages = {2770-2787},
          doi = {10.1093/mnras/sty2878},
archivePrefix = {arXiv},
       eprint = {1810.09581},
 primaryClass = {astro-ph.SR},
       adsurl = {https://ui.adsabs.harvard.edu/abs/2019MNRAS.482.2770C}
}

@ARTICLE{Lu2024,
       author = {{Lu}, Xue and {Yuan}, Haibo and {Xu}, Shuai and {Zhang}, Ruoyi and {Xiao}, Kai and {Huang}, Yang and {Beers}, Timothy C. and {Hong}, Jihye},
        title = "{Stellar Loci. VII. Photometric Metallicities of 5 Million FGK Stars Based on GALEX GR6+7 AIS and Gaia EDR3}",
      journal = {\apjs},
         year = 2024,
        month = mar,
       volume = {271},
       number = {1},
          eid = {26},
        pages = {26},
          doi = {10.3847/1538-4365/ad1eea},
       adsurl = {https://ui.adsabs.harvard.edu/abs/2024ApJS..271...26L}
}

@article{lu_stellar_2025,
	title = {Stellar Loci. {IX}. Estimation of Stellar Parameters from {CSST}-like Photometry},
	volume = {995},
	issn = {0004-637X},
	url = {https://ui.adsabs.harvard.edu/abs/2025ApJ...995...53L},
	doi = {10.3847/1538-4357/ae17b7},
	pages = {53},
	journal = {The Astrophysical Journal},
	publisher = {{IOP}},
	author = {Lu, Xue and Yuan, Haibo and Xiao, Kai and Huang, Bowen and Zhang, Ruoyi and Yang, Lin and Beers, Timothy C. and Xu, Shuai},
	urldate = {2026-05-29},
	date = {2025-12-01},
  year = {2025},
	note = {{ADS} Bibcode: 2025ApJ...995...53L},
}

@INPROCEEDINGS{Yuan2020,
       author = {{Yuan}, Xiangyan and {Li}, Zhengyang and {Liu}, Xiaowei and {Niu}, Dongsheng and {Lu}, Qishui and {Jiang}, Fanghua and {Wang}, Yuefei and {Li}, Xiaoyan and {Liang}, YongJun and {Wang}, Hai and {Zhang}, Chao and {Wang}, Jinfeng and {Li}, Bo and {Tian}, Jie and {Lu}, Haiping and {Chen}, Bingqiu and {Huang}, Yang and {Liu}, Xiangkun and {Yao}, Zhengqiu and {Cui}, Xiangqun and {Li}, Guoping},
        title = "{Development of the Multi-channel Photometric Survey telescope}",
    booktitle = {Ground-based and Airborne Telescopes VIII},
         year = 2020,
       editor = {{Marshall}, Heather K. and {Spyromilio}, Jason and {Usuda}, Tomonori},
       series = {Society of Photo-Optical Instrumentation Engineers (SPIE) Conference Series},
       volume = {11445},
        month = dec,
          eid = {114457M},
        pages = {114457M},
          doi = {10.1117/12.2562334},
       adsurl = {https://ui.adsabs.harvard.edu/abs/2020SPIE11445E..7MY}
}

@ARTICLE{Kurucz2005,
       author = {{Kurucz}, Robert L.},
        title = "{ATLAS12, SYNTHE, ATLAS9, WIDTH9, et cetera}",
      journal = {Memorie della Societa Astronomica Italiana Supplementi},
         year = 2005,
        month = jan,
       volume = {8},
        pages = {14},
       adsurl = {https://ui.adsabs.harvard.edu/abs/2005MSAIS...8...14K}
}

@ARTICLE{Kurucz1979,
       author = {{Kurucz}, R.~L.},
        title = "{Model atmospheres for G, F, A, B, and O stars.}",
      journal = {\apjs},
         year = 1979,
        month = may,
       volume = {40},
        pages = {1-340},
          doi = {10.1086/190589},
       adsurl = {https://ui.adsabs.harvard.edu/abs/1979ApJS...40....1K}
}

@ARTICLE{2025arXiv250402924M,
       author = {{Medina}, Gustavo E. and {Li}, Ting S. and {Koposov}, Sergey E. and {Riley}, A.~H. and {Beraldo e Silva}, L. and {Valluri}, M. and {Wang}, W. and {Bystr{\"o}m}, A. and {Gnedin}, O.~Y. and {Carlberg}, R.~G. and {Kizhuprakkat}, N. and {Weaver}, B.~A. and {Aguilar}, J. and {Ahlen}, S. and {Bianchi}, D. and {Brooks}, D. and {Claybaugh}, T. and {Cooper}, A.~P. and {de la Macorra}, A. and {Dey}, A. and {Doel}, P. and {Font-Ribera}, A. and {Forero-Romero}, J.~E. and {Gazta{\~n}aga}, E. and {Gontcho}, S. Gontcho A and {Gutierrez}, G. and {Guy}, J. and {Honscheid}, K. and {Ishak}, M. and {Kisner}, T. and {Landriau}, M. and {Le Guillou}, L. and {Meisner}, A. and {Miquel}, R. and {Myers}, A.~D. and {Nadathur}, S. and {Poppett}, C. and {Prada}, F. and {P{\'e}rez-R{\`a}fols}, I. and {Rossi}, G. and {Sanchez}, E. and {Seo}, H. and {Sprayberry}, D. and {Tarl{\'e}}, G. and {Wechsler}, R.~H. and {Zhou}, R. and {Zou}, H.},
        title = "{The DESI Y1 RR Lyrae catalog I: Empirical modeling of the cyclic variation of spectroscopic properties and a chemodynamical analysis of the outer halo}",
      journal = {arXiv e-prints},
         year = 2025,
        month = apr,
          eid = {arXiv:2504.02924},
        pages = {arXiv:2504.02924},
          doi = {10.48550/arXiv.2504.02924},
archivePrefix = {arXiv},
       eprint = {2504.02924},
 primaryClass = {astro-ph.GA},
       adsurl = {https://ui.adsabs.harvard.edu/abs/2025arXiv250402924M}
}

@ARTICLE{2023ApJ...947...37C,
       author = {{Cooper}, Andrew P. and {Koposov}, Sergey E. and {Allende Prieto}, Carlos and {Manser}, Christopher J. and {Kizhuprakkat}, Namitha and {Myers}, Adam D. and {Dey}, Arjun and {G{\"a}nsicke}, Boris T. and {Li}, Ting S. and {Rockosi}, Constance and {Valluri}, Monica and {Najita}, Joan and {Deason}, Alis and {Raichoor}, Anand and {Wang}, M.-Y. and {Ting}, Y.-S. and {Kim}, Bokyoung and {Carrillo}, Andreia and {Wang}, Wenting and {Beraldo e Silva}, Leandro and {Han}, Jiwon Jesse and {Ding}, Jiani and {S{\'a}nchez-Conde}, Miguel and {Aguilar}, Jessica N. and {Ahlen}, Steven and {Bailey}, Stephen and {Belokurov}, Vasily and {Brooks}, David and {Cunha}, Katia and {Dawson}, Kyle and {de la Macorra}, Axel and {Doel}, Peter and {Eisenstein}, Daniel J. and {Fagrelius}, Parker and {Fanning}, Kevin and {Font-Ribera}, Andreu and {Forero-Romero}, Jaime E. and {Gazta{\~n}aga}, Enrique and {Gontcho a Gontcho}, Satya and {Guy}, Julien and {Honscheid}, Klaus and {Kehoe}, Robert and {Kisner}, Theodore and {Kremin}, Anthony and {Landriau}, Martin and {Levi}, Michael E. and {Martini}, Paul and {Meisner}, Aaron M. and {Miquel}, Ramon and {Moustakas}, John and {Nie}, Jundan J.~D. and {Palanque-Delabrouille}, Nathalie and {Percival}, Will J. and {Poppett}, Claire and {Prada}, Francisco and {Rehemtulla}, Nabeel and {Schlafly}, Edward and {Schlegel}, David and {Schubnell}, Michael and {Sharples}, Ray M. and {Tarl{\'e}}, Gregory and {Wechsler}, Risa H. and {Weinberg}, David H. and {Zhou}, Zhimin and {Zou}, Hu},
        title = "{Overview of the DESI Milky Way Survey}",
      journal = {\apj},
         year = 2023,
        month = apr,
       volume = {947},
       number = {1},
          eid = {37},
        pages = {37},
          doi = {10.3847/1538-4357/acb3c0},
archivePrefix = {arXiv},
       eprint = {2208.08514},
 primaryClass = {astro-ph.GA},
       adsurl = {https://ui.adsabs.harvard.edu/abs/2023ApJ...947...37C}
}

\clearpage
\appendix
\raggedbottom
\setlength{\intextsep}{6pt plus 2pt minus 2pt}
\setlength{\textfloatsep}{8pt plus 2pt minus 4pt}
\setlength{\floatsep}{6pt plus 2pt minus 2pt}
\section{Results for the Other Selected RR Lyrae Stars}
\label{app:other_stars}

\begin{figure}
    \centering
    \includegraphics[width=\columnwidth]{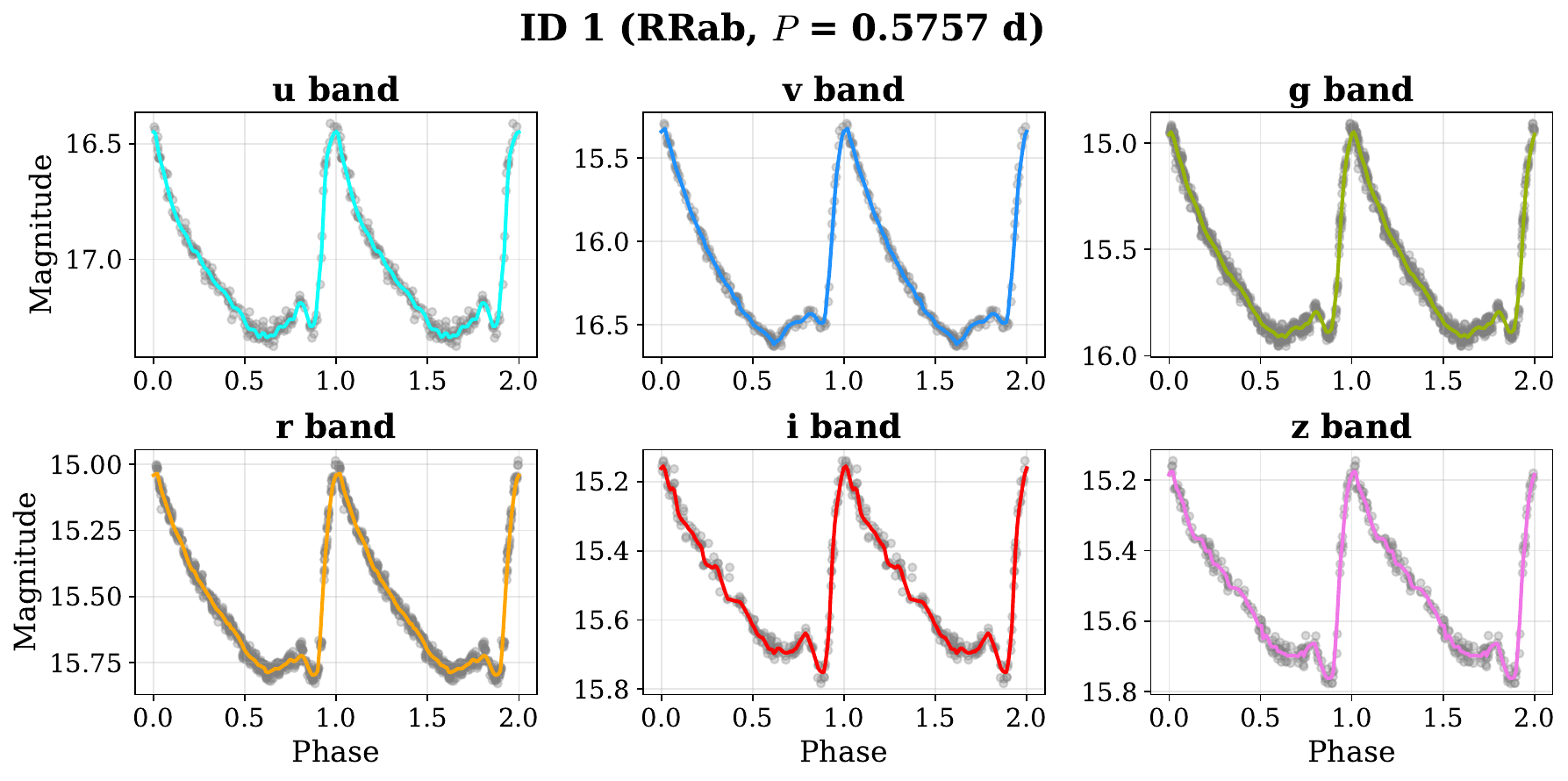}
    \caption{RRab star ID~1 (M15): multi-band light-curve templates.}
    \label{fig:app_lc_1}
\end{figure}

\begin{figure}
    \centering
    \includegraphics[width=\columnwidth]{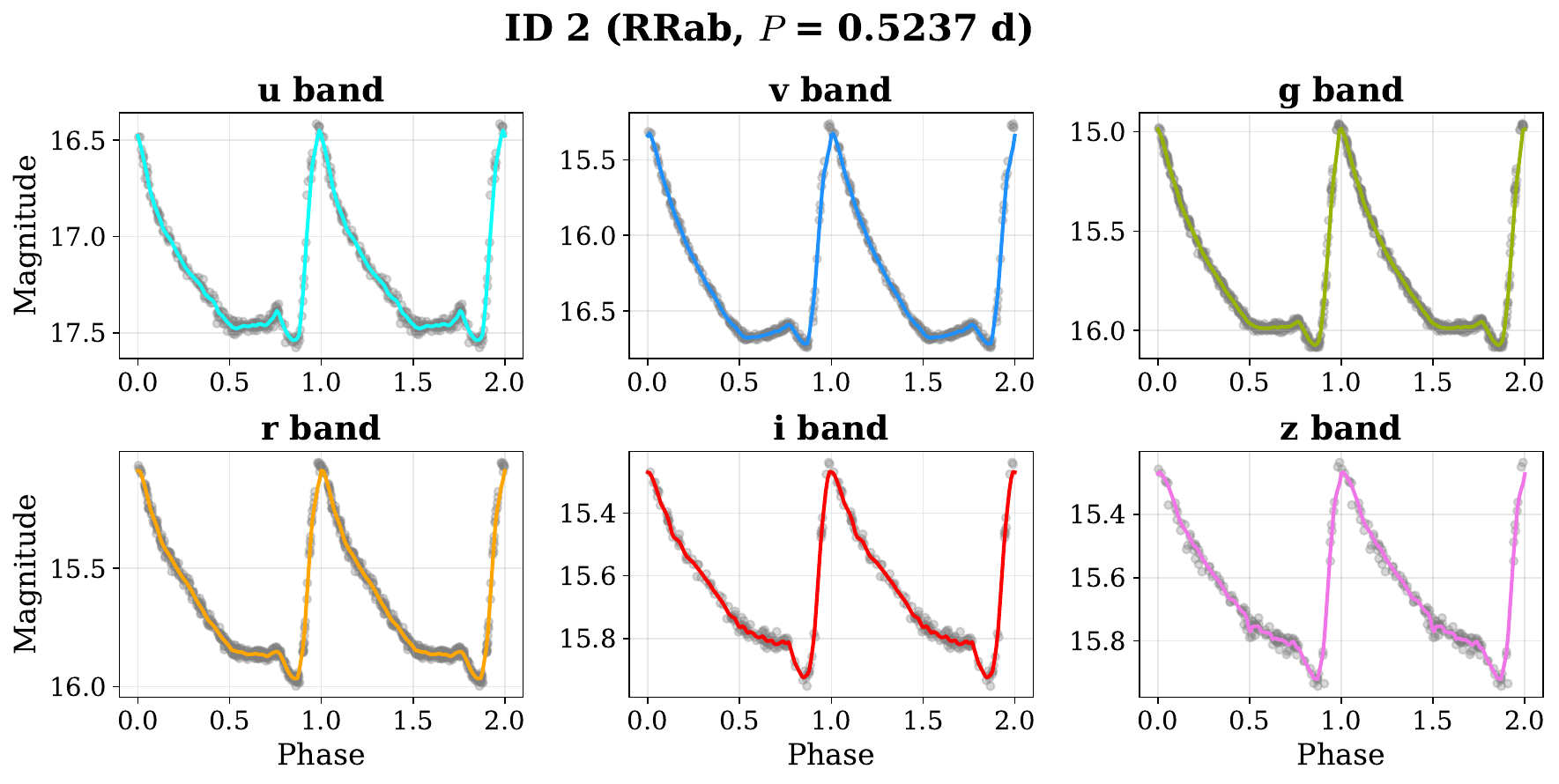}
    \caption{RRab star ID~2 (M3): multi-band light-curve templates.}
    \label{fig:app_lc_2}
\end{figure}

\begin{figure}
    \centering
    \includegraphics[width=\columnwidth]{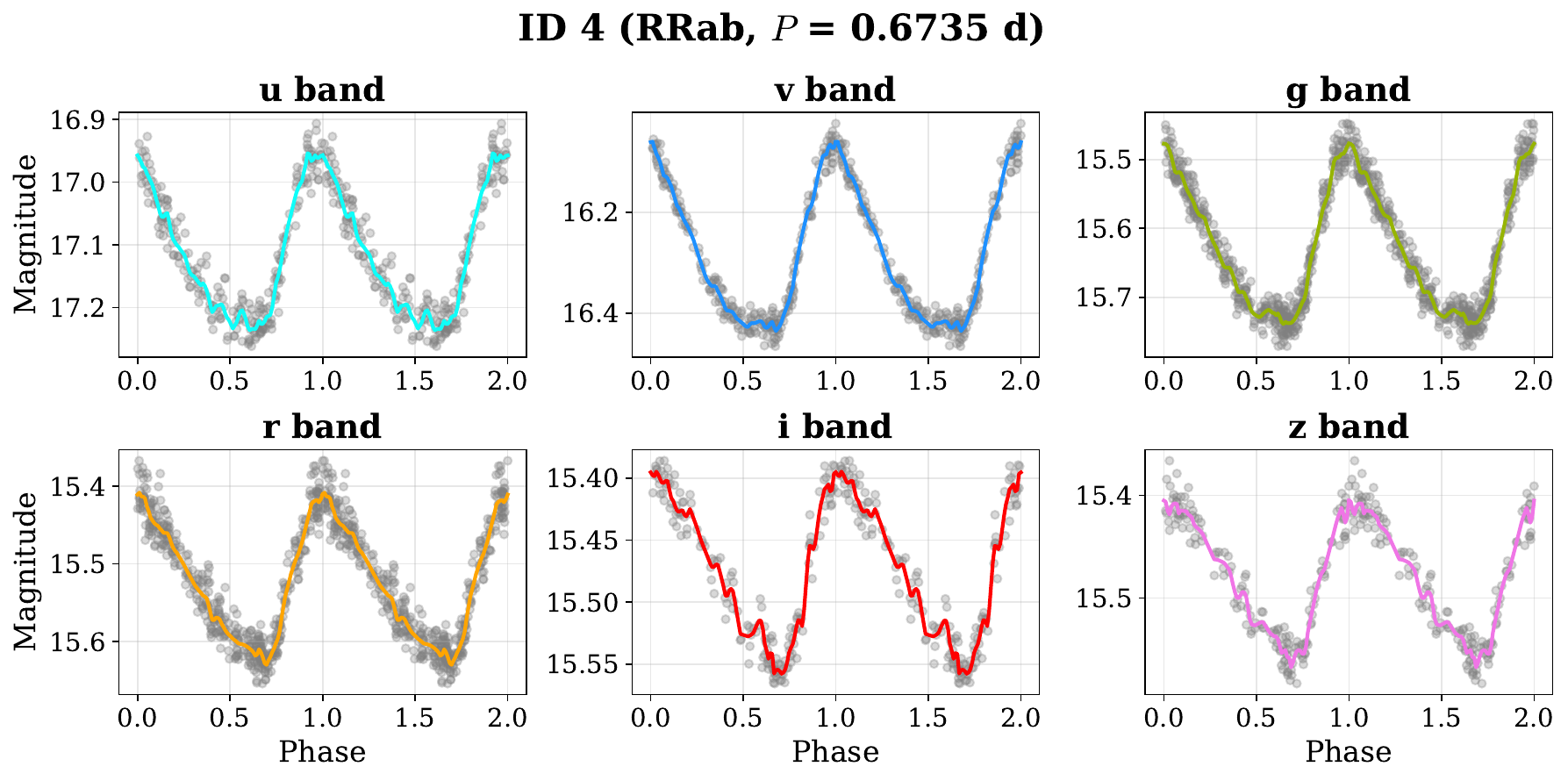}
    \caption{RRab star ID~4 (M3): multi-band light-curve templates.}
    \label{fig:app_lc_4}
\end{figure}

\begin{figure}
    \centering
    \includegraphics[width=\columnwidth]{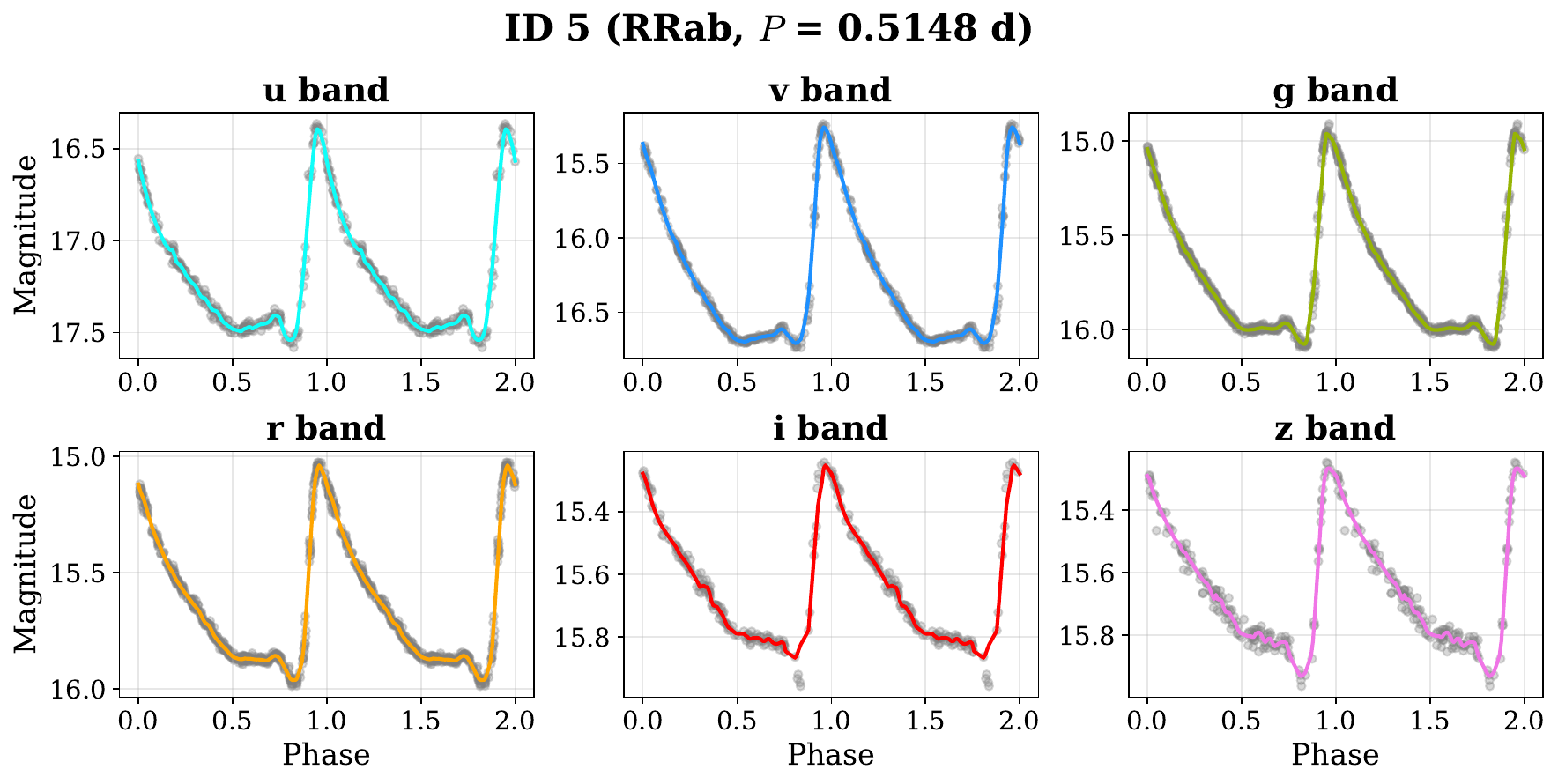}
    \caption{RRab star ID~5 (M3): multi-band light-curve templates.}
    \label{fig:app_lc_5}
\end{figure}

\begin{figure}
    \centering
    \includegraphics[width=\columnwidth]{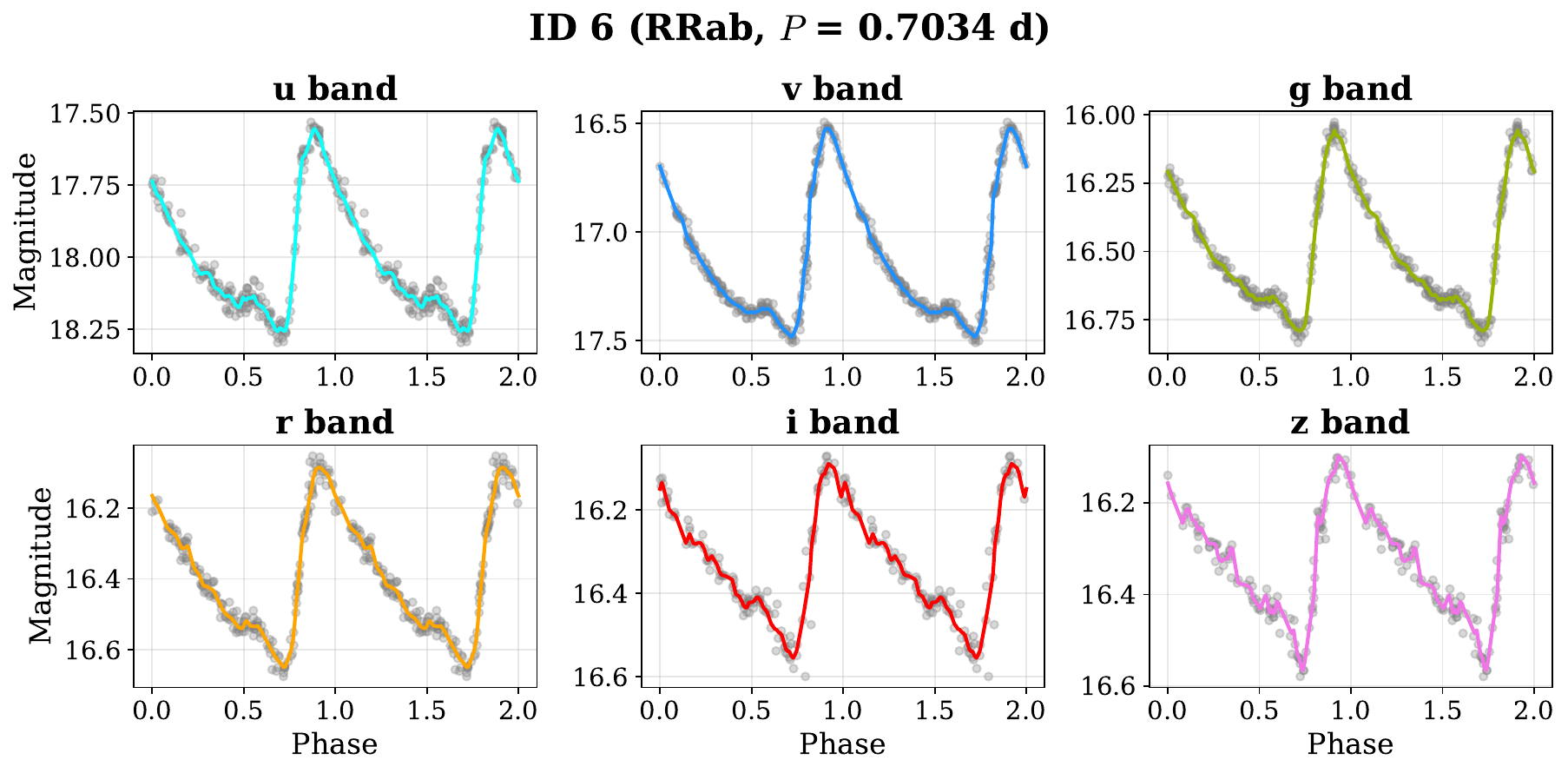}
    \caption{RRab star ID~6 (NGC~5466): multi-band light-curve templates.}
    \label{fig:app_lc_6}
\end{figure}

\begin{figure}
    \centering
    \includegraphics[width=\columnwidth]{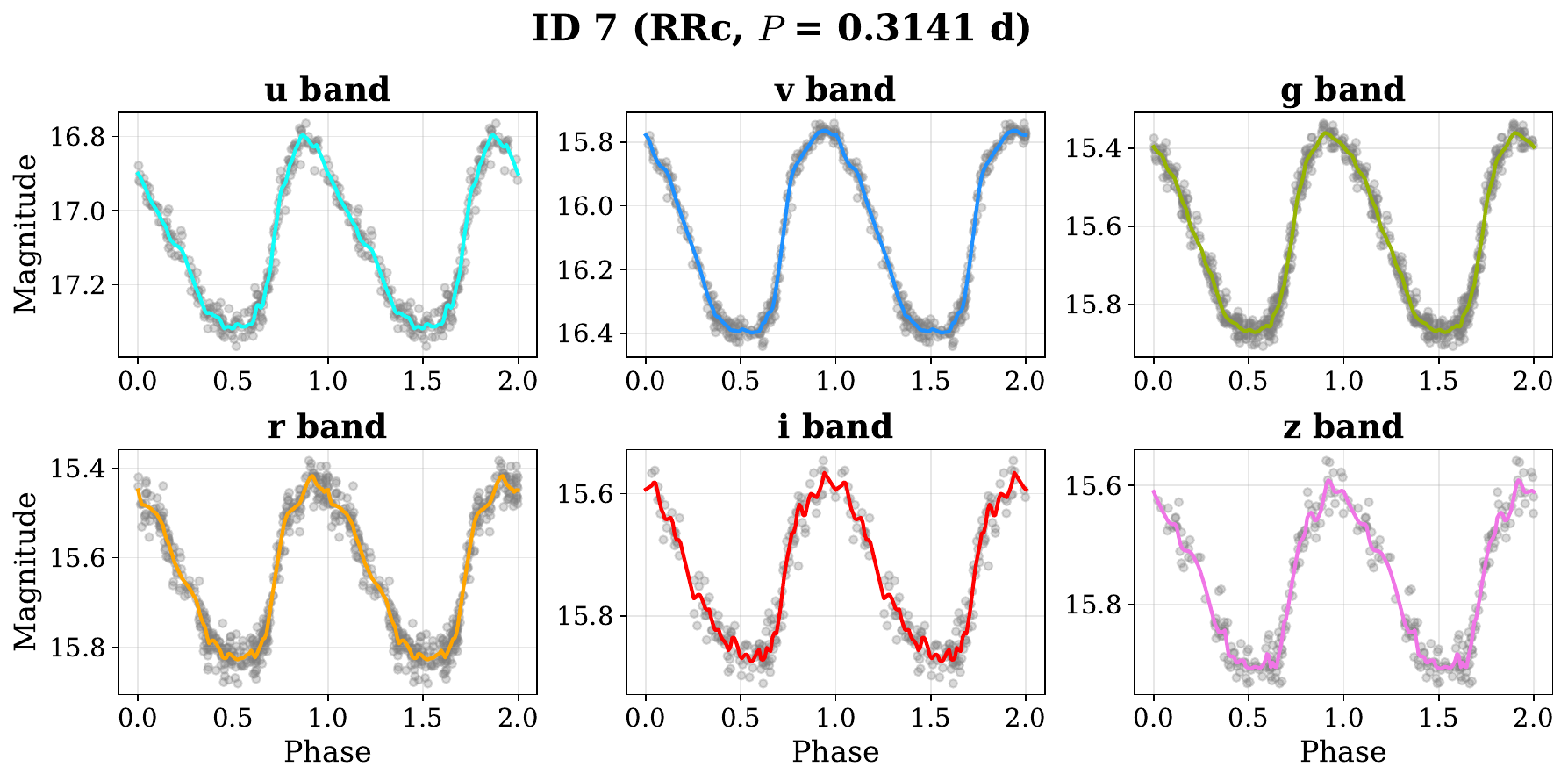}
    \caption{RRc star ID~7 (M3): multi-band light-curve templates.}
    \label{fig:app_lc_7}
\end{figure}

\begin{figure}
    \centering
    \includegraphics[width=\columnwidth]{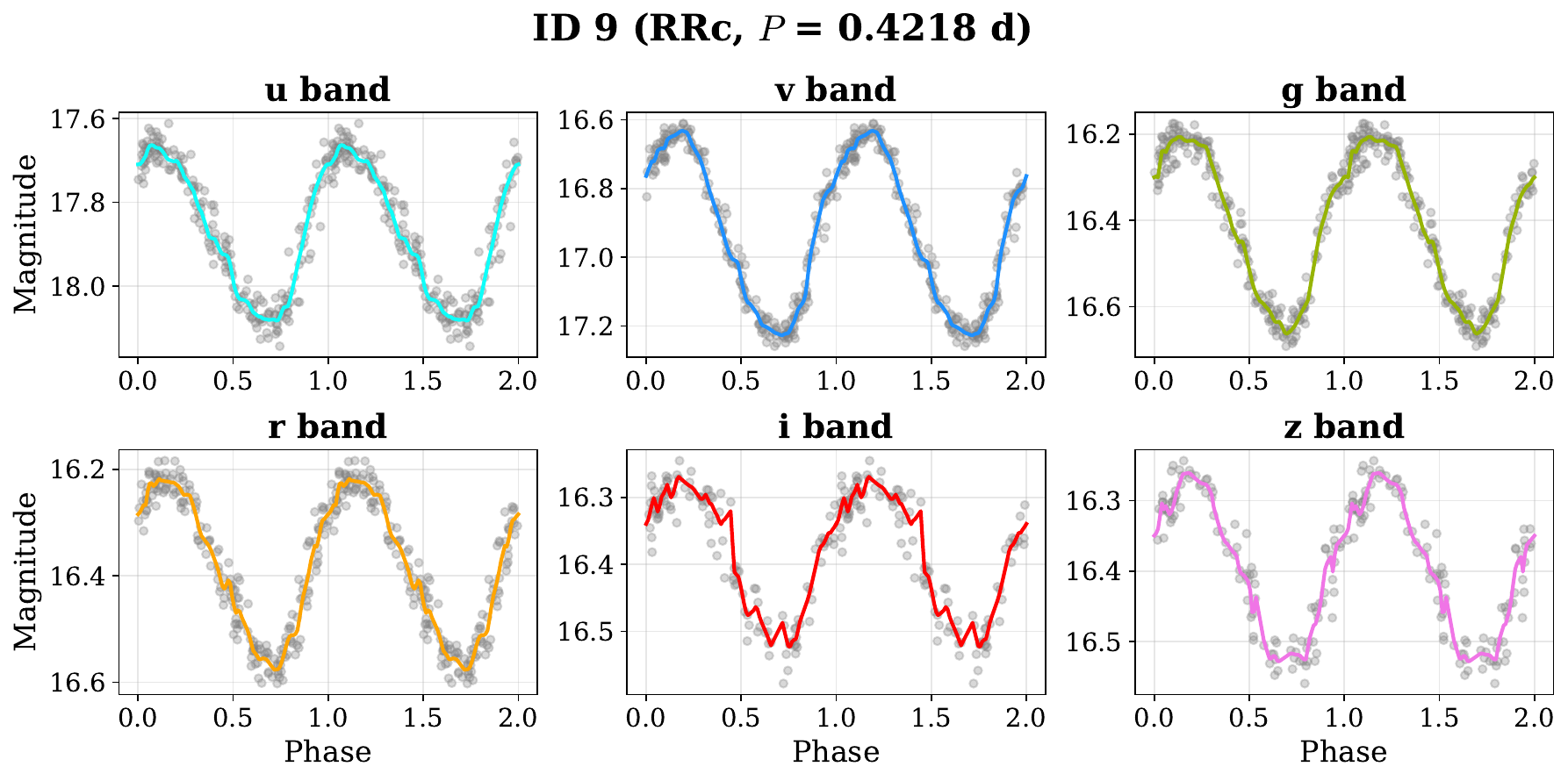}
    \caption{RRc star ID~9 (NGC~5466): multi-band light-curve templates.}
    \label{fig:app_lc_9}
\end{figure}

\begin{figure}
    \centering
    \includegraphics[width=\columnwidth]{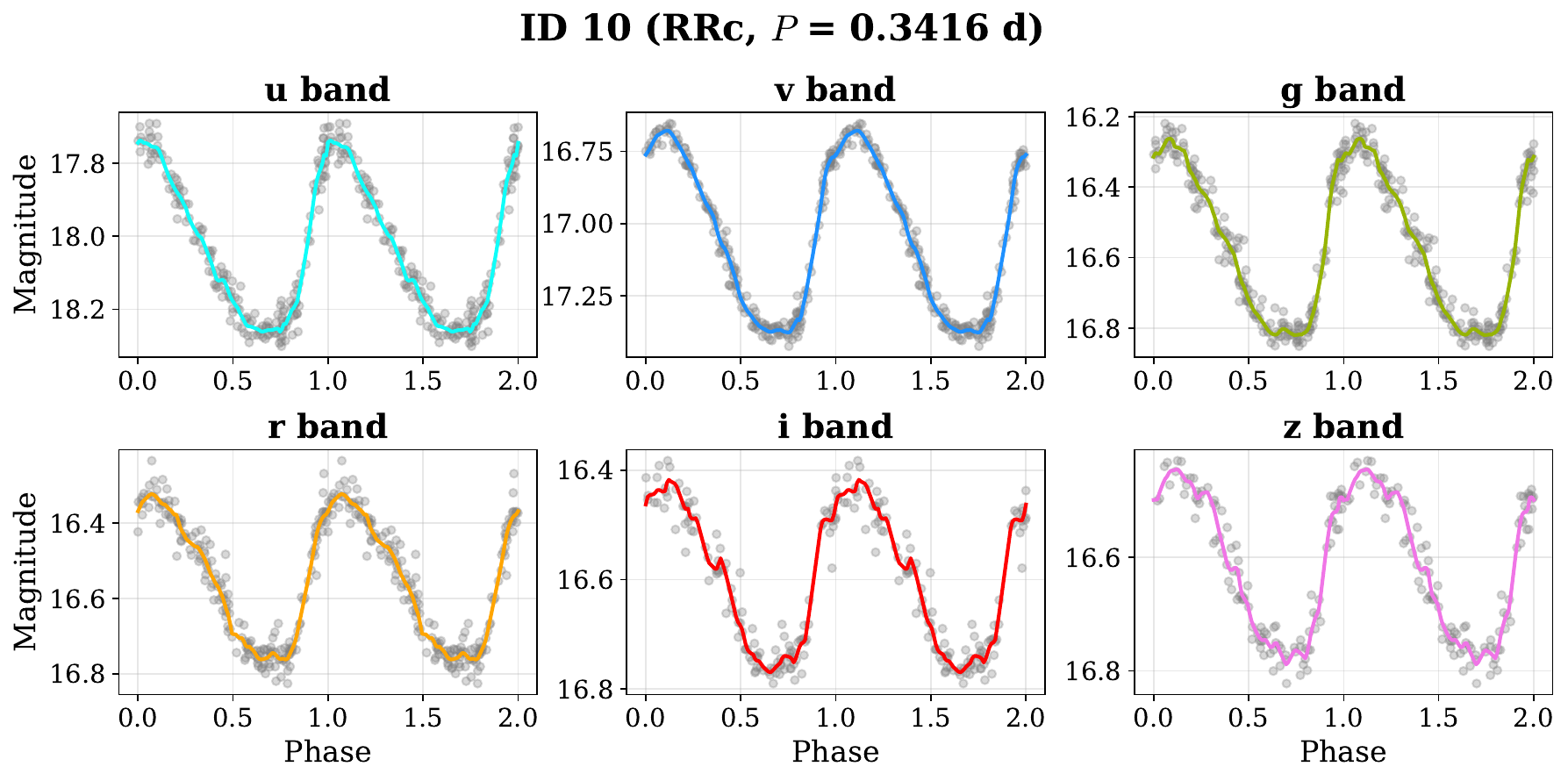}
    \caption{RRc star ID~10 (NGC~5466): multi-band light-curve templates.}
    \label{fig:app_lc_10}
\end{figure}

\begin{figure}
    \centering
    \includegraphics[width=\columnwidth]{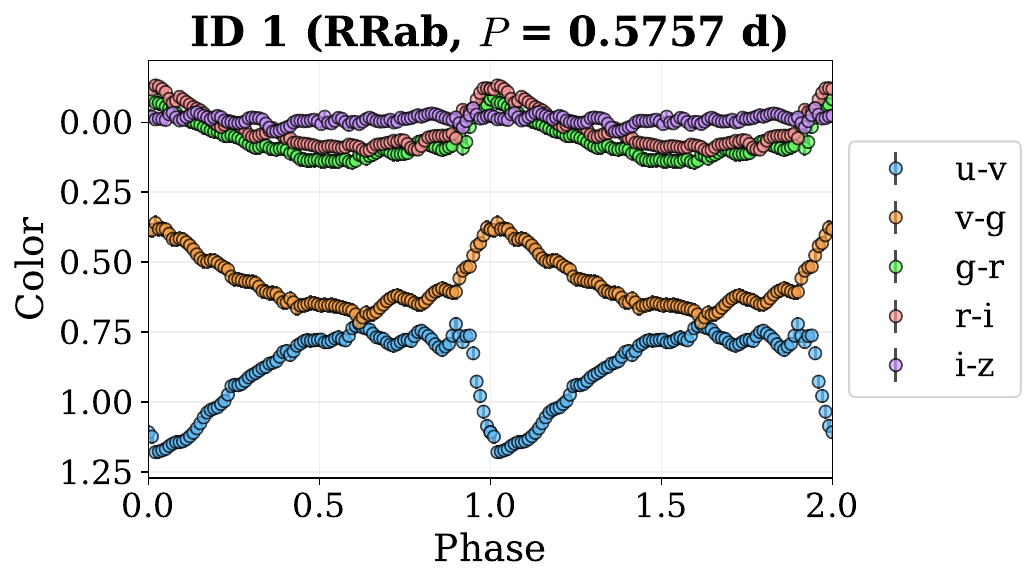}
    \caption{RRab star ID~1 (M15): phase-dependent color curves.}
    \label{fig:app_color_1}
\end{figure}

\begin{figure}
    \centering
    \includegraphics[width=\columnwidth]{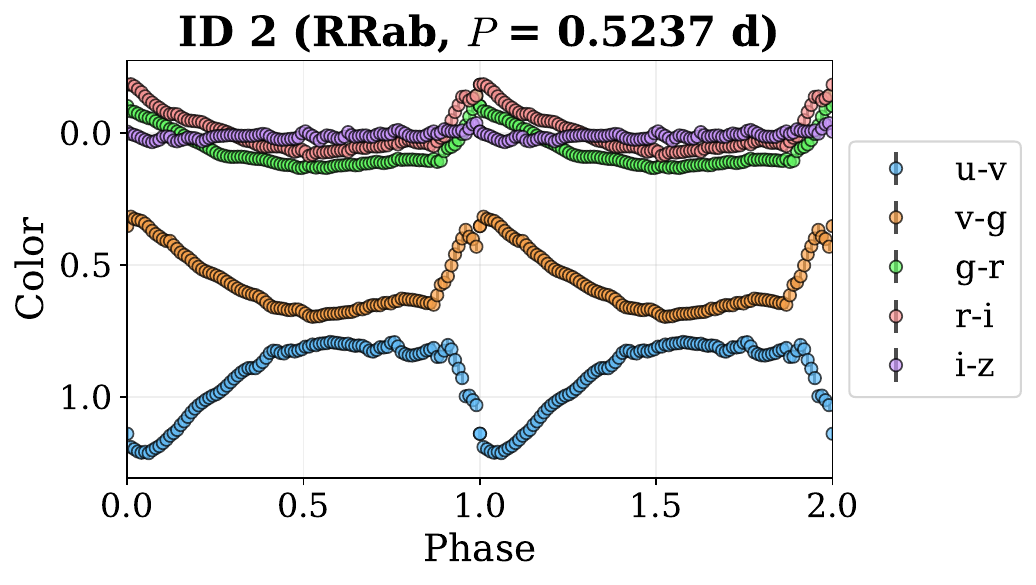}
    \caption{RRab star ID~2 (M3): phase-dependent color curves.}
    \label{fig:app_color_2}
\end{figure}

\begin{figure}
    \centering
    \includegraphics[width=\columnwidth]{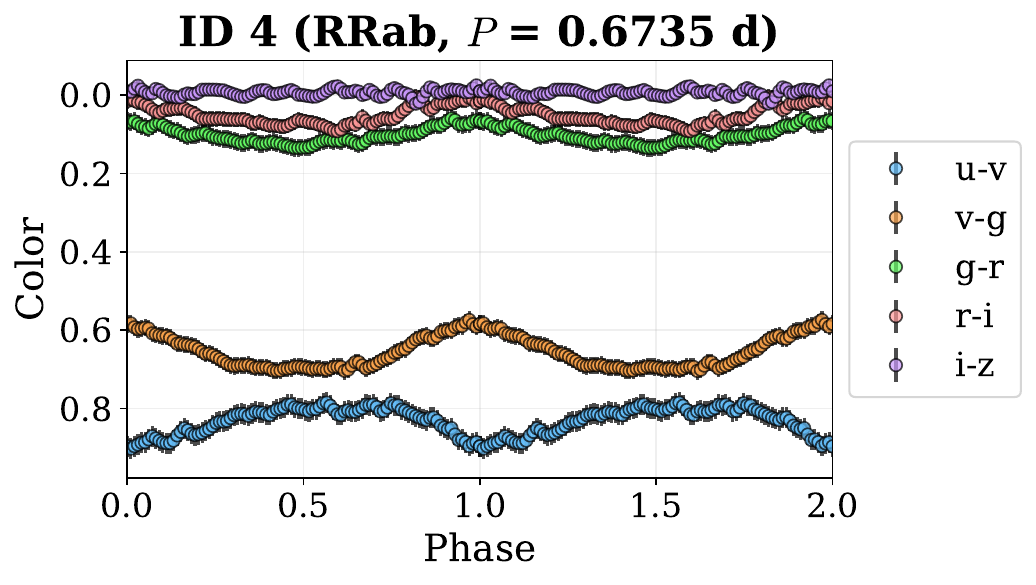}
    \caption{RRab star ID~4 (M3): phase-dependent color curves.}
    \label{fig:app_color_4}
\end{figure}

\begin{figure}
    \centering
    \includegraphics[width=\columnwidth]{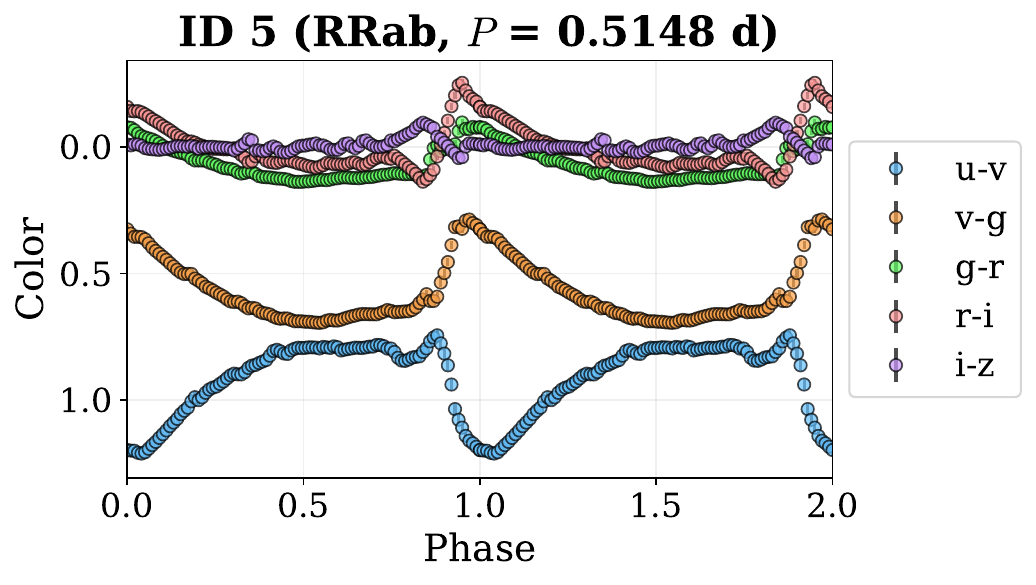}
    \caption{RRab star ID~5 (M3): phase-dependent color curves.}
    \label{fig:app_color_5}
\end{figure}

\begin{figure}
    \centering
    \includegraphics[width=\columnwidth]{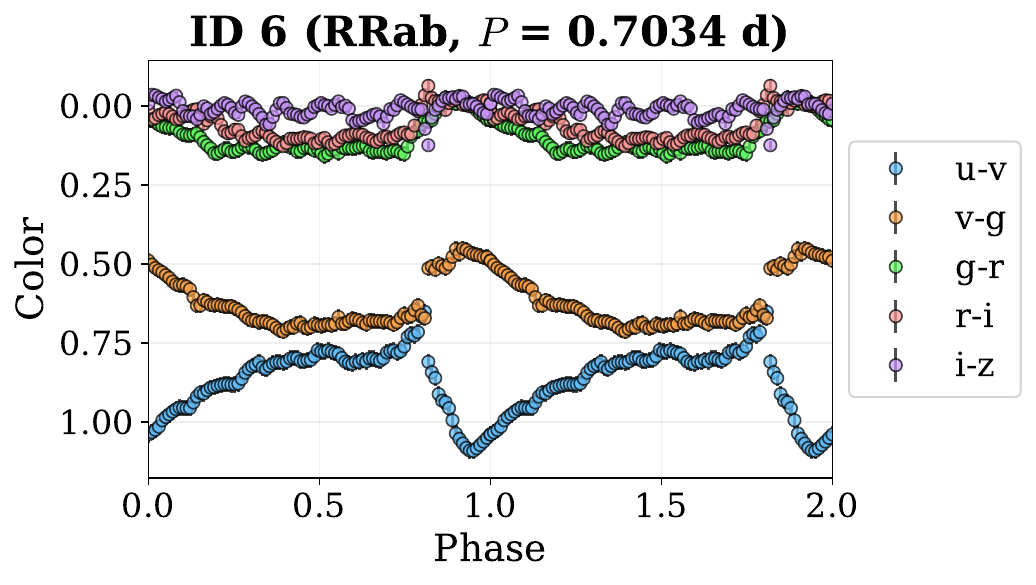}
    \caption{RRab star ID~6 (NGC~5466): phase-dependent color curves.}
    \label{fig:app_color_6}
\end{figure}

\begin{figure}
    \centering
    \includegraphics[width=\columnwidth]{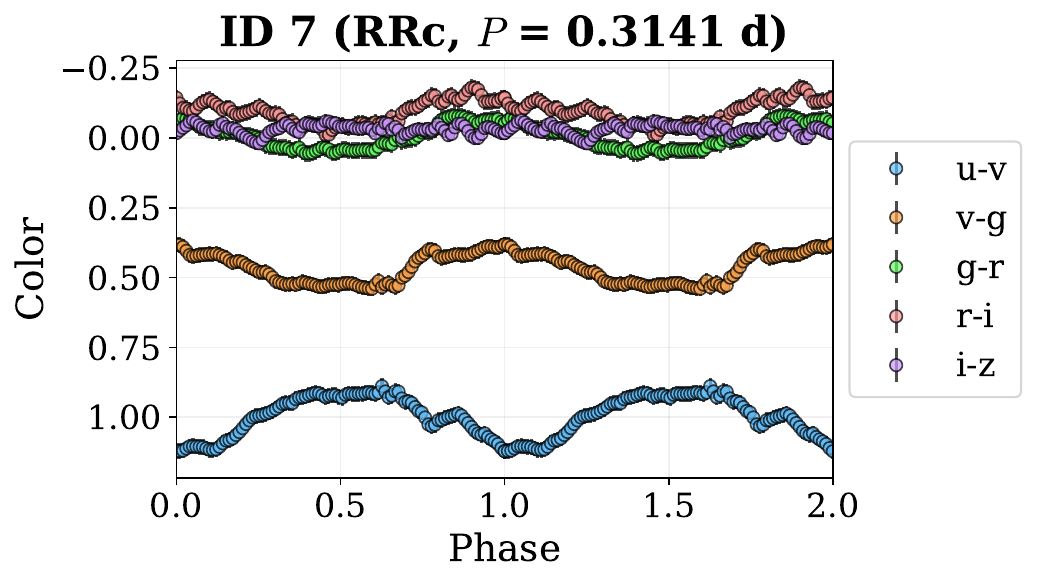}
    \caption{RRc star ID~7 (M3): phase-dependent color curves.}
    \label{fig:app_color_7}
\end{figure}

\begin{figure}
    \centering
    \includegraphics[width=\columnwidth]{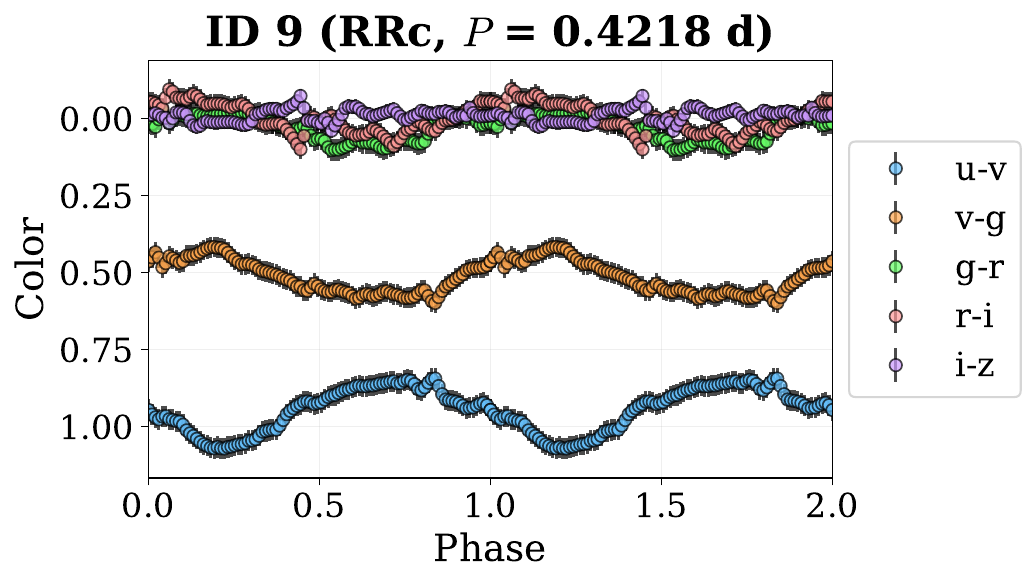}
    \caption{RRc star ID~9 (NGC~5466): phase-dependent color curves.}
    \label{fig:app_color_9}
\end{figure}

\begin{figure}
    \centering
    \includegraphics[width=\columnwidth]{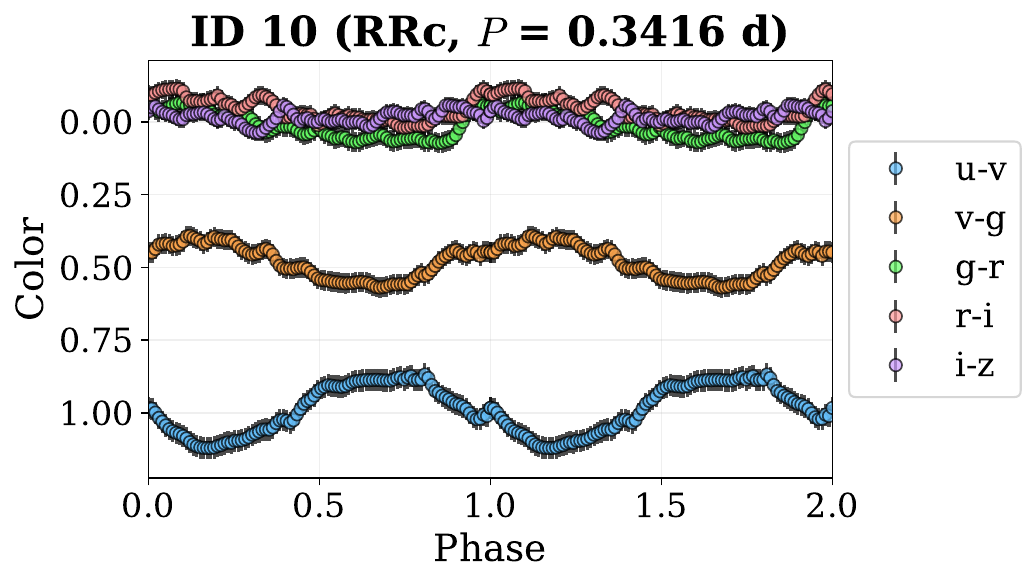}
    \caption{RRc star ID~10 (NGC~5466): phase-dependent color curves.}
    \label{fig:app_color_10}
\end{figure}

\begin{figure}
    \centering
    \includegraphics[width=\columnwidth]{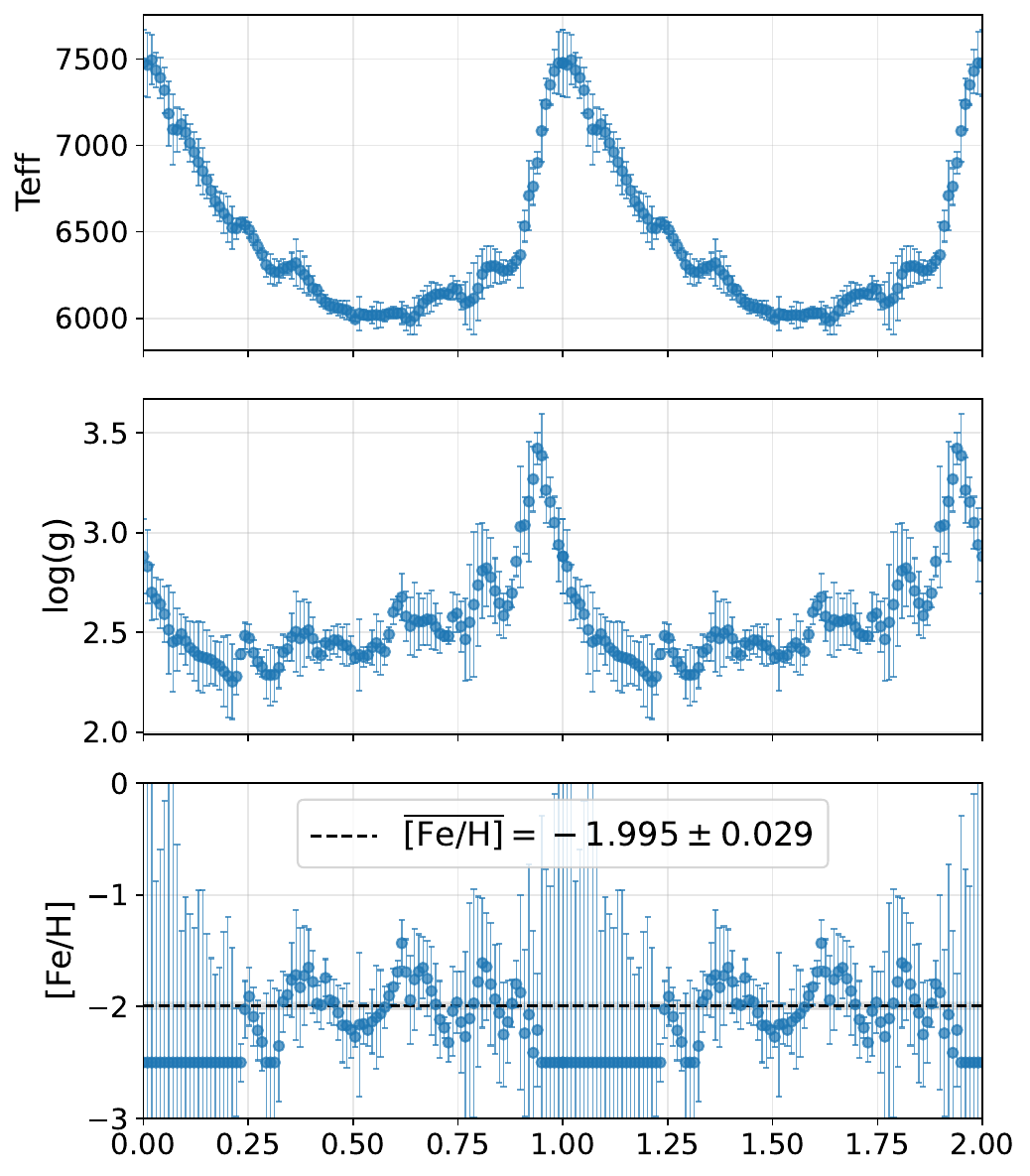}
    \caption{RRab star ID~1 (M15): phase-dependent atmospheric parameters.}
    \label{fig:app_param_1}
\end{figure}

\begin{figure}
    \centering
    \includegraphics[width=\columnwidth]{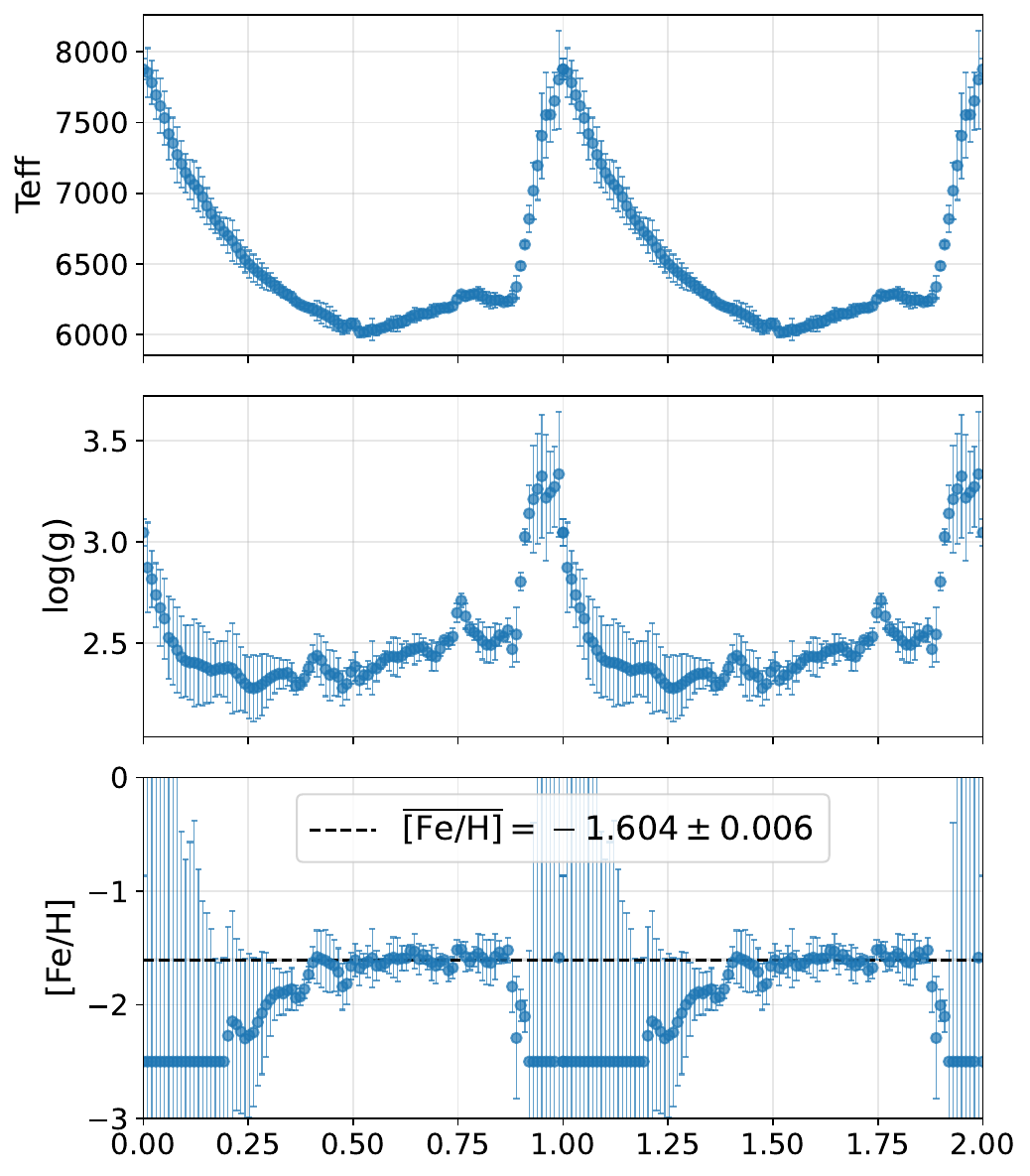}
    \caption{RRab star ID~2 (M3): phase-dependent atmospheric parameters.}
    \label{fig:app_param_2}
\end{figure}

\begin{figure}
    \centering
    \includegraphics[width=\columnwidth]{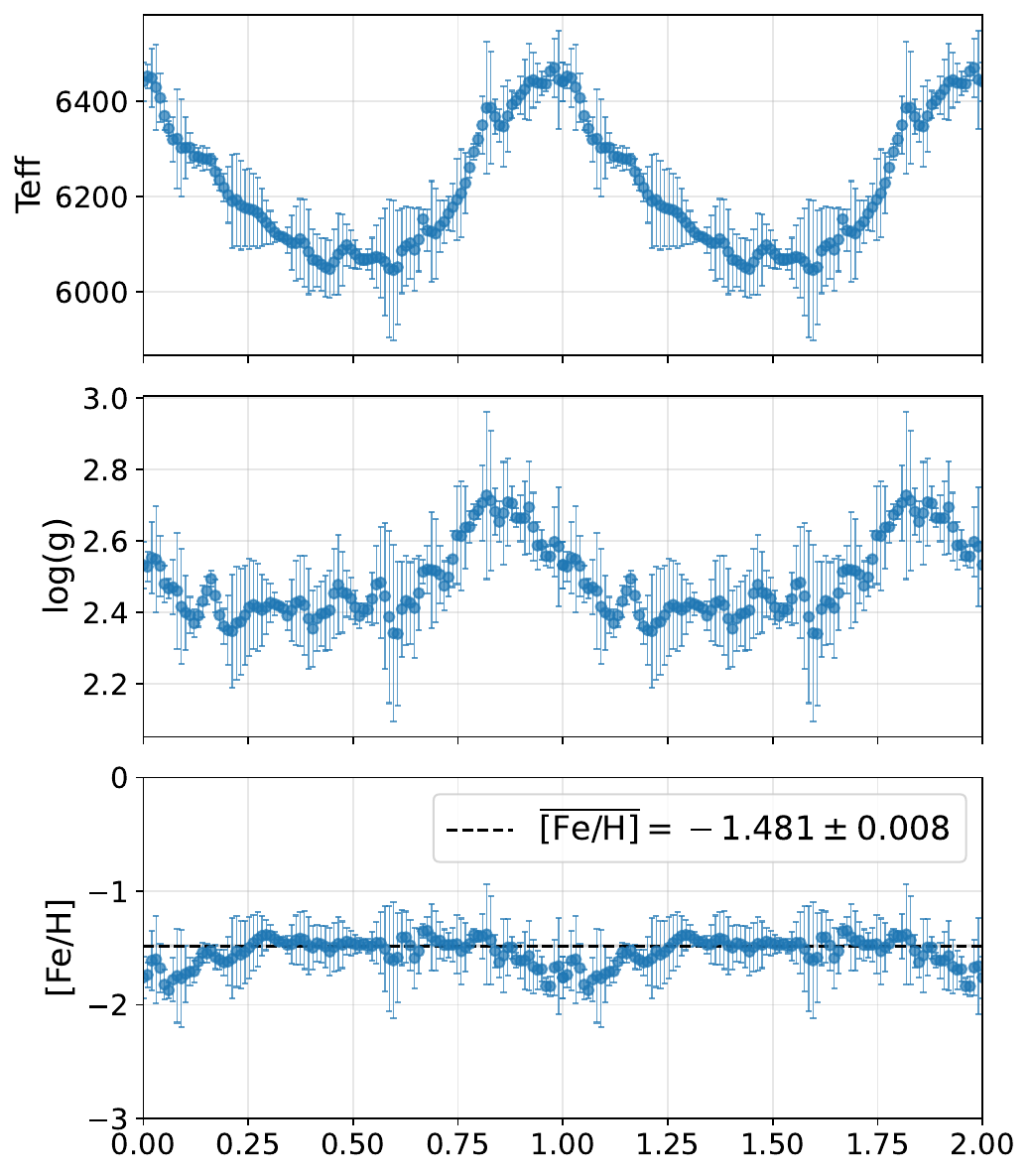}
    \caption{RRab star ID~4 (M3): phase-dependent atmospheric parameters.}
    \label{fig:app_param_4}
\end{figure}

\begin{figure}
    \centering
    \includegraphics[width=\columnwidth]{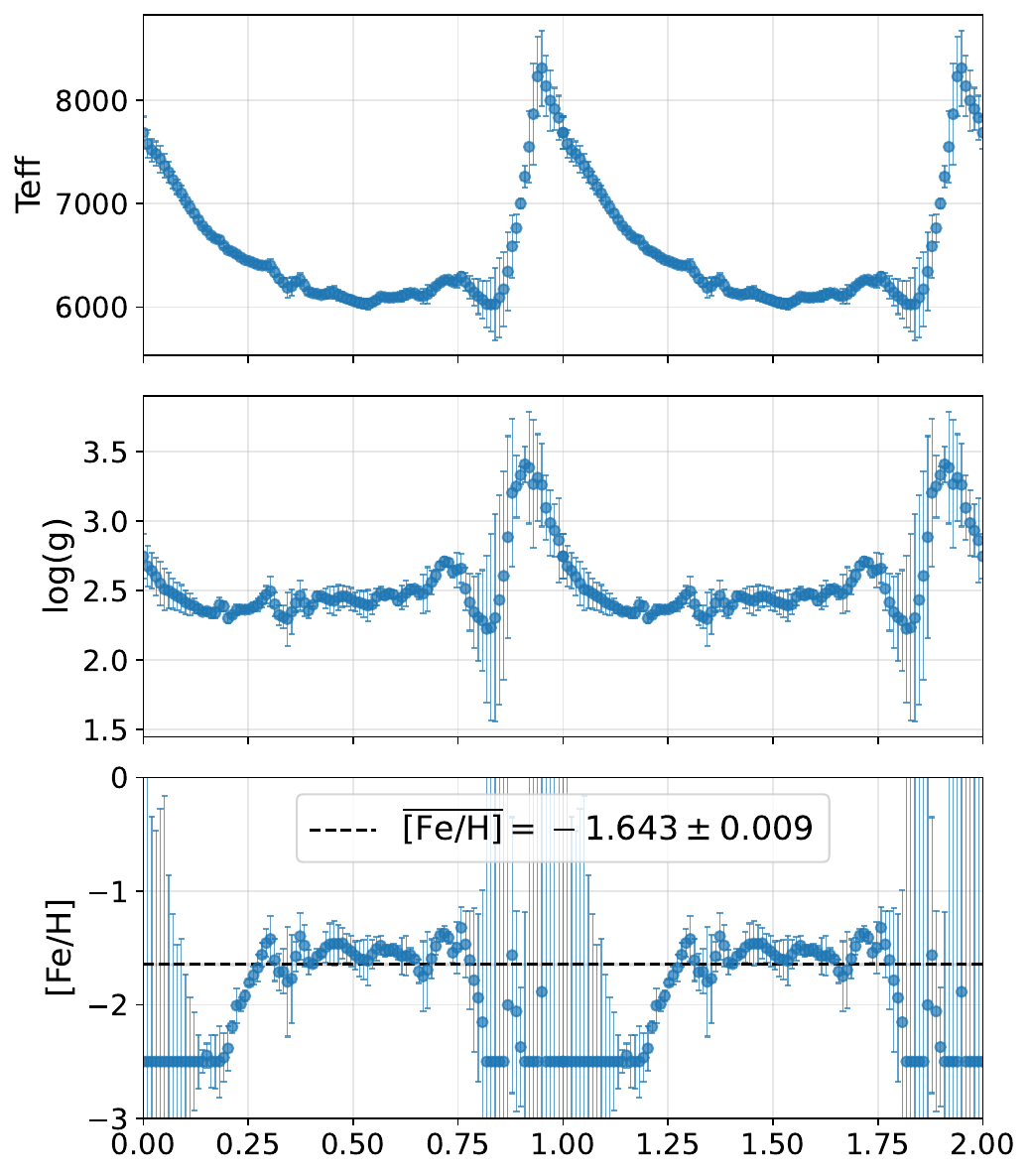}
    \caption{RRab star ID~5 (M3): phase-dependent atmospheric parameters.}
    \label{fig:app_param_5}
\end{figure}

\begin{figure}
    \centering
    \includegraphics[width=\columnwidth]{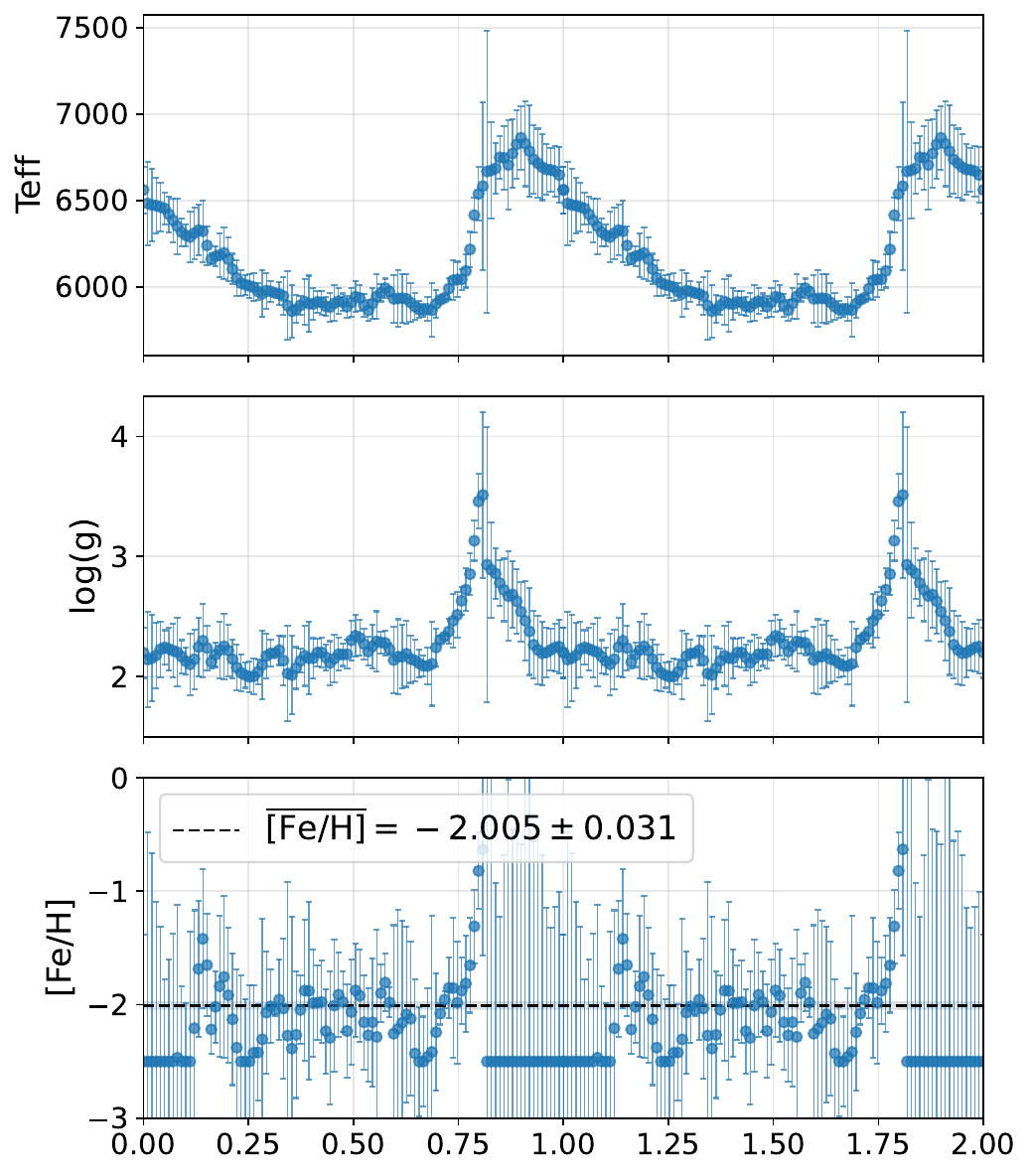}
    \caption{RRab star ID~6 (NGC~5466): phase-dependent atmospheric parameters.}
    \label{fig:app_param_6}
\end{figure}

\begin{figure}
    \centering
    \includegraphics[width=\columnwidth]{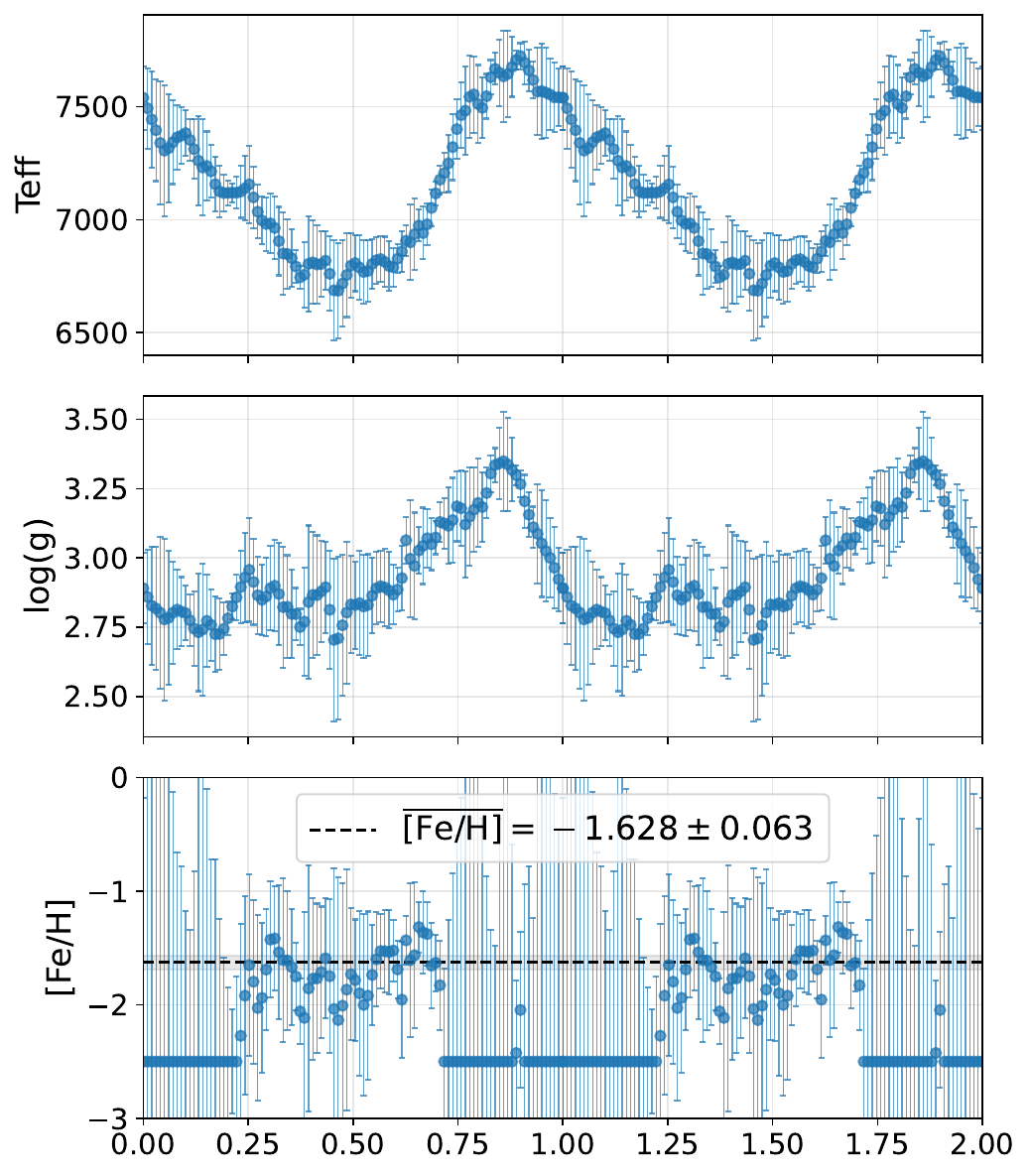}
    \caption{RRc star ID~7 (M3): phase-dependent atmospheric parameters.}
    \label{fig:app_param_7}
\end{figure}

\begin{figure}
    \centering
    \includegraphics[width=\columnwidth]{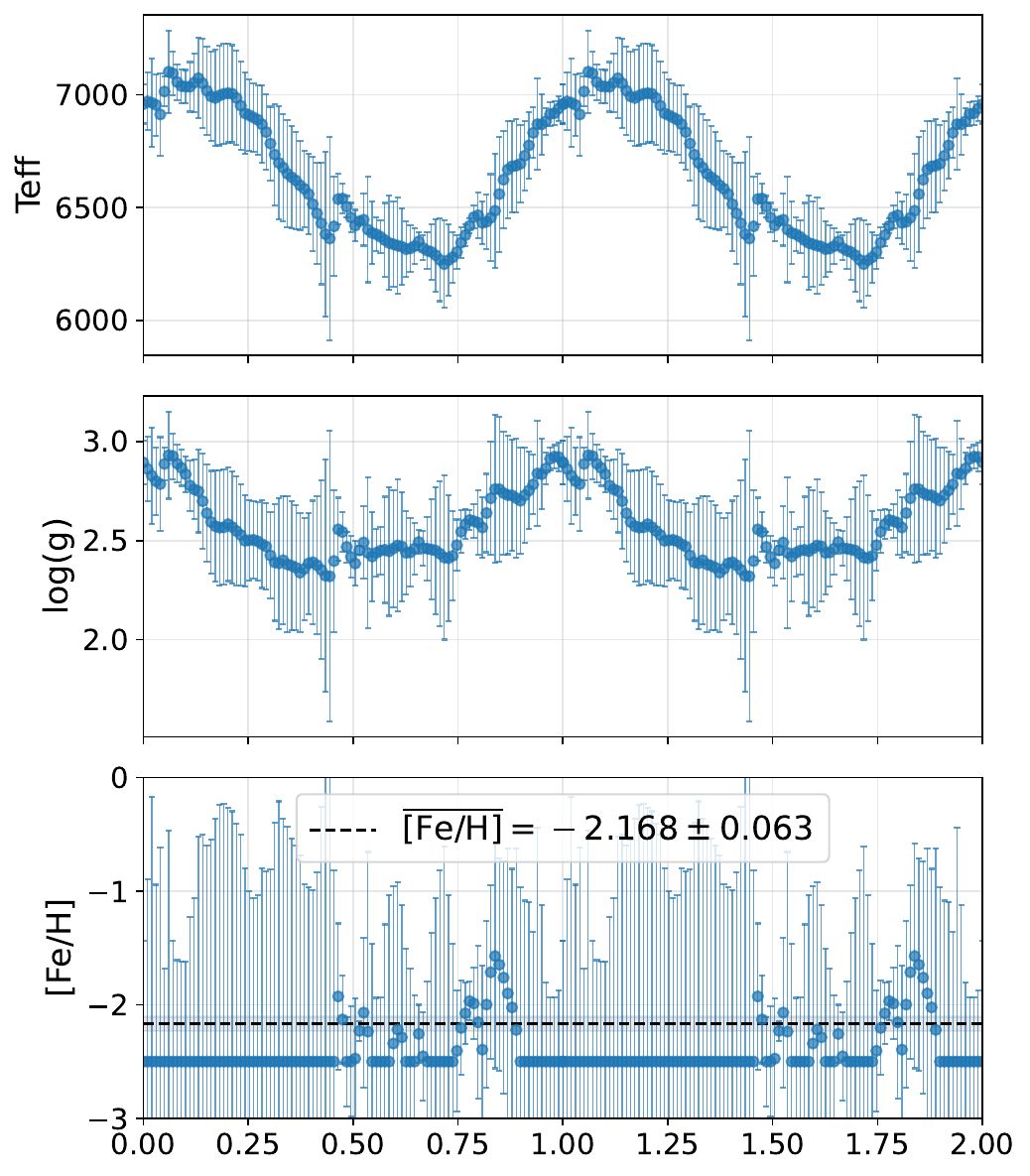}
    \caption{RRc star ID~9 (NGC~5466): phase-dependent atmospheric parameters.}
    \label{fig:app_param_9}
\end{figure}

\begin{figure}
    \centering
    \includegraphics[width=\columnwidth]{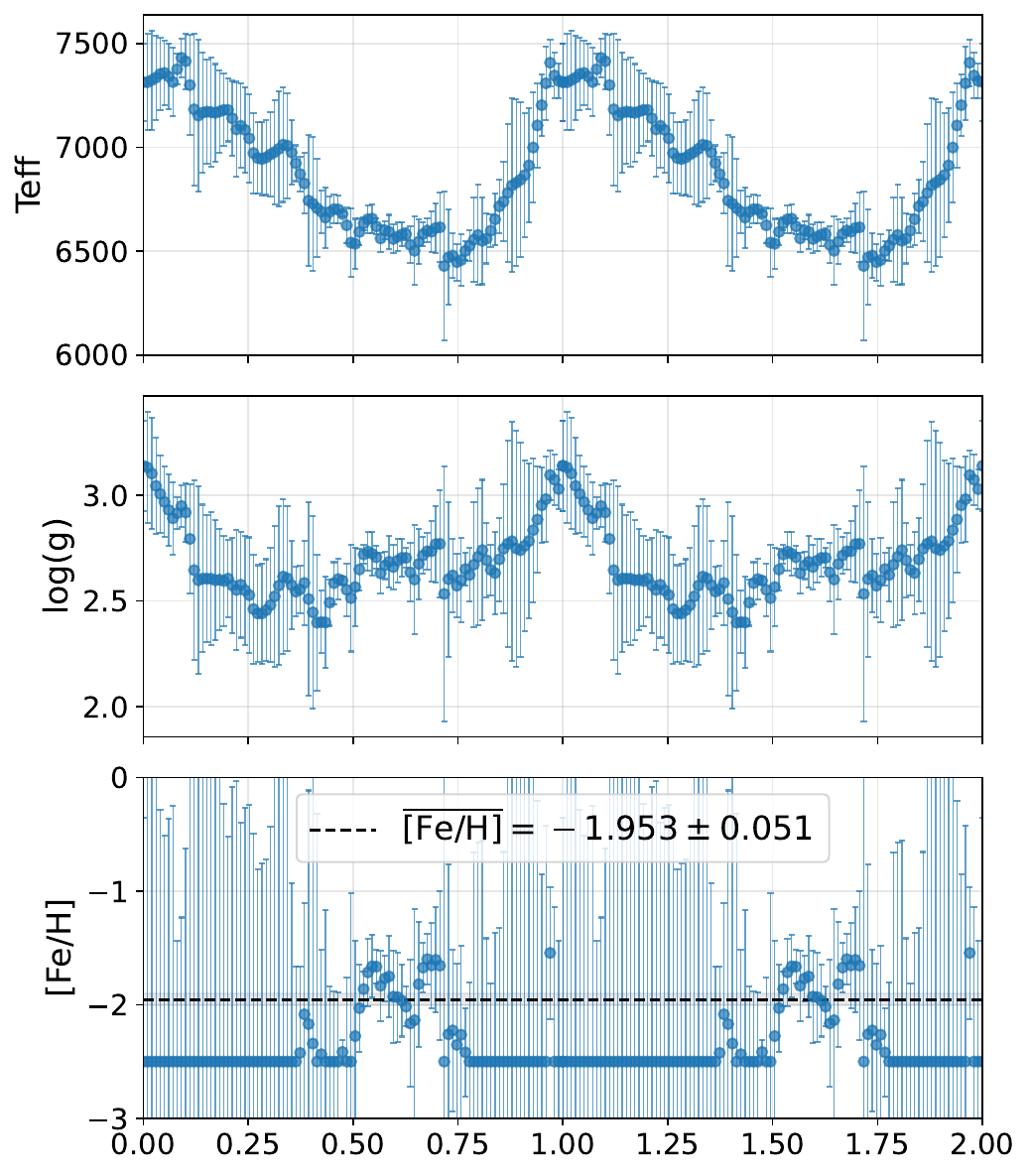}
    \caption{RRc star ID~10 (NGC~5466): phase-dependent atmospheric parameters.}
    \label{fig:app_param_10}
\end{figure}

Figures~\ref{fig:app_lc_1}--\ref{fig:app_param_10} present the multi-band light-curve templates, phase-dependent color curves, and phase-dependent atmospheric-parameter curves for the eight stars not shown in the main text. The star IDs and their corresponding cluster and Gaia DR3 source information are listed in Table~\ref{tab:rrl_sample}.

\subsection{Multi-band Light-Curve Templates}
The figures in this section show the extinction-corrected light curves and LOWESS templates in the six Mephisto bands.

\subsection{Phase-dependent Color Curves}
The figures in this section show the five phase-dependent color indices, $(u-v)$, $(v-g)$, $(g-r)$, $(r-i)$, and $(i-z)$.

\subsection{Phase-dependent Atmospheric Parameters}
The figures in this section show the inferred $T_{\rm eff}$, $\log g$, and [Fe/H] as functions of pulsation phase, together with their corresponding uncertainties.

\label{lastpage}

\end{document}